\RequirePackage{fix-cm}
\documentclass[twocolumn,epjc3]{svjour3}  
\smartqed  
\usepackage{array, amsmath, amssymb, booktabs, xcolor, eurosym, graphicx, longtable, pbox, siunitx, subcaption, url, verbatim}
\usepackage{hyperref}
\hypersetup{colorlinks,allcolors=blue}
\usepackage{orcidlink}
\usepackage[switch]{lineno}
\usepackage{balance,cite}
\usepackage{flushend}
\usepackage{lineno}
\usepackage[normalem]{ulem} 
\usepackage{soul}

\graphicspath{{./figures/}}
\journalname{Eur. Phys. J. C}
\begin{document}

\title{The XLZD Design Book: Towards the Next-Generation Liquid Xenon Observatory for Dark Matter and Neutrino Physics}

\titlerunning{The XLZD Design Book}        

\author{J.~Aalbers\thanksref{addr:groningen}
\and
K.~Abe\thanksref{addr:tokyo}
\and
M.~Adrover\thanksref{addr:zurich}
\and
S.~Ahmed Maouloud\thanksref{addr:paris}
\and
D.~S.~Akerib\thanksref{addr:SLACNationalAcceleratorLaboratory,addr:KavliInstituteforParticleAstrophysicsandCosmologyStanfordUniversity}
\and
A.~K.~Al Musalhi\thanksref{addr:UniversityCollegeLondonUCL}
\and
F.~Alder\thanksref{addr:UniversityCollegeLondonUCL}
\and
L.~Althueser\thanksref{addr:munster}
\and
D.~W.~P.~Amaral\thanksref{addr:rice}
\and
C.~S.~Amarasinghe\thanksref{addr:UniversityofCaliforniaSantaBarbara}
% ...
(XLZD Collaboration\thanksref{email1}). }
\newcommand{\banjaluka}{University of Banja Luka, 78000 Banja Luka, Bosnia and Herzegovina}
\newcommand{\barcelona}{{Departament de F\'{i}sica Qu\`{a}ntica i Astrof\'{i}sica and Institut de Ci\`{e}ncies del Cosmos, Universitat de Barcelona, 08028 Barcelona, Spain} }
\newcommand{\belgrade}{Vinca Institute of Nuclear Science, University of Belgrade, Mihajla Petrovica Alasa 12-14. Belgrade, Serbia}
\newcommand{\bern}{Albert Einstein Center for Fundamental Physics, Institute for Theoretical Physics, University of Bern, Sidlerstrasse 5, 3012 Bern, Switzerland}
\newcommand{\BlackHillsStateUniversity}{School of Natural Sciences, Black Hills State University, Spearfish, SD 57799-0002, USA}
\newcommand{\bologna}{Department of Physics and Astronomy, University of Bologna and INFN-Bologna, 40126 Bologna, Italy}
\newcommand{\BrandeisUniversity}{Department of Physics, Brandeis University, Waltham, MA 02453, USA}
\newcommand{\BrookhavenNationalLaboratoryBNL}{Brookhaven National Laboratory (BNL), Upton, NY 11973-5000, USA}
\newcommand{\BrownUniversity}{Department of Physics, Brown University, Providence, RI 02912-9037, USA}
\newcommand{\bucknell}{Department of Physics \& Astronomy, Bucknell University, Lewisburg, PA, USA}
\newcommand{\CaseWesternReserveUniversity}{Department of Physics, Case Western Reserve University, Cleveland, OH 44106, USA}
\newcommand{\CenterforUndergroundPhysicsCUP}{IBS Center for Underground Physics (CUP), Yuseong-gu, Daejeon, Korea}
\newcommand{\chicago}{Department of Physics, Enrico Fermi Institute \& Kavli Institute for Cosmological Physics, University of Chicago, Chicago, IL 60637, USA}
\newcommand{\coimbra}{LIBPhys, Department of Physics, University of Coimbra, 3004-516 Coimbra, Portugal}
\newcommand{\coimbrapoli}{Coimbra Polytechnic - ISEC, 3030-199 Coimbra, Portugal}
\newcommand{\columbia}{Physics Department, Columbia University, New York, NY 10027, USA}
\newcommand{\darmstadt}{Department of Physics, Technische Universit\"at Darmstadt, 64289 Darmstadt, Germany}
\newcommand{\dresden}{Institut f\"ur Kern und Teilchenphysik, Technische Universit\"at Dresden, 01069 Dresden, Germany}
\newcommand{\FermiNationalAcceleratorLaboratoryFNAL}{Fermi National Accelerator Laboratory (FNAL), Batavia, IL 60510-5011, USA}
\newcommand{\ferrara}{INFN-Ferrara and Dip. di Fisica e Scienze della Terra, Universit\`a di Ferrara, 44122 Ferrara, Italy}
\newcommand{\freiburg}{Physikalisches Institut, Universit\"at Freiburg, 79104 Freiburg, Germany}
\newcommand{\grenada}{University of Grenada}
\newcommand{\groningen}{Nikhef and the University of Groningen, Van Swinderen Institute, 9747AG Groningen, Netherlands}
\newcommand{\heidelberg}{Physikalisches Institut, Universit\"at Heidelberg, Heidelberg, Germany}
\newcommand{\heidelbergki}{Kirchhoff-Institut f\"ur Physik, Universit\"at Heidelberg, Heidelberg, Germany}
\newcommand{\ImperialCollegeLondon}{Department of Physics, Imperial College London, Blackett Laboratory, London SW7 2AZ, UK}
\newcommand{\KavliInstituteforParticleAstrophysicsandCosmologyStanfordUniversity}{Kavli Institute for Particle Astrophysics and Cosmology, Stanford University, Stanford, CA  94305-4085 USA}
\newcommand{\KingsCollegeLondon}{Department of Physics, King's College London, London WC2R 2LS, UK}
\newcommand{\kit}{Institute for Astroparticle Physics, Karlsruhe Institute of Technology, 76021 Karlsruhe, Germany}
\newcommand{\kitetp}{Institute of Experimental Particle Physics, Karlsruhe Institute of Technology, 76021 Karlsruhe, Germany}
\newcommand{\kitipe}{Institute for Data Processing and Electronics, Karlsruhe Institute of Technology, 76021 Karlsruhe, Germany}
\newcommand{\kobe}{Department of Physics, Kobe University, Kobe, Hyogo 657-8501, Japan}
\newcommand{\LaboratriodeInstrumentaoeFsicaExperimentaldePartculasLIP}{{Laborat\'orio de Instrumenta\c c\~ao e F\'isica Experimental de Part\'iculas (LIP)}, University of Coimbra, P-3004 516 Coimbra, Portugal}
\newcommand{\laquila}{Department of Physics and Chemistry, University of L'Aquila, 67100 L'Aquila, Italy}
\newcommand{\LawrenceBerkeleyNationalLaboratoryLBNL}{Lawrence Berkeley National Laboratory (LBNL), Berkeley, CA 94720-8099, USA}
\newcommand{\LawrenceLivermoreNationalLaboratoryLLNL}{Lawrence Livermore National Laboratory (LLNL), Livermore, CA 94550-9698, USA}
\newcommand{\lngs}{INFN-Laboratori Nazionali del Gran Sasso and Gran Sasso Science Institute, 67100 L'Aquila, Italy}
\newcommand{\mainz}{Institut f\"ur Physik \& Exzellenzcluster PRISMA$^{+}$, Johannes Gutenberg-Universit\"at Mainz, 55099 Mainz, Germany}
\newcommand{\melbourne}{ARC Centre of Excellence for Dark Matter Particle Physics, School of Physics, The University of Melbourne, VIC 3010, Australia}
\newcommand{\mpik}{Max-Planck-Institut f\"ur Kernphysik, 69117 Heidelberg, Germany}
\newcommand{\munster}{Institute for Nuclear Physics, University of M\"unster, 48149 M\"unster, Germany}
\newcommand{\nagoya}{Kobayashi-Maskawa Institute for the Origin of Particles and the Universe, and Institute for Space-Earth Environmental Research, Nagoya University, Furo-cho, Chikusa-ku, Nagoya, Aichi 464-8602, Japan}
\newcommand{\napels}{Department of Physics ``Ettore Pancini'', University of Napoli and INFN-Napoli, 80126 Napoli, Italy}
\newcommand{\nikhef}{Nikhef and the University of Amsterdam, Science Park, 1098XG Amsterdam, Netherlands}
\newcommand{\paris}{LPNHE, Sorbonne Universit\'{e}, CNRS/IN2P3, 75005 Paris, France}
\newcommand{\PennsylvaniaStateUniversity}{Department of Physics, Pennsylvania State University, University Park, PA 16802-6300, USA}
\newcommand{\purdue}{Department of Physics and Astronomy, Purdue University, West Lafayette, IN 47907, USA}
\newcommand{\rice}{Department of Physics and Astronomy, Rice University, Houston, TX 77005, USA}
\newcommand{\roma}{INFN-Roma Tre, 00146 Roma, Italy}
\newcommand{\RoyalHollowayUniversityofLondon}{Department of Physics, Royal Holloway, University of London, Egham, TW20 0EX, UK}
\newcommand{\shenzhen}{School of Science and Engineering, The Chinese University of Hong Kong (Shenzhen), Shenzhen, Guangdong, 518172, P.R. China}
\newcommand{\sissa}{Theoretical and Scientific Data Science, Scuola Internazionale Superiore di Studi Avanzati (SISSA), 34136 Trieste, Italy}
\newcommand{\SLACNationalAcceleratorLaboratory}{SLAC National Accelerator Laboratory, Menlo Park, CA 94025-7015, USA}
\newcommand{\SouthDakotaSchoolofMinesandTechnology}{South Dakota School of Mines and Technology, Rapid City, SD 57701-3901, USA}
\newcommand{\SouthDakotaScienceandTechnologyAuthoritySDSTA}{South Dakota Science and Technology Authority (SDSTA), Sanford Underground Research Facility, Lead, SD 57754-1700, USA}
\newcommand{\STFCRutherfordAppletonLaboratoryRAL}{STFC Rutherford Appleton Laboratory (RAL), Didcot, OX11 0QX, UK}
\newcommand{\stockholm}{Oskar Klein Centre, Department of Physics, Stockholm University, AlbaNova, Stockholm SE-10691, Sweden}
\newcommand{\subatech}{SUBATECH, IMT Atlantique, CNRS/IN2P3,  Nantes Universit\'e, Nantes 44307, France}
\newcommand{\sydney}{School of Physics, The University of Sydney, Camperdown, Sydney, NSW 2006, Australia}
\newcommand{\tokyo}{Kamioka Observatory, Institute for Cosmic Ray Research, and Kavli Institute for the Physics and Mathematics of the Universe (WPI), University of Tokyo, Higashi-Mozumi, Kamioka, Hida, Gifu 506-1205, Japan}
\newcommand{\torino}{INAF-Astrophysical Observatory of Torino, Department of Physics, University  of  Torino and  INFN-Torino,  10125  Torino,  Italy}
\newcommand{\tsinghua}{Department of Physics \& Center for High Energy Physics, Tsinghua University, Beijing 100084, P.R. China}
\newcommand{\ucsd}{Department of Physics, University of California San Diego, La Jolla, CA 92093, USA}
\newcommand{\UniversityCollegeLondonUCL}{Department of Physics and Astronomy, University College London (UCL), London WC1E 6BT, UK}
\newcommand{\UniversityofAlabama}{Department of Physics \& Astronomy, University of Alabama, Tuscaloosa, AL 34587-0324, USA}
\newcommand{\UniversityofBristol}{H.H. Wills Physics Laboratory, University of Bristol, Bristol, BS8 1TL, UK}
\newcommand{\UniversityofCaliforniaBerkeley}{Department of Physics, University of California, Berkeley,  Berkeley, CA 94720-7300, USA}
\newcommand{\UniversityofCaliforniaDavis}{Department of Physics, University of California, Davis, Davis, CA 95616-5270, USA}
\newcommand{\UniversityofCaliforniaLosAngeles}{Department of Physics \& Astronomy, University of Califonia, Los Angeles, Los Angeles, CA 90095-1547}
\newcommand{\UniversityofCaliforniaSantaBarbara}{Department of Physics, University of California, Santa Barbara,  Santa Barbara, CA 93106-9530, USA}
\newcommand{\UniversityofEdinburgh}{SUPA, School of Physics and Astronomy, University of Edinburgh, Edinburgh,  EH9 3FD, UK}
\newcommand{\UniversityofLiverpool}{Department of Physics, University of Liverpool, Liverpool L69 7ZE, UK}
\newcommand{\UniversityofMaryland}{Department of Physics, University of Maryland, College Park, MD 20742-4111, USA}
\newcommand{\UniversityofMassachusetts}{Department of Physics, University of Massachusetts, Amherst, MA 01003-9337, USA}
\newcommand{\UniversityofMichigan}{Randall Laboratory of Physics, University of Michigan, Ann Arbor, MI 48109-1040, USA}
\newcommand{\UniversityofOxford}{Department of Physics, University of Oxford, Oxford OX1 3RH, UK}
\newcommand{\UniversityofRochester}{Department of Physics and Astronomy, University of Rochester, Rochester, NY 14627-0171, USA}
\newcommand{\UniversityofSheffield}{School of Mathematical and Physical Sciences, University of Sheffield, Sheffield S3 7RH, UK}
\newcommand{\UniversityofTexasatAustin}{Department of Physics, University of Texas at Austin, Austin, TX 78712-1192, USA}
\newcommand{\UniversityofWisconsinMadison}{Physical Sciences Laboratory, University of Wisconsin-Madison, Madison, WI 53589-3034, USA}
\newcommand{\VaticanObservatory}{Vatican Observatory, Castel Gandolfo, V-00120, Vatican City State}
\newcommand{\westlake}{Department of Physics, School of Science, Westlake University, Hangzhou 310030, P.R. China}
\newcommand{\wis}{Department of Particle Physics and Astrophysics, Weizmann Institute of Science, Rehovot 7610001, Israel}
\newcommand{\zurich}{Physik-Institut, University of Z\"urich, 8057  Z\"urich, Switzerland}
\authorrunning{XLZD Collaboration}
\thankstext{addr:banjaluka}{Also at \banjaluka}\hypertarget{addr:banjaluka}{}
\thankstext{addr:roma}{Also at \roma}\hypertarget{addr:roma}{}
\thankstext{addr:coimbrapoli}{Also at \coimbrapoli}\hypertarget{addr:coimbrapoli}{}
\thankstext{addr:grenada}{Also at \grenada}\hypertarget{addr:grenada}{}

\thankstext{email1}{\texttt{contact@xlzd.org}}\hypertarget{email1}{}

\institute{\hypertarget{addr:groningen}{\groningen}\label{addr:groningen}
\and
\hypertarget{addr:tokyo}{\tokyo}\label{addr:tokyo}
\and
\hypertarget{addr:zurich}{\zurich}\label{addr:zurich}
\and
\hypertarget{addr:paris}{\paris}\label{addr:paris}
\and
\hypertarget{addr:SLACNationalAcceleratorLaboratory}{\SLACNationalAcceleratorLaboratory}\label{addr:SLACNationalAcceleratorLaboratory}
\and
\hypertarget{addr:KavliInstituteforParticleAstrophysicsandCosmologyStanfordUniversity}{\KavliInstituteforParticleAstrophysicsandCosmologyStanfordUniversity}\label{addr:KavliInstituteforParticleAstrophysicsandCosmologyStanfordUniversity}
\and
\hypertarget{addr:UniversityCollegeLondonUCL}{\UniversityCollegeLondonUCL}\label{addr:UniversityCollegeLondonUCL}
\and
\hypertarget{addr:munster}{\munster}\label{addr:munster}
\and
\hypertarget{addr:rice}{\rice}\label{addr:rice}
\and
\hypertarget{addr:UniversityofCaliforniaSantaBarbara}{\UniversityofCaliforniaSantaBarbara}\label{addr:UniversityofCaliforniaSantaBarbara}
\and
\hypertarget{addr:ImperialCollegeLondon}{\ImperialCollegeLondon}\label{addr:ImperialCollegeLondon}
\and
\hypertarget{addr:torino}{\torino}\label{addr:torino}
\and
\hypertarget{addr:lngs}{\lngs}\label{addr:lngs}
\and
\hypertarget{addr:belgrade}{\belgrade}\label{addr:belgrade}
\and
\hypertarget{addr:columbia}{\columbia}\label{addr:columbia}
\and
\hypertarget{addr:UniversityofMaryland}{\UniversityofMaryland}\label{addr:UniversityofMaryland}
\and
\hypertarget{addr:UniversityofAlabama}{\UniversityofAlabama}\label{addr:UniversityofAlabama}
\and
\hypertarget{addr:KingsCollegeLondon}{\KingsCollegeLondon}\label{addr:KingsCollegeLondon}
\and
\hypertarget{addr:kitipe}{\kitipe}\label{addr:kitipe}
\and
\hypertarget{addr:BrownUniversity}{\BrownUniversity}\label{addr:BrownUniversity}
\and
\hypertarget{addr:melbourne}{\melbourne}\label{addr:melbourne}
\and
\hypertarget{addr:subatech}{\subatech}\label{addr:subatech}
\and
\hypertarget{addr:LawrenceBerkeleyNationalLaboratoryLBNL}{\LawrenceBerkeleyNationalLaboratoryLBNL}\label{addr:LawrenceBerkeleyNationalLaboratoryLBNL}
\and
\hypertarget{addr:bologna}{\bologna}\label{addr:bologna}
\and
\hypertarget{addr:UniversityofWisconsinMadison}{\UniversityofWisconsinMadison}\label{addr:UniversityofWisconsinMadison}
\and
\hypertarget{addr:mpik}{\mpik}\label{addr:mpik}
\and
\hypertarget{addr:kit}{\kit}\label{addr:kit}
\and
\hypertarget{addr:UniversityofEdinburgh}{\UniversityofEdinburgh}\label{addr:UniversityofEdinburgh}
\and
\hypertarget{addr:sydney}{\sydney}\label{addr:sydney}
\and
\hypertarget{addr:BrookhavenNationalLaboratoryBNL}{\BrookhavenNationalLaboratoryBNL}\label{addr:BrookhavenNationalLaboratoryBNL}
\and
\hypertarget{addr:LaboratriodeInstrumentaoeFsicaExperimentaldePartculasLIP}{\LaboratriodeInstrumentaoeFsicaExperimentaldePartculasLIP}\label{addr:LaboratriodeInstrumentaoeFsicaExperimentaldePartculasLIP}
\and
\hypertarget{addr:wis}{\wis}\label{addr:wis}
\and
\hypertarget{addr:STFCRutherfordAppletonLaboratoryRAL}{\STFCRutherfordAppletonLaboratoryRAL}\label{addr:STFCRutherfordAppletonLaboratoryRAL}
\and
\hypertarget{addr:kitetp}{\kitetp}\label{addr:kitetp}
\and
\hypertarget{addr:freiburg}{\freiburg}\label{addr:freiburg}
\and
\hypertarget{addr:UniversityofSheffield}{\UniversityofSheffield}\label{addr:UniversityofSheffield}
\and
\hypertarget{addr:UniversityofLiverpool}{\UniversityofLiverpool}\label{addr:UniversityofLiverpool}
\and
\hypertarget{addr:tsinghua}{\tsinghua}\label{addr:tsinghua}
\and
\hypertarget{addr:PennsylvaniaStateUniversity}{\PennsylvaniaStateUniversity}\label{addr:PennsylvaniaStateUniversity}
\and
\hypertarget{addr:heidelberg}{\heidelberg}\label{addr:heidelberg}
\and
\hypertarget{addr:RoyalHollowayUniversityofLondon}{\RoyalHollowayUniversityofLondon}\label{addr:RoyalHollowayUniversityofLondon}
\and
\hypertarget{addr:SouthDakotaSchoolofMinesandTechnology}{\SouthDakotaSchoolofMinesandTechnology}\label{addr:SouthDakotaSchoolofMinesandTechnology}
\and
\hypertarget{addr:UniversityofBristol}{\UniversityofBristol}\label{addr:UniversityofBristol}
\and
\hypertarget{addr:nikhef}{\nikhef}\label{addr:nikhef}
\and
\hypertarget{addr:stockholm}{\stockholm}\label{addr:stockholm}
\and
\hypertarget{addr:UniversityofRochester}{\UniversityofRochester}\label{addr:UniversityofRochester}
\and
\hypertarget{addr:SouthDakotaScienceandTechnologyAuthoritySDSTA}{\SouthDakotaScienceandTechnologyAuthoritySDSTA}\label{addr:SouthDakotaScienceandTechnologyAuthoritySDSTA}
\and
\hypertarget{addr:mainz}{\mainz}\label{addr:mainz}
\and
\hypertarget{addr:UniversityofOxford}{\UniversityofOxford}\label{addr:UniversityofOxford}
\and
\hypertarget{addr:laquila}{\laquila}\label{addr:laquila}
\and
\hypertarget{addr:nagoya}{\nagoya}\label{addr:nagoya}
\and
\hypertarget{addr:purdue}{\purdue}\label{addr:purdue}
\and
\hypertarget{addr:napels}{\napels}\label{addr:napels}
\and
\hypertarget{addr:UniversityofCaliforniaBerkeley}{\UniversityofCaliforniaBerkeley}\label{addr:UniversityofCaliforniaBerkeley}
\and
\hypertarget{addr:chicago}{\chicago}\label{addr:chicago}
\and
\hypertarget{addr:UniversityofMichigan}{\UniversityofMichigan}\label{addr:UniversityofMichigan}
\and
\hypertarget{addr:UniversityofMassachusetts}{\UniversityofMassachusetts}\label{addr:UniversityofMassachusetts}
\and
\hypertarget{addr:bern}{\bern}\label{addr:bern}
\and
\hypertarget{addr:ucsd}{\ucsd}\label{addr:ucsd}
\and
\hypertarget{addr:UniversityofCaliforniaLosAngeles}{\UniversityofCaliforniaLosAngeles}\label{addr:UniversityofCaliforniaLosAngeles}
\and
\hypertarget{addr:bucknell}{\bucknell}\label{addr:bucknell}
\and
\hypertarget{addr:UniversityofTexasatAustin}{\UniversityofTexasatAustin}\label{addr:UniversityofTexasatAustin}
\and
\hypertarget{addr:heidelbergki}{\heidelbergki}\label{addr:heidelbergki}
\and
\hypertarget{addr:CenterforUndergroundPhysicsCUP}{\CenterforUndergroundPhysicsCUP}\label{addr:CenterforUndergroundPhysicsCUP}
\and
\hypertarget{addr:dresden}{\dresden}\label{addr:dresden}
\and
\hypertarget{addr:westlake}{\westlake}\label{addr:westlake}
\and
\hypertarget{addr:shenzhen}{\shenzhen}\label{addr:shenzhen}
\and
\hypertarget{addr:coimbra}{\coimbra}\label{addr:coimbra}
\and
\hypertarget{addr:LawrenceLivermoreNationalLaboratoryLLNL}{\LawrenceLivermoreNationalLaboratoryLLNL}\label{addr:LawrenceLivermoreNationalLaboratoryLLNL}
\and
\hypertarget{addr:barcelona}{\barcelona}\label{addr:barcelona}
\and
\hypertarget{addr:kobe}{\kobe}\label{addr:kobe}
\and
\hypertarget{addr:VaticanObservatory}{\VaticanObservatory}\label{addr:VaticanObservatory}
\and
\hypertarget{addr:BlackHillsStateUniversity}{\BlackHillsStateUniversity}\label{addr:BlackHillsStateUniversity}
\and
\hypertarget{addr:sissa}{\sissa}\label{addr:sissa}
\and
\hypertarget{addr:darmstadt}{\darmstadt}\label{addr:darmstadt}
\and
\hypertarget{addr:UniversityofCaliforniaDavis}{\UniversityofCaliforniaDavis}\label{addr:UniversityofCaliforniaDavis}
\and
\hypertarget{addr:ferrara}{\ferrara}\label{addr:ferrara}
}

\date{Received: date / Accepted: date}
\onecolumn 
\maketitle
\twocolumn

\abstract {This report describes the experimental strategy and technologies for XLZD, the next-generation xenon observatory sensitive to dark matter and neutrino physics. In the baseline design, the detector will have an active liquid xenon target of 60\,tonnes, which could be increased to 80\,tonnes if the market conditions for xenon are favorable. It is based on the mature liquid xenon time projection chamber technology used in current-generation experiments, LZ and XENONnT. The report discusses the baseline design and opportunities for further optimization of the individual detector components. The experiment envisaged here has the capability to explore parameter space for Weakly Interacting Massive Particle (WIMP) dark matter down to the neutrino fog, with a 3$\sigma$ evidence potential for WIMP-nucleon cross sections as low as $3\times10^{-49}\rm\,cm^2$ (at 40\,GeV/c$^2$ WIMP mass). The observatory will also have leading sensitivity to a wide range of alternative dark matter models. It is projected to have a 3$\sigma$ observation potential of neutrinoless double beta decay of $^{136}$Xe at a half-life of up to $5.7\times 10^{27}$~years. Additionally, it is sensitive to astrophysical neutrinos from the sun and galactic supernovae.}

\newcommand{\tonneyear}{\ensuremath{\mathrm{t}\times\mathrm{y}}}
\newcommand{\cp}[2]{\vspace{-3mm}\color{purple}\textbf{\underline{$\ggg$Volunteers: #1 | Max Pages: #2 $\lll$}}\color{black}}
\newcommand{\cpg}[2]{\vspace{-3mm}\color{teal}\textbf{\underline{$\ggg$Volunteers: #1 | Max Pages: #2 $\lll$}}\color{black}}
\newcommand{\FIXME}[1]{\textcolor{red}{{\bf \,FIXME:} #1}}

\section{Introduction}\label{sec:executivesummary}

The nature of dark matter is one of the most important unsolved questions in physics today. Liquid xenon time projection chambers (LXe-TPCs) have been leading the search for Weakly Interacting Massive Particle (WIMP) dark matter candidates above a few GeV/c$^2$ for over a decade, building on many generations of experiments with ever-increasing target mass~\cite{Alner:2007ja,Lebedenko:2008gb,Angle:2007uj,Aprile:2016swn,Aprile:2017iyp,Akerib:2016vxi,Xiao:2014xyn,Wang:2020coa,PandaX-4T:2021bab}. The latest iterations of these experiments have demonstrated world-leading sensitivity and unprecedented low background levels with detector masses of several tonnes~\cite{LZ:2022ufs,LZ_2024_DM_results,XENON:2023sxq,aprile2025wimpdarkmattersearch,PandaX-4T:2024dm}. The XLZD (XENON-LUX-ZEPLIN-DARWIN) collaboration brings together the collective expertise from these experiments and the leading R\&D efforts to build a dual-phase LXe-TPC. This XLZD experiment will have the capability to explore the WIMP parameter space down to cross sections where the expected number of neutrino events in the signal region would be equal to or larger than WIMP events - often referred to as the neutrino fog or floor~\cite{Billard:2013qya,O'Hare:2015mda,OHare:2021utq}. The envisioned detector could readily accomplish a 200\,t$\cdot$y fiducial exposure within its initial operating period while being designed to allow for a sensitivity improvement with an exposure of up to 1000\,t$\cdot$y to deliver a 3$\sigma$ discovery at the background-limited neutrino fog. Due to the large number of naturally occurring xenon isotopes, XLZD's sensitivity will extend to spin-dependent and a range of effective WIMP-nuclear couplings. The XLZD experiment will also be competitive in the search for the neutrinoless double-beta decay ($0\nu\beta\beta$) of $^{136}$Xe and precision measurements of astrophysical neutrinos, making it the definitive rare event observatory with considerable impact on particle, nuclear and astrophysics~\cite{Aalbers:2022dzr}.

In this report, we present the experimental strategy and provisional design of the XLZD detector with 60--80\,tonnes xenon target mass depending on the acquisition rate of the xenon. The detector design is driven by the requirement to minimize the particle detection threshold while optimizing the self-shielding of external backgrounds from the laboratory environment and detector construction materials, mitigation of internal trace contaminants, mostly radon and krypton, in the liquid xenon target, as well as instrument-related accidental coincidence backgrounds. The mature technology of current-generation LXe-TPCs provides a reliable base for our design, profiting from the often complementary strengths and expertise of the merging collaborations. Existing screening techniques and purification mechanisms are in place or are currently being developed to ensure that backgrounds due to trace radioactivity fall below the level of irreducible neutrino backgrounds in the form of neutrino-electron and neutrino-nucleus scattering. Technical risks exist, as is to be expected for such an ambitious program~\cite{Kopec:2023uii}. However, these risks are well understood, and mitigation strategies such as early, long-term, and large-scale testing are planned.

\section{Science Sensitivity and Design Drivers}\label{sec:science}

\begin{figure*}[!h]
\begin{center}
\includegraphics[width=0.9\textwidth]{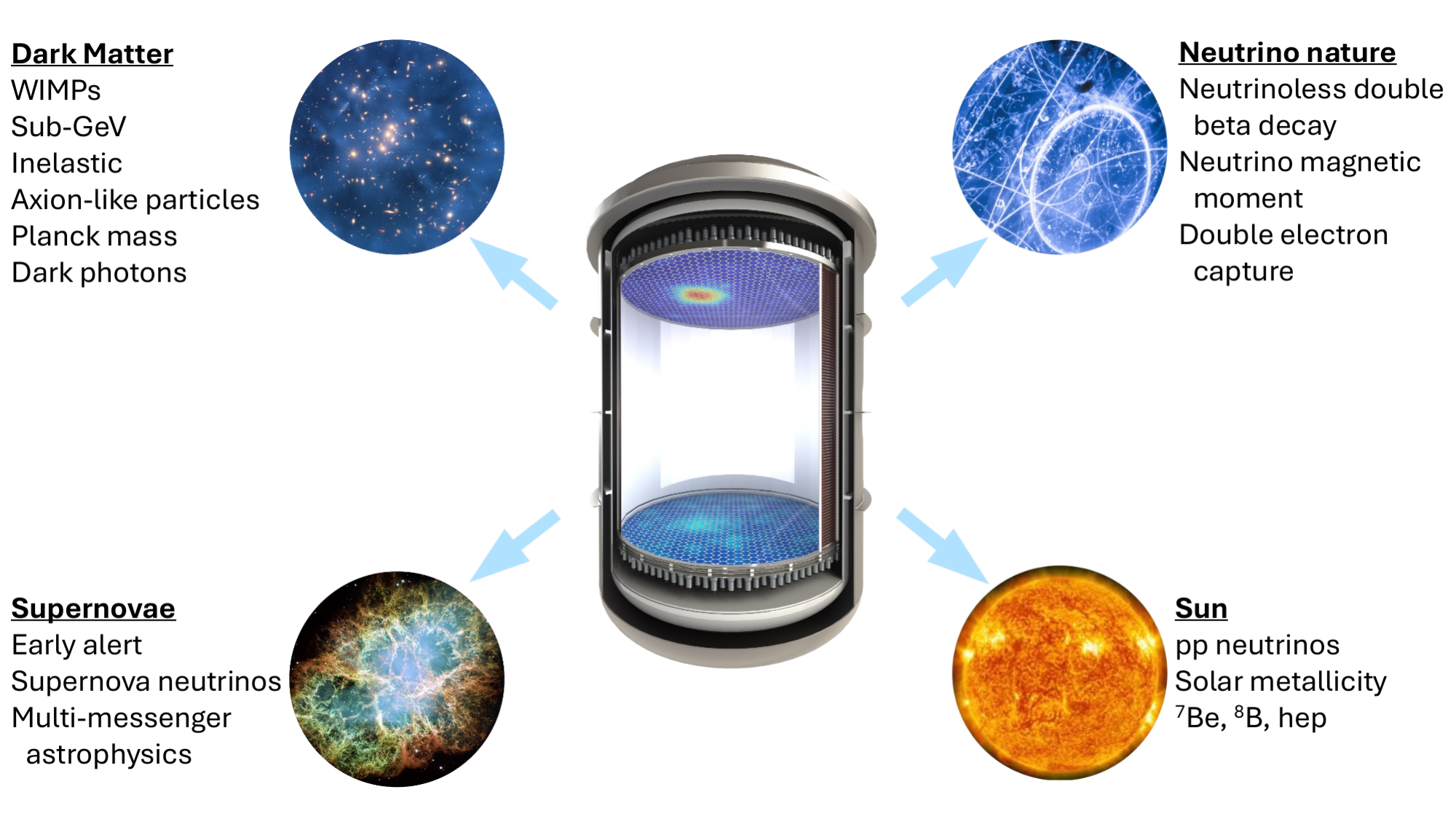}
\caption{\small The science channels of the proposed LXe observatory for rare events span many areas and are of interest to particle physics, nuclear physics, astrophysics, solar physics, and cosmology.}\label{fig:sciencechannels}
\end{center}
\end{figure*}

Advances in liquid xenon (LXe) detectors open the possibility to design, commission, and operate an observatory for several extremely rare processes with far-reaching impact for particle-, astro- and nuclear physics and associated fields~\cite{Aalbers:2022dzr} -- see~Fig.~\ref{fig:sciencechannels}. Such a detector is ideal for probing some of the most popular dark matter models, with the potential for a momentous discovery allowing characterization of the particle nature of galactic dark matter. The same detector can also search for neutrinoless double-beta decay of $^{136}$Xe to determine if the neutrino is its own antiparticle, i.e., a Majorana particle. It can also function as an observatory for several astrophysical neutrino sources, including solar {\it pp} and $^8$B neutrinos, and neutrinos from a galactic supernova explosion. 
 The XLZD observatory thus will be pivotal to tackle some of the biggest mysteries in physics.

\subsection{Dark Matter}\label{sec:darkmatter}

\begin{figure*}[!htbp]
\centering
\includegraphics[width=.5\textwidth]{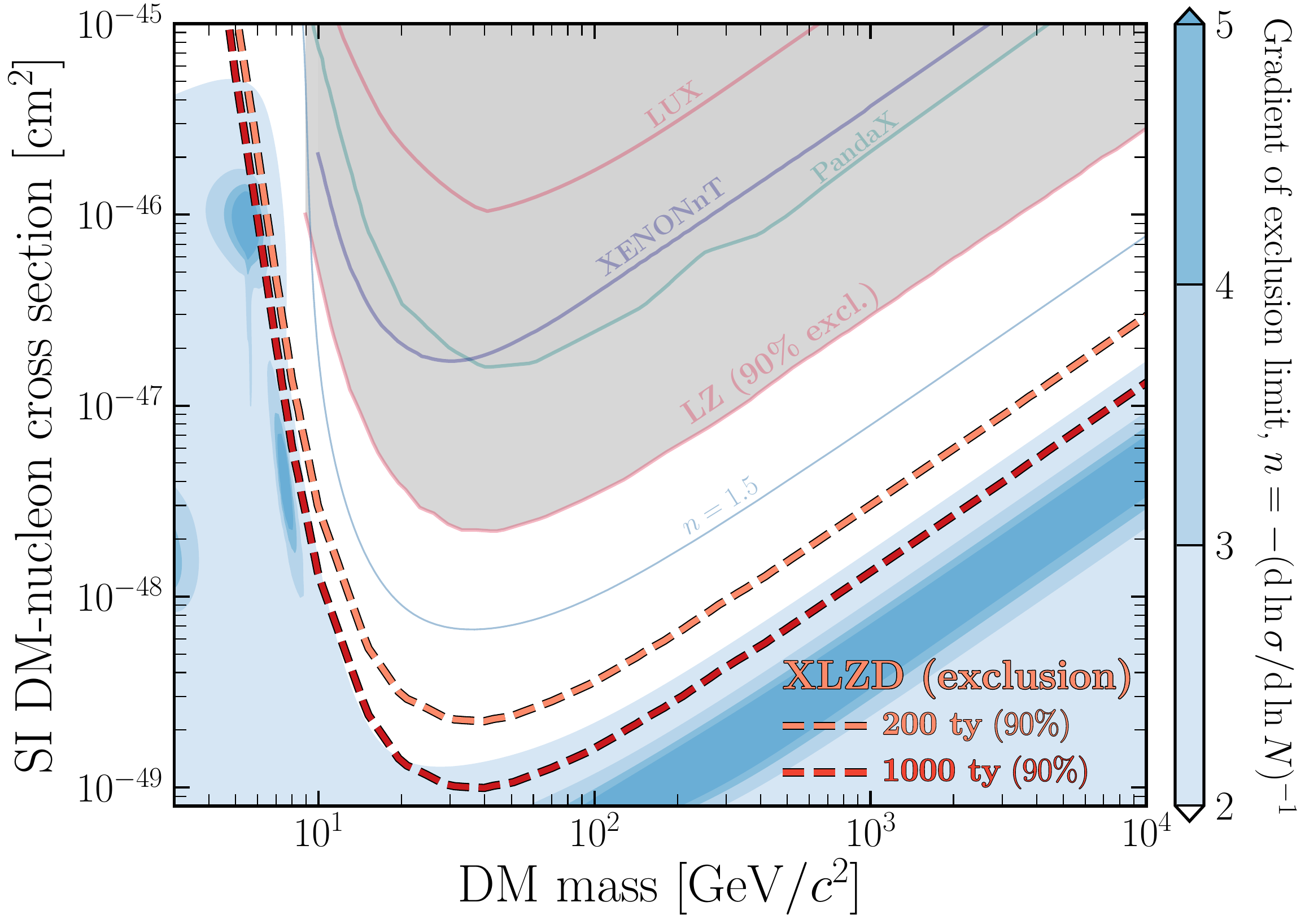}
\hfill
\includegraphics[width=.45\textwidth]{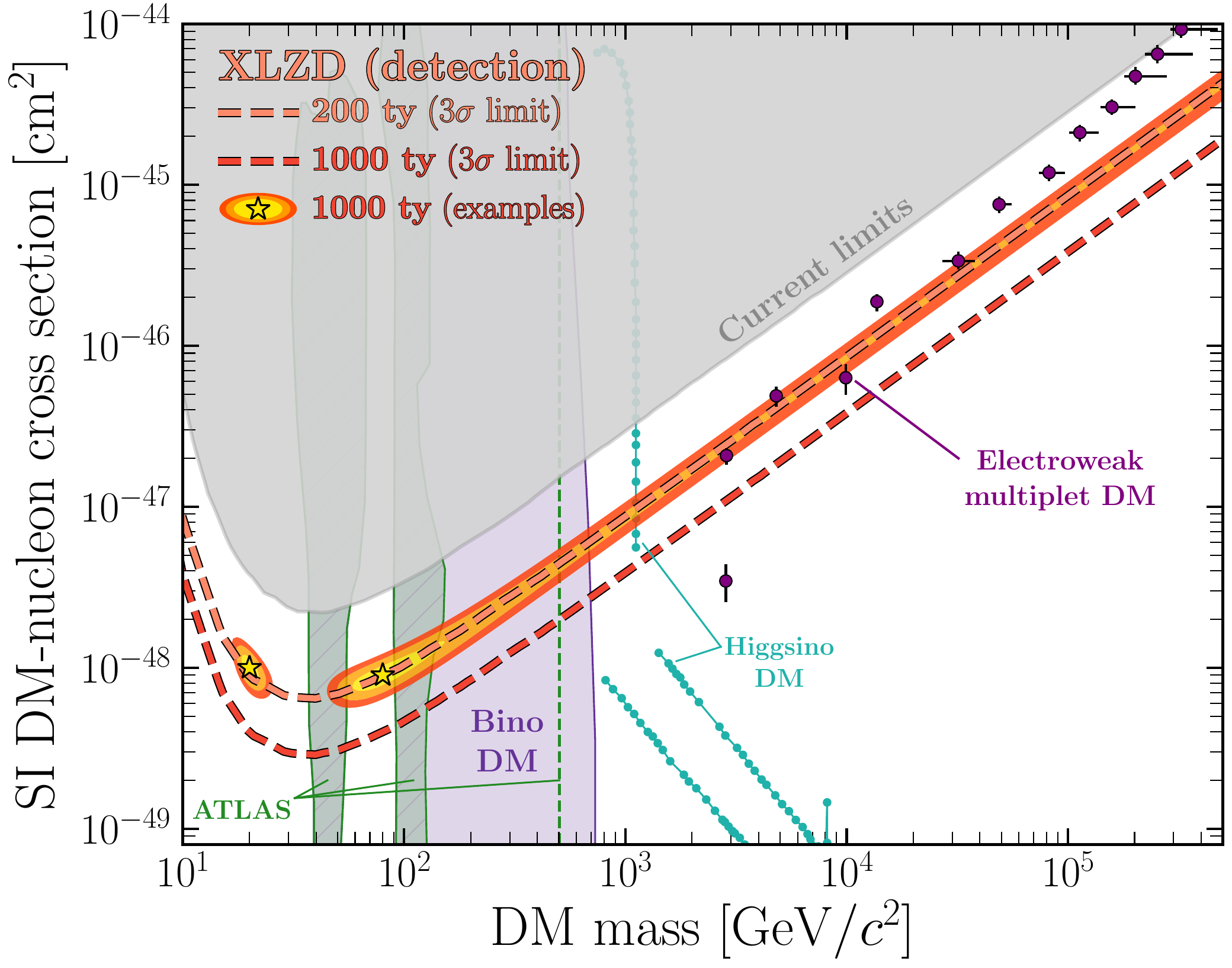}
\caption{\small \textit{Left:} Projected $90$\%~C.L. upper limits on the spin-independent WIMP-nucleon cross section for 200 and 1000\,tonne-year\,(t$\cdot$y) exposures of the XLZD detector, along with current upper limits~\cite{LZ_2024_DM_results,aprile2025wimpdarkmattersearch,PandaX-4T:2024dm,Akerib:2015rjg}. 
The blue shaded regions illustrate the neutrino fog as defined in\cite{OHare:2021utq}. \textit{Right:} The dashed contours indicate the median 3$\sigma$ detection limit for 200 and 1000\,t$\cdot$y exposure. Example evidence contours for 20 and 80\,GeV WIMPs are shown with confidence intervals of 1, 2, and 3$\sigma$ (yellow, orange, and red, respectively). These illustrate that extending exposure from 200\,t$\cdot$y to 1000\,t$\cdot$y significantly improves our ability to constrain dark matter properties after initial detection. The figure displays several well-motivated dark matter candidates within XLZD's reach: Electroweak multiplet DM~\cite{Bloch:2024} is an example of a minimal dark matter model still largely unexplored, while Higgsino~\cite{Ellis_2023} and Bino DM~\cite{ATLAS2024} candidates arise from supersymmetry. In green, we highlight that XLZD is highly 
complementary to collider experiments, with the shaded region showing Bino DM exclusion limits from ATLAS~\cite{ATLAS2024} and the 
vertical green line indicating the maximum mass testable with ATLAS electroweak searches for this DM model.}
\label{fig:wimps}
\end{figure*}

The gravitational effects of dark matter are evident on astrophysical and cosmic scales~\cite{Clowe:2006eq,aghanim2020planck}, but the fundamental nature of dark matter remains a mystery. One of the most compelling hypotheses is that dark matter consists of new subatomic particles, one of the most prominent candidates being the Weakly Interacting Massive Particles (WIMPs)
~\cite{gelmini2010dm,Jungman:1995df}. The relic abundance of dark matter is easily reproducible by adding a coupling to Standard Model (SM) particles at the electroweak scale~\cite{Bertone:2010zza}. 
Xenon, with an average atomic mass of $\sim$130\,GeV/c$^2$, is an ideal target to search for kinematic collisions from such WIMPs. In the simplest case, the interaction would be a spin-independent (SI) interaction between a WIMP and a xenon nucleus. For the last two decades, experiments utilizing LXe-TPCs have led the search for WIMPs~\cite{Alner:2007ja,Lebedenko:2008gb,Angle:2007uj,Aprile:2016swn,Akerib:2016vxi,PandaX-4T:2021bab,AKIMOV201214,Aprile:2018dbl}, with current most-competitive results derived from the results of the still running LUX-ZEPLIN (LZ), PandaX-4T and XENONnT detectors~\cite{LZ_2024_DM_results,aprile2025wimpdarkmattersearch,PandaX-4T:2024dm}.
While these experiments have ruled out various WIMP candidates, many well motivated WIMPs remain, which can be realized in simple extensions to the SM. For example, electroweak multiplet DM~\cite{Bloch:2024} and $Z^\prime$ mediated models~\cite{Blanco:2019hah} provide minimal WIMP candidates that satisfy experimental constraints. More complex extensions to the SM, such as supersymmetry, can also offer viable WIMPs~\cite{ATLAS2024,Ellis_2023}. In these benchmark cases, significant parts of the remaining parameter space will be explored by XLZD (see Fig.~\ref{fig:wimps}), probing higher WIMP masses than accessible at the LHC~\cite{ATLAS2024}. 

XLZD will also have leading sensitivity to spin-dependent WIMP-nucleon scattering via the naturally occurring $^{129}$Xe (spin 1/2, 26.4\% natural abundance) and $^{131}$Xe (spin 3/2, 21.2\% natural abundance) isotopes. Current generation liquid xenon detectors have demonstrated this for both WIMP-neutron and WIMP-proton scattering~\cite{LZ_2024_DM_results,aprile2025wimpdarkmattersearch,PandaX-4T:2024dm}. In case of a discovery, varying the isotopic abundance of $^{129}$Xe and $^{131}$Xe would allow testing the spin-dependent character of the WIMP-nuclear response and, together with other handles such as the dependence of nuclear responses on momentum transfer~\cite{Fieguth:2018vob}, provide information on the nature of the WIMP.

Though designed to target WIMP-nuclear scattering, the XLZD observatory has exceptional sensitivity to other types of WIMP interactions (for example, WIMP-pion scattering~\cite{Aprile:2018cxk}, inelastic WIMP-nucleus scattering~\cite{Baudis:2013bba}, and Effective Field Theory analysis~\cite{LZ:2023lvz,Hoferichter:2016nvd,Hoferichter:2018acd}) and additional well-motivated dark-matter candidates: sub-GeV~\cite{Essig:2012yx}, dark photon~\cite{Pospelov:2008jk,Essig:2013lka}, axion-like particle~\cite{Abbott:1982af,raffelt2002axions,Preskill:1982cy,duffy2009axions}, and Planck mass dark matter~\cite{XENON:2023iku}. Prominent examples are shown in Fig.~\ref{fig:sciencechannels} with sensitivity studies summarised in Ref.~\cite{Aalbers:2022dzr}.

Figure~\ref{fig:wimps} illustrates the decisive progress achievable with the XLZD detector in the flagship spin-independent WIMP-nucleon search channel.
As the definitive WIMP discovery instrument, the experiment must reach into the neutrino fog~\cite{Billard:2013qya,O'Hare:2015mda,OHare:2021utq}, where its discovery potential becomes systematically limited by the coherent nuclear scattering of astrophysical neutrinos (CEvNS). In the nuclear recoil (NR) signal region, non-neutrino background events are maintained at the order of one event within the entire exposure. However, the expected reach into the neutrino fog predicts more than one coherent elastic neutrino-nucleus scatter. The other backgrounds arise from neutrons, which can be reduced by low-background material selection, vetoed using a set of nested outer detectors surrounding the LXe-TPC, and discriminated against with potential multiple scatterings inside the detector. In addition, leakage from the more numerous electronic recoil (ER) backgrounds into the nuclear recoil signal region will need to be controlled. With a target suppression of $^{222}$Rn to $\SI{0.1}{\mu Bq /kg}$ and $^{nat}$Kr to 0.03\,ppt, the processes dominating electron recoils will be those from solar (mostly {\it pp}) neutrinos scattering off electrons and decays of naturally occurring $^{136}$Xe in the LXe~\cite{Schumann:2015cpa}. Background reduction strategies to achieve these goals are further discussed in Section~\ref{sec:backgrounds}.

With a conservative 200\,t$\cdot$y exposure, assuming a 
3~keV energy threshold, successful suppression of radioactive backgrounds below the irreducible neutrino backgrounds and negligible leakage of electron recoil events from solar neutrinos into the signal region, the experiment will deliver an order of magnitude improvement in exclusion sensitivity and discovery capability compared to current experiments. For a 40\,GeV/c$^2$ WIMP it will reach 90\% exclusion sensitivity down to a cross-section of $2 \times 10^{-49}\rm\,cm^{2}$ and 3$\sigma$ evidence capability at a cross-section of $7\times 10^{-49}\rm\,cm^{2}$. In order to be the definitive xenon experiment, the detector must be capable of running up to a 1000\,t$\cdot$y exposure without becoming limited by backgrounds from radioactive impurities, making it sensitive to a potential 3$\sigma$ evidence at a cross-section of $3\times 10^{-49}\rm\,cm^{2}$ for a 40\,GeV/c$^2$ WIMP as shown in Fig.~\ref{fig:wimps}. 

\subsection{Neutrinoless Double Beta Decay}\label{sec:doublebeta}
\begin{figure*}[!h]
    \centering
    \includegraphics[height=5.8cm]{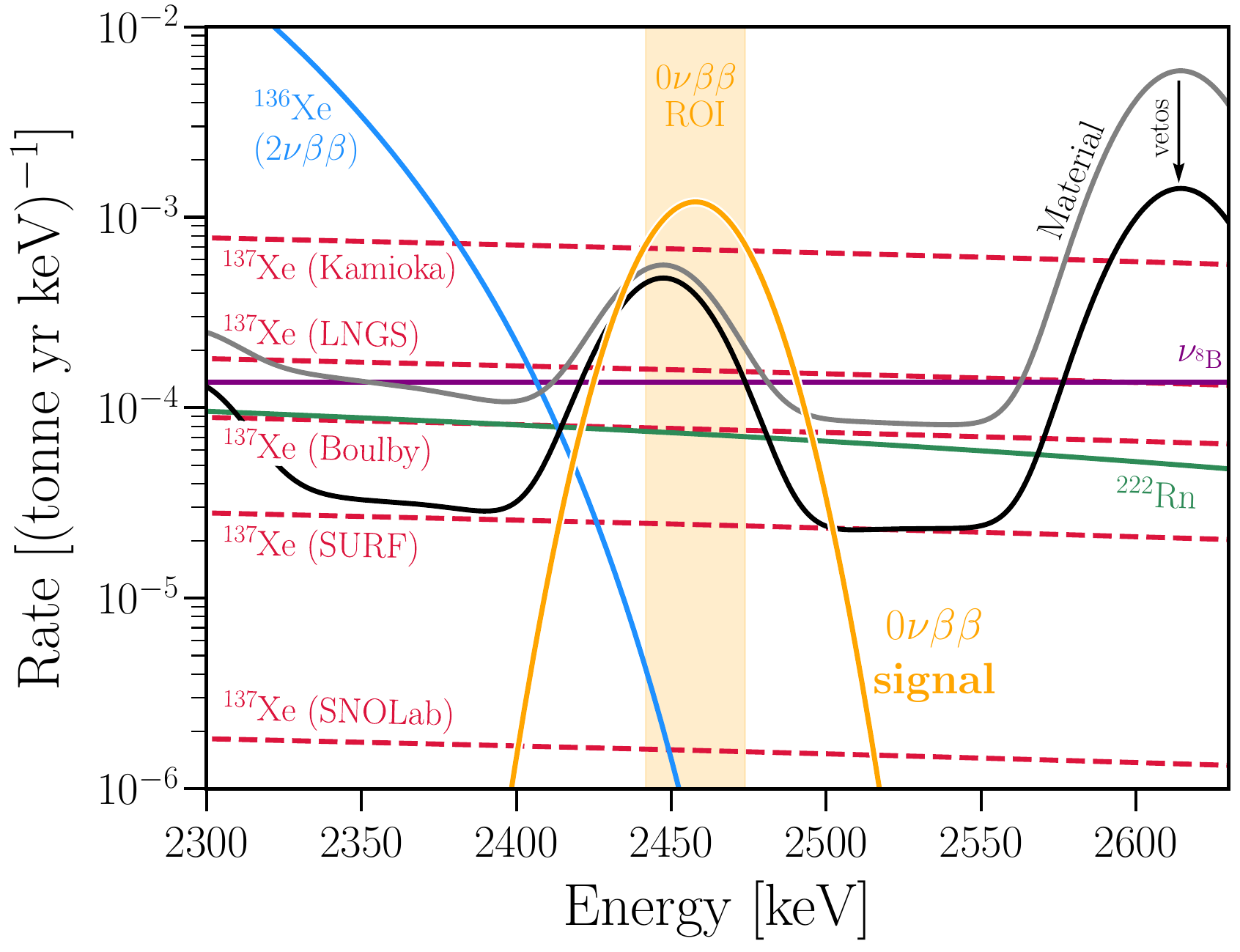}
    \hfill
    \includegraphics[height=5.8cm]{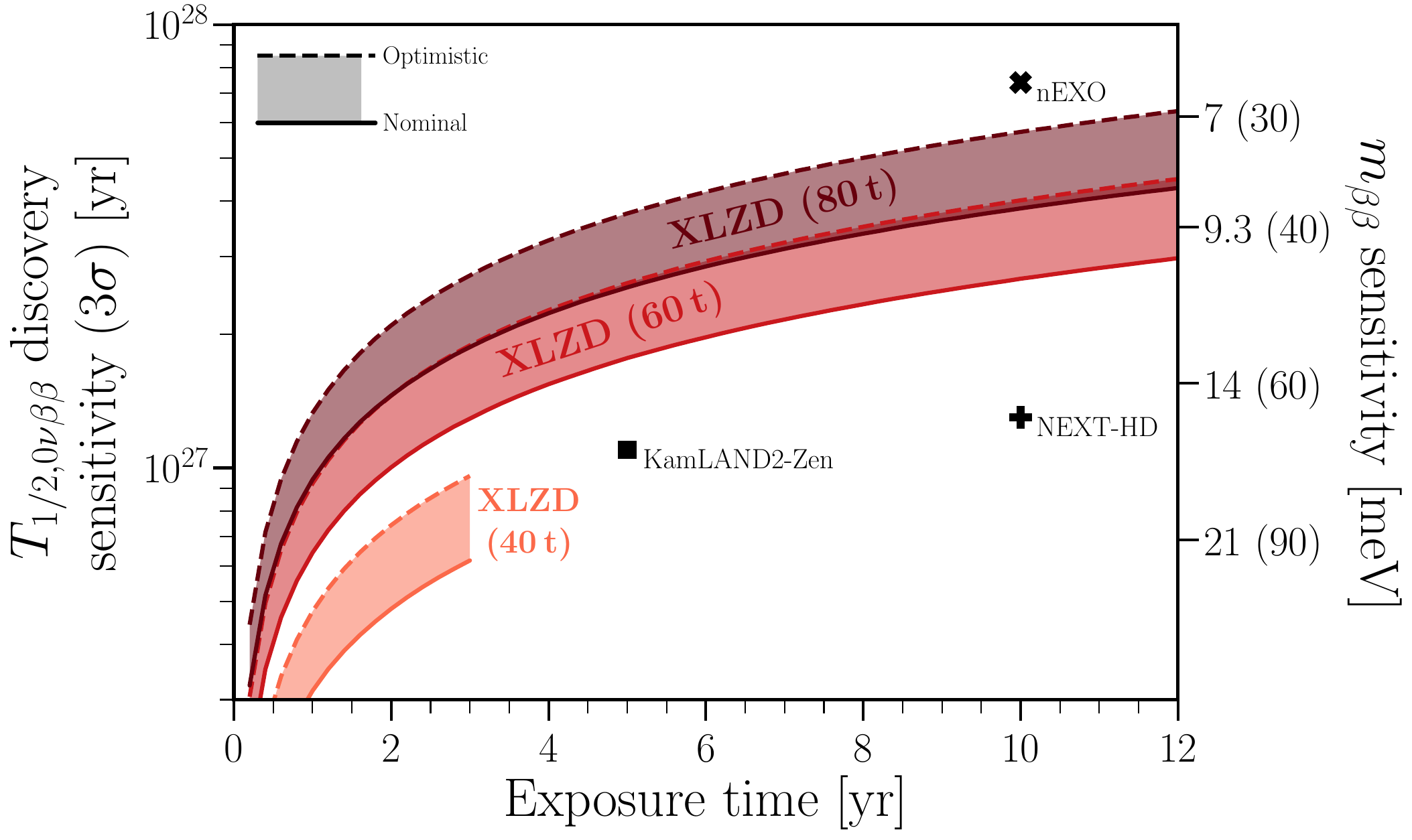}
    \caption{\small \textit{Left:} Energy spectra of a hypothetical \SI{5e27}{yr} $^{136}$Xe $0\nu\beta\beta$ signal (yellow) and the dominant backgrounds to this search: $\gamma$-rays from materials  before (grey)  and after (black) rejection by the veto systems (see text), $2\nu\beta\beta$-decays of $^{136}$Xe (blue), $^{222}$Rn induced $\beta$-decays of $^{214}$Bi (green), $\nu$-$e^-$ scattering of $^8$B neutrinos (purple), and $\beta$-decays of $^{137}$Xe (red) for examples of different host laboratories. \textit{Right:} Projected evidence sensitivity at $3\sigma$ significance to the $0\nu\beta\beta$ decay of $^{136}$Xe as a function of exposure time for the two final target mass scenarios: \SI{60}{t} and \SI{80}{t}, and an interim \SI{40}{t} configuration further discussed in Section~\ref{sec:project}. For each target mass, the band represents the range of detector performance parameters and background assumptions between the nominal (lower limit) and optimistic (upper limit) scenarios, as discussed in the text. The right axis shows the sensitivity in effective Majorana mass $m_{\beta\beta},$ assuming a maximum (minimum) nuclear matrix element $M_{136\rm{Xe}}^{0\nu}$ of 4.77 (1.11)~\cite{Agostini:2023}. The projected sensitivity of other proposed $^{136}$Xe $0\nu\beta\beta$ experiments is shown for comparison~\cite{Agostini:2023,nEXO:2021ujk,NEXT:2020amj}.}
    \label{fig:0vbb_search}
\end{figure*}

The lepton number violating neutrinoless double beta ($0\nu\beta\beta$) decay process is another critical signature for Beyond Standard Model physics. Its detection would reveal the nature of neutrinos, and under the assumption of light-neutrino exchange and SM interactions, the half-life will yield insights into their mass ordering. With 8.9$\%$ abundance of $^{136}$Xe in natural, non-enriched xenon, the observatory will instrument 5.3\,tonnes of this isotope in its 60\,tonnes LXe target, aiming to observe $0\nu\beta\beta$-decays above the background spectrum shown on the left of Fig.~\ref{fig:0vbb_search}. 
The corresponding half-life sensitivity projections for the $0\nu\beta\beta$-decay, derived using a Figure-of-Merit estimator~\cite{Agostini:2017fom} in corresponding optimal fiducial volumes, are shown on the right of Fig.~\ref{fig:0vbb_search} for the XLZD detector with either 60 or 80\,tonnes target mass. The sensitivity is estimated for a range of detector performance parameters and background assumptions, as discussed below and in~\cite{xlzd:2024}, represented by a band for each target mass between a nominal (lower bound) and an optimistic (upper bound) scenario.
The nominal scenario considers detector performance parameters (such as energy resolution and SS/MS separation) as already achieved in current generation detectors and installation at LNGS, the optimistic scenario assumes a slight improvement in these performance parameters and installation at SURF. The projection of the external background in the nominal scenario assumes the material budget of a dimensionally-scaled LZ design, with a 75\% reduction in $\gamma$-ray backgrounds based on the radiopurity of materials already identified. In the optimistic scenario, a further reduction of the external background by a factor of 2.5 is assumed.

The backgrounds to a $0\nu\beta\beta$-decay signal, shown on the left of Fig.~\ref{fig:0vbb_search}, are caused by $\gamma$-rays emitted from detector materials and electron-induced signals, with the latter being uniformly distributed in the detector volume. Two of these uniform backgrounds are irreducible: the continuous spectrum induced by solar $^8$B neutrinos scattering off electrons and leakage of the $2\nu\beta\beta$ decay spectrum of $^{136}$Xe into the $0\nu\beta\beta$ energy region of interest (ROI). The latter is highly suppressed due to the excellent energy resolution of $\sigma_E =0.65\%~Q_{\beta\beta}$ (0.60\% $Q_{\beta\beta}$) in the nominal (optimistic) scenario~\cite{Pereira_2023,Aprile:2020yad}, which corresponds to a $Q_{\beta\beta}\pm1\sigma$ ROI of \SI{32}{keV} (\SI{29}{keV}) illustrated by the vertical yellow band on the left of Fig.~\ref{fig:0vbb_search}. Additionally, two isotopes present in the target cause $\beta$-decays with energies extending over the $0\nu\beta\beta$ ROI: $^{214}$Bi, a progeny of $^{222}$Rn, and $^{137}$Xe, which is produced in the TPC by the capture of neutrons on $^{136}$Xe. The rate of the latter scales with the muon flux and thus primarily depends on the depth of the host laboratory~\cite{DARWIN:2023uje}. 
We consider Laboratori Nazionali del Gran Sasso (LNGS), Italy (29.7\,$\mu$~m$^{-2}$d$^{-1}$) and the Sanford Underground Research Facility (SURF), USA (4.6\,$\mu$~m$^{-2}$d$^{-1}$) as representative of reasonable nominal and optimistic scenarios, respectively. A natural abundance xenon target limits $^{137}$Xe production, and both facilities provide a sufficient muon flux reduction for it not to be a dominant background. As $^{137}$Xe has a short half-life (3.8 min), only $^{137}$Xe production directly inside the TPC is considered an inevitable source of background: xenon activation outside the shielding water tank, e.g.~during xenon purification, can be avoided by adequate shielding of the xenon handling infrastructure (discussed in Section~\ref{sec:fluid}). 
A low level of $^{222}$Rn in the target of $\SI{0.1}{\mu Bq /kg}$ of natural xenon is a common requirement for the dark matter science goal \cite{Schumann:2015cpa,Baudis:2013qla}. Additionally, the tagging of the subsequent $^{214}$Po $\alpha$-decay in the $^{222}$Rn decay chain allows for vetoing $^{214}$Bi decays with high efficiency, from 99.95\% (nominal) to 99.99\% (optimistic).

Suppression of the $^{208}$Tl contribution in the external background (shown by the rate decrease from the gray to the black line in Fig.~\ref{fig:0vbb_search}, left) requires an outer detector surrounding the LXe cryostat, as well as instrumentation of the LXe region surrounding the TPC.   
The background from $\gamma$-rays emitted by radioactivity in the laboratory rock walls is assumed to be suppressed by a water shield (\SI{3.5}{m} minimum thickness) to a negligible level compared to the detector material-induced $\gamma$-ray flux. 

As shown on the right of Fig.~\ref{fig:0vbb_search}, after an initial essentially background-free period, XLZD becomes background limited, with the uniform backgrounds (dominated by solar $^8$B neutrinos) and the external $\gamma$-ray backgrounds having similar contributions in the optimized fiducial volumes. Running the detector twice as long (20 years) would result in a 50\% higher half-life sensitivity. The $0\nu\beta\beta$ science reach is primarily determined by the instrumented target mass, and within the sensitivity range bands, the reduction of the external background by improved material selection has the strongest impact on the achievable sensitivity. Partial enrichment in $^{136}$Xe could be deployed at a future stage to confirm at 3$\sigma$ level a putative signal in XLZD or another experiment.

\subsection{Astrophysical Neutrinos}\label{sec:neutrinos}

\begin{figure}[t]
\centering
\includegraphics[width=\columnwidth]{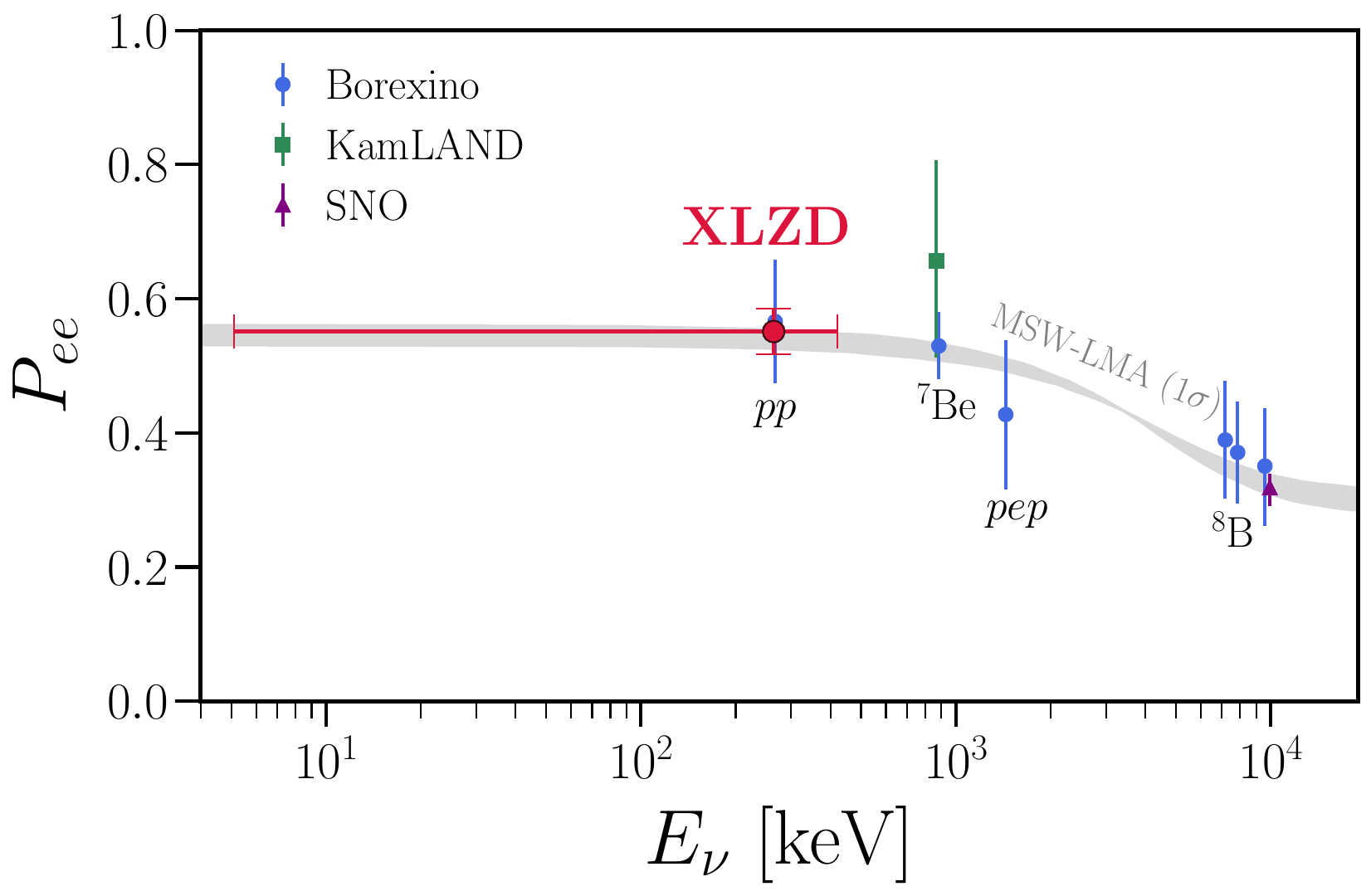} \caption{The $\nu_e$ survival probability versus neutrino energy, assuming the high-$Z$ Solar Standard Model (SSM). The blue dots represent the solar measurements from Borexino (above an energy threshold of 190~keV)~\cite{Agostini:2018uly,Agostini_2020}. The green (purple) point shows a measurement of $^7$Be ($^8$B) from KamLAND (SNO)~\cite{KamLAND:2011fld,Aharmim:2011vm}. The red point indicates that XLZD, with 600\,t$\cdot$y exposure, could enhance the precision of the $\nu_e$ survival probability in the low energy region to $\sim5\%$, using solar pp neutrino events. The point is set at the mean neutrino energy and the range bars in x indicate the full energy range (5.1~keV to 420~keV) accessible with XLZD for this measurement. The grey band represents the 1$\sigma$ prediction of the MSW-LMA solution~\cite{Capozzi:2017}.}
\label{fig:nusurvival}
\end{figure}

The XLZD detector is a prime observatory for low-energy, MeV-scale, astrophysical neutrinos through nuclear and electronic recoil signatures of a few keV of energy. Primarily, several solar neutrino flux components can be measured. Current generation experiments have measured first solar neutrinos through coherent elastic neutrino-nucleus scattering (CE$\nu$NS) of $^8$B neutrinos~\cite{Freedman:1973yd,aprile2024measurementsolar8bneutrinos,pandaxcollaboration2024indicationsolar8bneutrino}. In such a channel, the XLZD detector expects an event rate of $\sim 90$ events/t/y above a nuclear recoil energy threshold of 1\,$\rm keV_{nr}$~\cite{Baudis:2013qla,Aalbers:2016jon}, providing an independent measurement of the solar $^8$B neutrino flux. Similar to WIMP-nucleus scattering, neutrino-nucleus scattering can also be treated consistently in Effective Field Theories of the SM~\cite{Hoferichter:2020osn}. In addition, by combining this measurement with neutrino-electron scattering data from other neutrino detectors, XLZD aims to constrain the $\nu_e$ survival probability in this energy range. The most critical background for this measurement comes from accidental coincidences (ACs), spurious events created by the incorrect pairing of detector signals, detailed in Section~\ref{sec:acbkg}. 

As for ER signals from solar neutrinos, XLZD aims to measure the {\it pp} solar neutrino spectrum via neutrino-electron scattering, improving the measurement of the neutrino luminosity of the Sun~\cite{Baudis:2013qla,Aalbers:2020gsn,PhysRevD.106.096017}. Furthermore, a high-statistics measurement of the solar {\it pp} neutrino flux in the XLZD detector will enable a direct measurement of the oscillation probability of the electron-type neutrinos emitted from the Sun in an energy range that is not accessible to any other experiment, as well as an independent measurement of the weak mixing angle $\sin^2\left(\theta_W\right)$. Fig.~\ref{fig:nusurvival} shows that with an exposure of 600\,t$\cdot$y XLZD will constrain the low-energy survival probability to $\sim 5\%$. Such a measurement would test models of neutrino oscillations and probe exotic neutrino properties and non-standard interactions.
The results in Fig.~\ref{fig:nusurvival} were obtained considering a $^{222}$Rn concentration of $\SI{0.1}{\mu Bq /kg}$, which is the main background for a pp-neutrino measurement. The XLZD detector's anticipated low-energy threshold, exposure, and energy resolution would also enable splitting the data into four bins for several measurements of the survival probability in narrower energy ranges. These measurements would achieve uncertainties comparable to the current best measurements, while extending to lower neutrino energies. 

Galactic supernova neutrinos may also be detected in the XLZD detector through CE$\nu$NS. In contrast to other neutrino detectors, such detection is flavor-independent and allows the reconstruction of the total neutrino flux. A Type II core-collapse supernova at a distance of 10\,kpc from the Earth would produce of $\mathcal{O}$(100)\,events in the detector within a 10 second window. As a result, XLZD can detect a supernova burst with 5$\sigma$ significance beyond the edge of the Milky Way and the Large Magellanic Cloud~\cite{Lang:2016zhv}. In this regard, XLZD will be able to actively contribute to the inter-experiment Supernova Early Warning System (SNEWS2.0) network~\cite{Antonioli:2004zb,Kharusi:2020ovw}, responsible for the prompt follow-up of supernovae detection to the general astrophysics community. Unlike solar-neutrino detection, the signal of supernova neutrinos is highly concentrated in time ($\sim$ 10 seconds). The measurement will be limited by single and few-electron background rates, both sporadic and correlated with large events within the supernova burst duration.

\subsection{Science Complementarity}\label{sec:context}

Throughout the last decades, the search for dark matter has grown into an extensive interdisciplinary and international effort with three main thrusts: direct detection using ultra-sensitive detectors operating underground, production of dark matter at high-energy colliders, and indirect detection of particle fluxes caused by dark matter annihilation in the cosmos. 

The dark matter search goals, as presented in Section~\ref{sec:darkmatter}, complement the next-generation astrophysical searches like the Cherenkov Telescope Array (CTA), which will probe for a thermal WIMP annihilation gamma-ray signal in the $\sim$ 0.2-20\,GeV/c$^2$ mass range~\cite{Ong:2017ihp}. When combined with lighter dark matter mass constraints (\textless 200\,GeV/c$^2$) from the currently operating Fermi-LAT~\cite{FermiLAT_2015}, CTA results could severely constrain the WIMP parameter space. Water Cherenkov detectors, like the future SWGO, complement CTA due to a wider field-of-view~\cite{abreu2019_swgo}. 

The production of dark matter at the electroweak energy scale will be tested at the high-luminosity LHC~\cite{Arduini:2016xsb}. Several generations of LXe-based WIMP-search experiments have successfully demonstrated how this interplay of experimental approaches can work by ruling out new dark matter candidates in the GeV/c$^2$ to TeV/c$^2$ mass range beyond the reach of the LHC in mass or production cross-section~\cite{LZ_2024_DM_results,aprile2025wimpdarkmattersearch,PandaX-4T:2024dm,Aprile:2018dbl,Akerib:2015rjg}. XLZD will continue this tradition and explore WIMP-nucleus scattering in the remaining well-motivated regions of parameter space down to the neutrino fog.

Besides XLZD, PandaX-xT is planning a liquid xenon detector of similar scope with a somewhat smaller target~\cite{PandaX:2024oxq}. The liquid argon (LAr) community is proposing a future dark matter detector, ARGO~\cite{Agnes_2021}, to also explore the WIMP cross-section space to the neutrino fog. Probing this phase space using both targets, LXe and LAr, would allow exploring degeneracies, as the comparison of recoil measurements performed on different mass nuclei allows determination of dark matter properties such as the dark matter mass, density, and velocity distribution~\cite{Pato:2010zk,Bozorgnia_2018}. In order to probe the same parameter space as XLZD, a LAr experiment will require significantly higher exposure due to the lower atomic mass and the higher energy threshold of LAr detectors, which would be an upscale by two to three orders of magnitude in target mass over existing detectors depending on the architecture that is chosen. XLZD and ARGO are also complementary in the sense that spin-dependent interactions can only be probed by LXe detectors due to the $\sim 50\%$ abundance of xenon isotopes with unpaired spins and that only LXe detectors can effectively search for new physics in low-energy electron recoils~\cite{Chepel:2012sj}.

Neutrinoless double beta decay can occur in different isotopes, with different $Q$-values, natural abundances, and material properties. This results in different detection techniques and experimental challenges. The dual-phase (liquid/gas) TPCs with natural xenon, XLZD and PandaX-xT~\cite{PandaX:2024oxq}, will be complementary to other experiments searching for $0\nu\beta\beta$ decay in $^{136}$Xe with different technologies, such as liquid-xenon-only TPCs with an enriched xenon target (nEXO~\cite{nEXO:2021ujk}), high-pressure gaseous TPCs (NEXT~\cite{NEXT:2020amj}) or $^{136}$Xe-loaded liquid scintillator (KamLAND-Zen~\cite{Nakamura:2020szx}). Conversion from the measured decay half-life to the physically relevant Majorana mass ($m_{\beta\beta}$) requires knowledge of the nuclear matrix element (NME) for each isotope. NME predictions provided by different nuclear models can vary by a factor of more than four~\cite{Agostini:2023}, resulting in significant uncertainties in $m_{\beta\beta}$ measurements,
see Ref.~\cite{xlzd:2024} for a more detailed discussion. Complementarity between experiments using different isotopes (e.g., SNO+~\cite{Albanese_2021}, LEGEND~\cite{legendcollaboration2021legend1000}, CUPID~\cite{CUPIDInterestGroup:2019inu}, AMoRE~\cite{amore2021}, SuperNEMO~\cite{Arnold_2010}) is therefore paramount to minimize the impact of NME uncertainties in the global effort to probe the neutrino mass ordering phase space as well as the parameter space of other potential $0\nu\beta\beta$ mechanisms.

With its sensitivity to solar neutrinos, XLZD will provide complementary measurements to dedicated neutrino efforts. Current and future CE$\nu$NS experiments either use stopped-pion beams as a source of electron, muon, and antimuon neutrinos or nuclear reactors as sources of electron antineutrinos~\cite{Akimov:2017ade,Akimov:2020pdx,Angloher:2019,ackermann2025observationreactorantineutrinoscoherent}. These experiments measure the CE$\nu$NS cross-section for energies of $\mathcal{O}$(10)\,MeV (stopped-pion beam) and $\mathcal{O}$(1)\,MeV (reactor). Solar neutrinos causing CE$\nu$NS in XLZD provide a unique source as the flux contains all three flavour components allowing XLZD to probe for flavour dependence beyond the tree level. The accessible energy range spans approximately 1-10\,MeV, bridging the gap in energy between the other experiments~\cite{Abdullah:2022zue,Strigari:2023}. The current generation of detectors have just published first measurements of CE$\nu$NS from $^8$B solar neutrinos~\cite{aprile2024measurementsolar8bneutrinos,pandaxcollaboration2024indicationsolar8bneutrino}. XLZD is also sensitive to neutrino-electron scattering at lower energies than typically detected at neutrino observatories~\cite{KamLAND:2011fld,Smirnov:2015lxy,Liao:2017awz} and, therefore, will also provide complementary measurements in this interaction channel.
\section{The XLZD Project}
\label{sec:project}
The XLZD project builds upon the legacy of several successful generations of liquid xenon dark matter searches. It brings together experts from the leading current-generation detectors and R\&D efforts. In this section, the experimental strategy based on this mature detector technology is introduced.

\subsection{Technical Heritage of Liquid Xenon TPCs}
\label{sec:heritage}

\begin{figure}[t]
\includegraphics[width=\columnwidth]{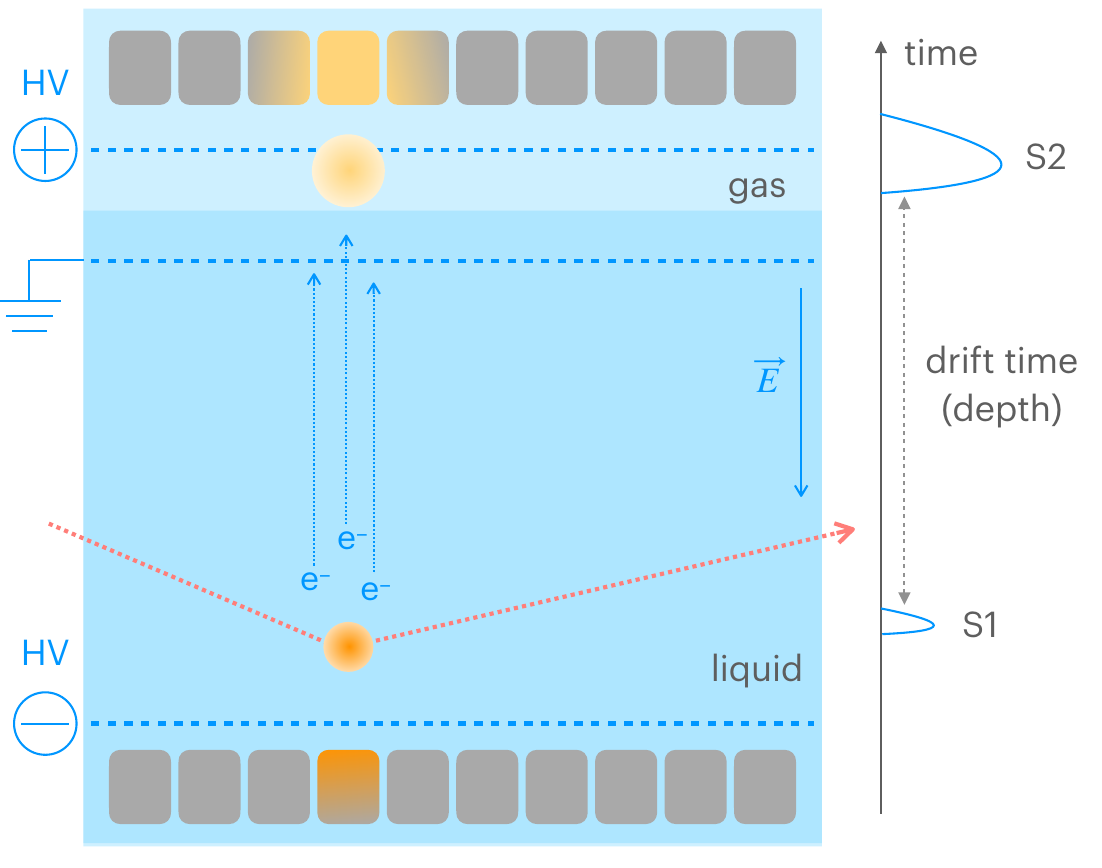}\hspace{1pc}%
\caption{\label{fig:tpc}\small Operating principle of the dual-phase (liquid/gas) xenon time projection chamber. Particle interactions in the active liquid xenon target produce prompt (S1) and delayed (S2) optical signals. These are detected by two arrays of VUV-sensitive photosensors located at the top and bottom of the TPC.}
\end{figure}

Dual-phase LXe-TPCs take advantage of several excellent properties of xenon~\cite{Chepel:2012sj,Aprile:2008bga}. Its high atomic mass enhances the scattering cross-section for the simplest elastic interactions of WIMPs. Natural xenon is free from problematic radioisotopes, and its dense liquid phase offers excellent properties for particle detection and for realizing extremely low backgrounds.
The detector technology exploits two signatures to reconstruct the location, energy, interaction multiplicity, and type of particle interactions: a prompt primary scintillation signal in the vacuum ultraviolet (VUV) at 175\,nm and delayed electroluminescence, also in the VUV, produced by the ionization electrons released at the interaction site, see Fig.~\ref{fig:tpc}. In the so-called `dual-phase' detector configuration, a thin layer of xenon vapor tops the liquid phase that constitutes the target for particle interactions. The ionization electrons from the event site drift upwards through the liquid in an applied electric field, which is established between transparent electrodes. Electrodes straddling the liquid-gas boundary establish a separate high-field region to extract electrons into the gas region, where they produce electroluminescence light. Each interaction in the active region thus generates one primary (S1) and one secondary (S2) scintillation signal, both of which can be detected by photosensors above and below the TPC. Combining information from these two signatures allows the accurate reconstruction of the interaction site and the discrimination between electronic and nuclear recoil interactions~\cite{Aprile:2006kx}. Owing to the large scintillation and ionization yields of liquid xenon~\cite{Doke:2002oab,Akerib:2016qlr,Aprile:2017xxh}, thresholds of order~keV can be realized for both electronic and nuclear recoils~\cite{XENON:2023sxq,Aalbers:2022dzr}.

Combining the ability to spatially resolve interactions with the high density and high atomic number of the medium enables excellent self-shielding properties of liquid xenon: An inner `fiducial' volume can be defined, which is largely free from external radiation. In the larger detectors operating now, the main remaining experimental backgrounds are thus from extremely long-lived second order weak decays of $^{136}$Xe and $^{124}$Xe, and from radioisotopes dispersed in the liquid bulk ($^{222}$Rn, $^{85}$Kr), and various techniques have been developed to effectively remove them (see Section~\ref{sec:backgrounds}). Finally, the ratio of S2 to S1 can be used to distinguish nuclear recoils from electronic recoils~\cite{Aprile:2006kx}. Electronic recoil rejection levels of 99.9\% at 50\% nuclear recoil acceptance have been achieved right down to the detection threshold~\cite{Akerib:2020lkv,Aprile:2017xxh}~(see a summary of electron recoil background rejection in \cite{Szydagis:2022ikv,Araujo:2020rwg}).
Pulse shape discrimination has been used to improve electron recoil background rejection by exploiting subtle differences in S1 pulse shapes, offering rejection capabilities that are independent of the drift field \cite{Akerib_2018}.

Many generations of such LXe-TPCs have successfully conducted searches for dark matter since the early 2000s through the ZEPLIN~\cite{Alner:2007ja,Lebedenko:2008gb}, XENON~\cite{Angle:2007uj,Aprile:2016swn,Aprile:2017iyp,XENON:2023sxq}, LUX/LZ~\cite{LZ:2019sgr,Aalbers:2022dzr} and Panda-X~\cite{Xiao:2014xyn,Wang:2020coa,PandaX-4T:2021bab} programs, leading to three currently running multi-tonne LXe-TPCs (PandaX-4T, XENONnT and LZ). The PandaX collaboration plans to expand to $\SI{43}{t}$ active mass in a staged-approach experiment at China Jinping Underground Laboratory (CJPL) \cite{PandaX:2024oxq}. 

\begin{figure}[t]
    \centering
    \begin{subfigure}[b]{0.49\columnwidth}
         \centering
         \includegraphics[width=\textwidth, trim={0 140 0 30}, clip]{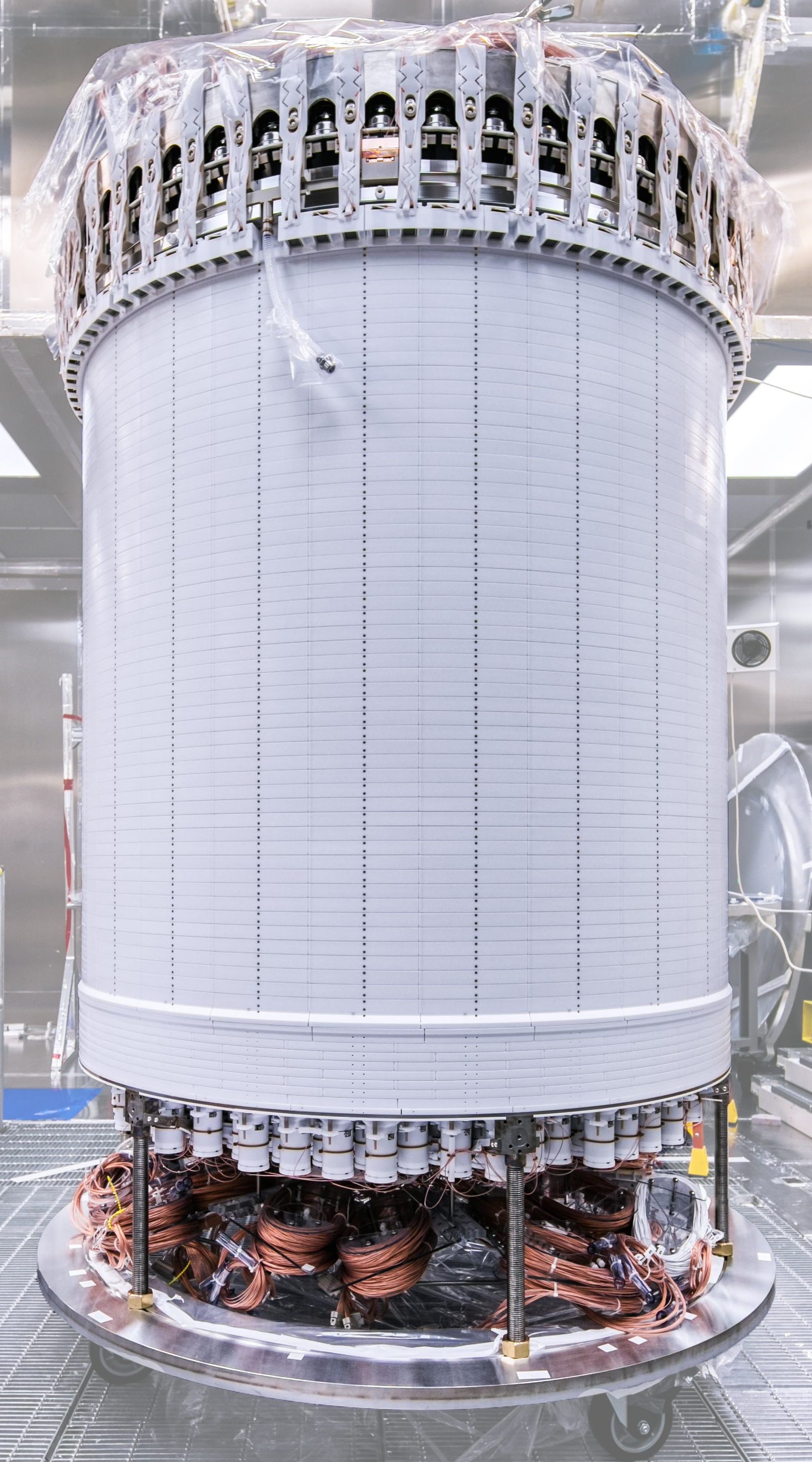}
    \end{subfigure}
    \begin{subfigure}[b]{0.49\columnwidth}
         \centering
         \includegraphics[width=\textwidth]{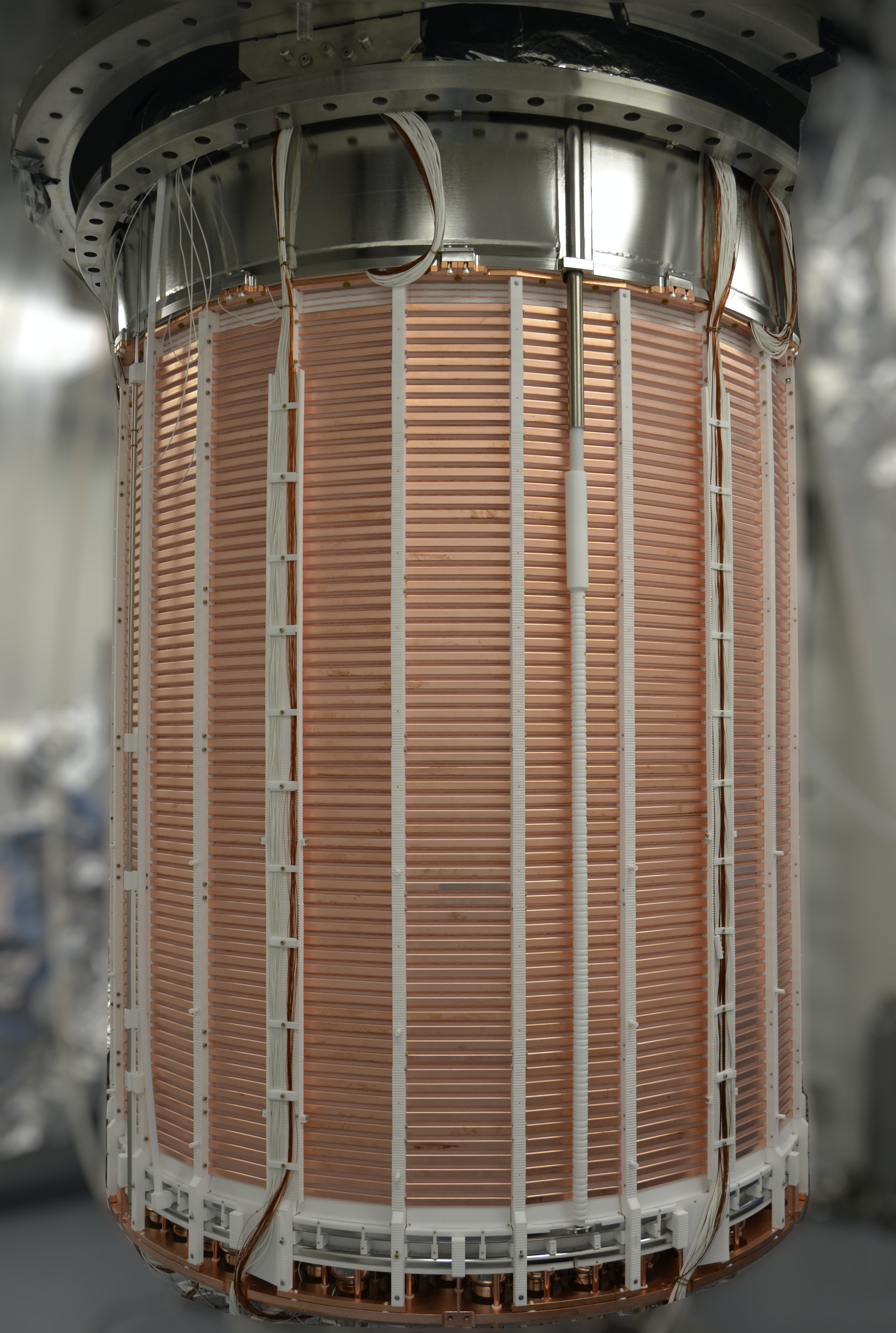}
    \end{subfigure}
    \caption{\small The LZ (left) and XENONnT (right) LXe-TPCs containing 7-tonnes and 5.9-tonnes of liquid xenon in the active detector volume, respectively. Both TPCs have a drift length of around 1.5\,m, while the diameter of the sensitive target is 1.5\,m in LZ and 1.3\,m in XENONnT.}
    \label{fig:TPCs}
\end{figure}

Since 2007, the direct search for dark matter of masses above a few GeV has been led by LXe-TPCs; while the target mass increased by over a factor of~1000, the achieved background levels decreased by four orders of magnitude. The competitive and incremental nature of these projects has been crucial to their success and places this technology in a unique position to probe the remaining parameter space well ahead of any other. Technical risks do exist -- as is natural for such an ambitious experiment as XLZD -- but they are now well understood as discussed in Section~\ref{sec:rnd}. Furthermore, the technical track record of the new collaboration, which operates two of the world-leading experiments of the day, LZ and XENONnT (shown in Fig.~\ref{fig:TPCs}), is critical to the experiment's success. 

The LZ collaboration is operating a TPC with 7\,tonnes of LXe in its active region (10\,tonnes LXe in the full detector)~\cite{LZ:2019sgr} at the Sanford Underground Research Facility (SURF), South Dakota, USA. The detector features a segmented-polytetrafluoroethylene (PTFE) field cage capped by woven electrode grids, with optical readout by two arrays of 3-inch Hamamatsu R11410-22 photomultiplier tubes (PMTs). The LXe-TPC is housed in a titanium cryostat. The thin layer of LXe ($\sim$2\,tonnes) between the TPC and the inner cryostat vessel, often referred to as LXe Skin or LXe Veto, is instrumented with scintillation readout to function as a veto detector. The cryostat is surrounded by an Outer Detector containing 17\,tonnes of Gd-loaded liquid scintillator in acrylic vessels, viewed by 120 8-inch~PMTs. Both detector systems are immersed in a water tank with a 7.6\,m diameter. Xenon cooling and purification are done separately; cold LXe is fed to a pipe manifold at the bottom of the detector; liquid is extracted from a weir system at the top and converted to gas for purification. Krypton removal was conducted using gas chromatography before deployment. LZ published world-leading results from its initial science runs in 2022 and 2024~\cite{LZ:2022ufs, LZ_2024_DM_results}.

The XENONnT collaboration is operating a TPC with 5.9-tonnes of LXe in the active region (8.5\,tonnes LXe in the full detector) \cite{XENON:2024wpa} at the INFN Laboratori Nazionali del Gran Sasso (LNGS), Italy. It is the upgrade of XENON1T~\cite{Aprile:2017aty}, the first LXe-TPC with a target above the tonne scale, which was operated at LNGS from 2016-2018. XENONnT features a lightweight TPC made of thin PTFE walls, two concentric sets of field-shaping electrodes, and high-transparency electrode grids made of individual parallel wires. Two arrays of Hamamatsu R11410-21 PMTs provide the optical readout. Although the LXe Skin concept was pioneered by XENON100~\cite{Aprile:2011dd}, it was not installed in XENON1T/nT to minimize backgrounds due to radon emanation from PTFE and maximize the active target. The TPC is housed in a stainless steel cryostat, placed in the center of a 9.6\,m diameter water shield operated as Cherenkov muon veto. The neutron veto has an inner volume of 33\,m$^3$ around the cryostat, defined by highly-reflective PTFE walls and instrumented with 120 8-inch PMTs: after the first runs with demineralized water \cite{xenoncollaboration2024neutronvetoxenonntexperiment}, Gd has been added to the shield to increase the neutron tagging efficiency. A diving bell controls the LXe level inside the TPC. Xenon cooling and purification are also done separately. For purification, LXe is extracted from the bottom of the cryostat and efficiently purified in the liquid phase \cite{Plante:2022}; an additional gas purification system cleans the warmer gas phase. Krypton removal is done via a cryogenic distillation column~\cite{XENON:2021fkt} installed on-site, allowing for online distillation. A second cryogenic distillation system constantly removes radon atoms from the liquid and gaseous xenon target~\cite{Murra:2022mlr}. XENONnT published results for a search for new physics with a world-leading electronic recoil background level in 2022~\cite{XENON:2022ltv} and for WIMP searches in 2023 and 2025~\cite{XENON:2023sxq,aprile2025wimpdarkmattersearch}.

Although the two experiments may appear similar, very different implementations have been adopted for most subsystems, with some differences highlighted above. These proven alternative implementations, 
already demonstrated at the multi-tonne scale in world-leading dark matter detectors, constitute a powerful tool for risk management. In each case, there are two solutions to choose from, and their performance is thoroughly evaluated in real dark matter search conditions over extended periods of time. 
Adding this diversity of options to the long track record of this technology and the members of the XLZD collaboration, it may be argued that the next step to the XLZD detector entails only modest technical risk -- and although the proposed $\sim$10-fold mass scale-up is significant, the increase in linear dimensions is relatively modest (factor $\sim$2).

\begin{figure}[h]
    \centering
    \includegraphics[width=0.35\textwidth]{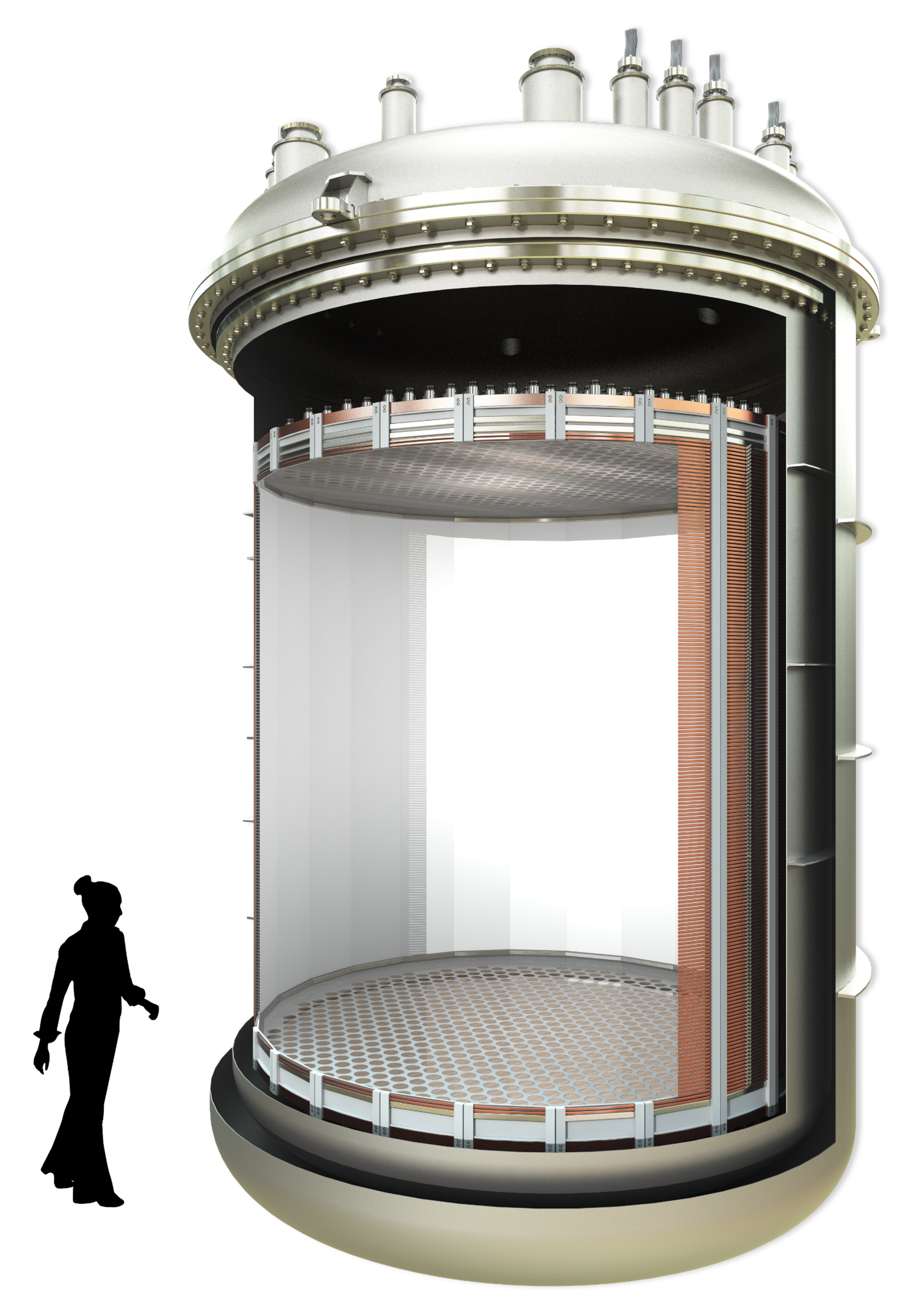}
    \caption{\small The XLZD nominal system features a LXe-TPC with a 1:1 aspect ratio for 60\,tonnes of active mass (2.98 m diameter and 2.97 m height) housed in a double-walled cryostat.}
    \label{fig:XLZD_detector}
\end{figure}

\subsection{XLZD Detector: Strategy and Xenon Acquisition}
\label{sec:strategy}

The nominal XLZD detector features a dual-phase LXe-TPC of almost 3\,m inner diameter and height, containing 60\,tonnes of active mass as depicted in Fig.~\ref{fig:XLZD_detector}. Such a detector enables dark matter searches down to the neutrino fog and a competitive search for $0\nu\beta\beta$ decay in $^{136}$Xe using a natural abundance target (see Section~\ref{sec:science} and \cite{xlzd:2024}). While the XLZD TPC diameter is fixed early in the detector design, adjusting the field cage height will allow for an interim detector of reduced mass (of around 40\,tonnes) and the flexibility to achieve a larger target mass of 80\,tonnes for the final detector. This strategy provides a central element of risk and opportunity management, allowing for early testing of detector components and adaptability to the xenon market.

The interim detector configuration of around 40\,tonnes at full field cage width and reduced height will be used for initial technical performance verification and risk mitigation. It also has the potential for competitive early science with a dark matter search exposure of up to 1000\,days, leading to more than five times the exposure expected for current-generation detectors. The $0\nu\beta\beta$-decay sensitivity for a 40\,tonnes detector running for up to 3 years is illustrated in Fig.~\ref{fig:0vbb_search}. The interim configuration will also maximize the time XLZD can monitor for a supernova burst with competitive sensitivity~\cite{Lang:2016zhv} and enable an early measurement of the solar neutrino flux via CE$\nu$NS and neutrino-electron scattering~\cite{Aalbers:2020gsn}.

Simultaneously, we will continue the xenon acquisition program until we acquire sufficient xenon to deploy a TPC  with the full nominal height with 60\,tonnes of active mass. At this point, all critical detector systems have been thoroughly tested in the running interim detector, and problems encountered can be addressed or components replaced. The 60\,tonnes baseline detector delivers the full science case outlined in Section~\ref{sec:science}.

Should the xenon market conditions permit a higher acquisition rate after the initial period, or if our initial phase already sees a hint of a signal, a more ambitious instrument with 80\,tonnes of active mass can be accommodated to acquire exposure faster. The larger target mass would require a taller field cage keeping the same diameter as in the baseline scenario. 
It would also approach the best discovery sensitivity for $0\nu\beta\beta$ decay searches since, at this target mass, we benefit more significantly from the self-shielding of external backgrounds from 2.5\,MeV $\gamma$-rays.

The outlined experimental strategy builds on a realistic xenon acquisition schedule based on conversations with the global rare-gas suppliers and the collaboration's experience in successfully procuring large amounts of xenon ($2 \times 10$\,t) for currently running experiments. The strategy allows for xenon procurement to spread over several years as not to disrupt the market. 
In-advance purchase planning, multi-year contracts, and early availability of funding will be essential for a successful xenon acquisition campaign. It should also be noted that the xenon will be retained over the experiment's lifetime and will remain an asset that does not deteriorate. 

\begin{table*}[t]
    \caption{The liquid xenon active target dimensions and mass for the envisaged TPC with 60 (nominal) or 80~tonne (opportunity) target mass with current detector dimensions as comparison. The fiducial mass estimate is based on performance of current generation detectors. The total mass of liquid xenon includes that in the Skin veto and other volumes, as discussed in the text.} 
    \begin{center}
    \begin{tabular}{ l  c   c   c  c }\hline 
    \textbf{TPC parameters} &  \textbf{XENONnT} \cite{aprile2025wimpdarkmattersearch} &  \textbf{LZ} \cite{LZ_2024_DM_results} & \textbf{XLZD Nominal}  &  \textbf{XLZD Opportunity} \\ \hline
    Target diameter [cm] & 130 & 150 & 298 & 298  \\ 
    Target drift length [cm] & 150 & 150 & 297 & 396  \\  
    Target Mass [tonne] & 5.9 & 7 & 60 & 80  \\ 
    Fiducial Mass [tonne] & $4.00 \pm 0.15$ & $5.5 \pm 0.2$ & 48 & 64 \\
    Total Mass [tonne] & 8.5 & 10 & 78 & 104  \\ \hline
    \end{tabular}
    \label{tab:tpc_specs}
    \end{center}
\end{table*}

\subsection{Siting}\label{sec:siting}

A suitable underground facility will be required to host this flagship observatory, featuring key characteristics (e.g.~depth, space, accessibility, services, support) appropriate for such an ambitious project. Five underground laboratories located at 1,000\,m.w.e. depth or more have expressed interest in hosting the detector planned by the XLZD collaboration; these are located either in deep mines or under mountains, and all offer atmospheric cosmic-ray muon fluxes attenuated by a factor of at least one million relative to the flux on the surface. These are the Boulby Underground Laboratory in the UK, the INFN Laboratori Nazionali del Gran Sasso in Italy, the Kamioka Observatory in Japan, the Sanford Underground Research Facility in the USA, and SNOLAB in Canada. All have experience hosting world-class dark matter and/or $0\nu\beta\beta$ decay research. 

In addition to the all-important depth requirement, other parameters must be considered, including the availability of suitable amount and quality of spaces underground and on the surface, the schedule for occupation, accessibility for people and materials, the radiation and cleanliness environments underground, and availability of power and other services.

The XLZD collaboration is currently evaluating hosting options thoroughly, working towards an inclusive down-selection of sites that meet key physical and infrastructure requirements, i.e.~those laboratories where such an experiment can deliver its science mission fully. The sites that will be shortlisted either meet these requirements now or will meet them with proposed developments. We have studied the cosmogenic backgrounds at potential host sites~\cite{DARWIN:2023uje}.  

\subsection{The XLZD Collaboration}\label{sec:xlders}

The XLZD collaboration was formed in September 2024 by a Memorandum of Understanding signed by over 100 senior scientists of the XENON, LZ, and DARWIN collaborations after working together as a consortium since June 2021. This is a global collaboration of scientists and engineers in 17 countries who have committed to working together to develop the definitive rare event observatory based on the world-leading liquid xenon technology. This endeavor benefits from two decades of expertise and a track record of the world's prominent collaborations now operating experiments at the 10-tonne scale, as well as from the DARWIN collaboration, which has been working on design studies for a next-generation experiment.

We anticipate that a construction project would receive substantial support from the major international funding agencies in the U.S.A., Europe (including the U.K. and Switzerland), Japan, and others. The science reach of this LXe observatory is very timely and has high priority in many major national and international roadmaps\footnote{Including the Astroparticle Physics European Consortium (APPEC) Report 2017 and the report of the 2023 Particle Physics Project Prioritization Panel (US).}, and the collaboration member institutions have the track record required to deliver that science. Coordinated proposals will be submitted to fund the experiment's design, construction, and operation.

\section{Experimental Architecture}\label{sec:detector}

Operating for up to 1000\,t$\cdot$y exposure requires the largest liquid xenon target yet and the ability to run stably for over two decades. Radioactive and electronegative impurities in materials and the LXe target must be minimized and controlled while maintaining adequate structural support and optimizing scintillation detection. Backgrounds from neutrons must be mitigated to the level of one event over the entire exposure. The TPC design, mitigation of fast neutrons, detection infrastructure such as electric fields and photosensors, materials selection, and outer detectors for coincidence vetoes are vital to this goal. This section describes the components of the XLZD observatory that are needed to meet these stringent requirements. 

\begin{figure*}[h]
    \centering
    \begin{subfigure}[b]{0.5\textwidth}
         \centering
         \includegraphics[width=\textwidth]{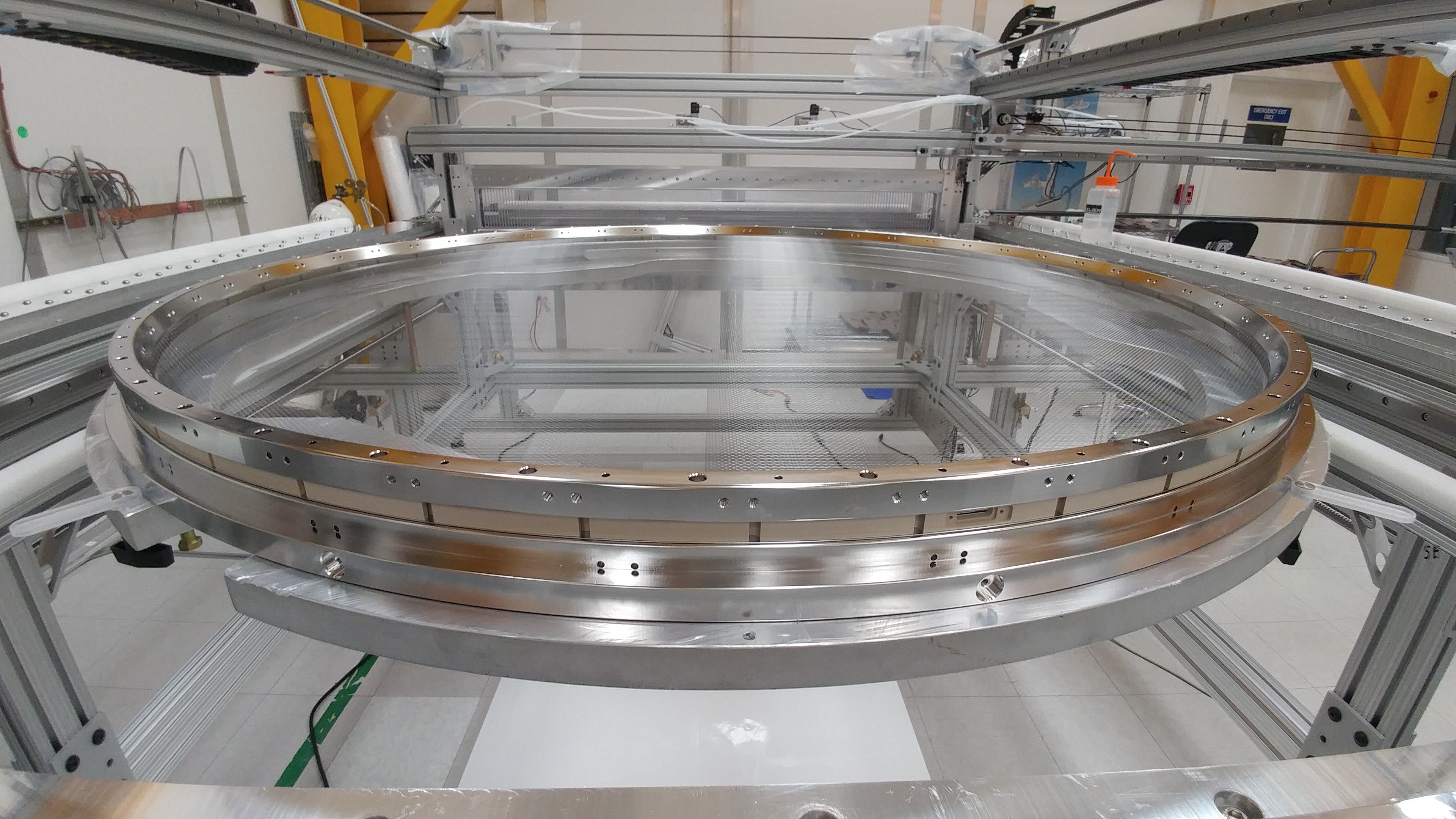}
    \end{subfigure}
    \begin{subfigure}[b]{0.425\textwidth}
         \centering
         \includegraphics[width=\textwidth]{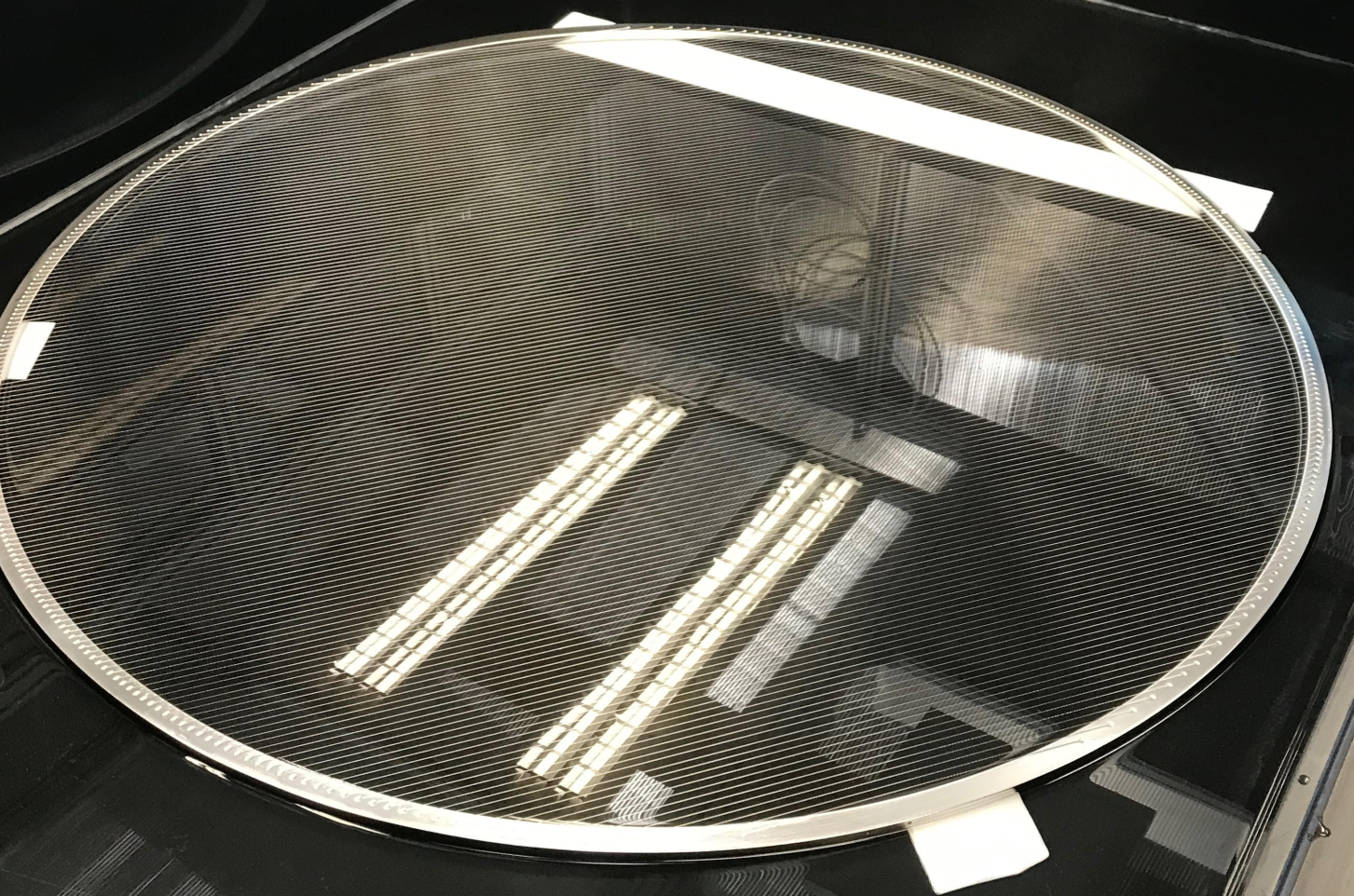}
    \end{subfigure}
    \caption{\small A finished LZ grid with crossed wires on the custom loom (left) and a XENONnT electrode with parallel wires (right).}
    \label{fig:Grids}
\end{figure*}

\subsection{The Xenon Time Projection Chamber}
The primary science detector is the cylindrical TPC (similar to those illustrated in Fig.~\ref{fig:TPCs}), containing the liquid xenon target. The TPC design needs to ensure high detection efficiency of VUV photons by choosing high reflectivity materials such as PTFE and efficient photosensors (see Section~\ref{sec:photosensors}). Extraction of ionization electrons to the top of the detector requires a homogeneous drift field established by field-shaping electrodes (see Section~\ref{sec:electrodes}) connected by high-value ohmic resistors. To maintain thermal equilibrium between the liquid and gas phase (in which ionization electrons create electroluminescence), the TPC will be housed in a vacuum-insulated cryostat. Between the walls of the TPC and the cryostat, a thin surrounding layer of liquid xenon can be instrumented to veto gamma-ray backgrounds. This is further discussed in Section~\ref{sec:outerdetectors}. Table~\ref{tab:tpc_specs} lists the nominal TPC dimensions and xenon requirements, along with a more ambitious detector if the opportunity arises as discussed in Section~\ref{sec:strategy}. The required total mass of xenon assumes a 10\,cm charge-insensitive reverse field region below the cathode, a LXe Skin of 7\,cm between the TPC walls and the cryostat, and takes into account the LXe around and below the bottom PMTs. 

\subsubsection{Electrodes and High Voltage}\label{sec:electrodes}

A dual-phase LXe-TPC contains three electrodes that create the drift (cathode to gate) and extraction (gate to anode) regions, typically consisting of parallel wires or meshes mounted on ring frames. Two additional electrodes can be placed to serve as electrical protection for the top and bottom photosensor arrays, which is the design found in the XENON TPCs. In the LZ TPC, only the bottom shielding electrode was placed, and a shield for the top array was omitted to optimize the electric field in the region around the anode.

Commonly considered TPC electrode designs include parallel wires, woven crossed wires, and etched sheets of hexagonal elements. Additional design variations are currently explored on smaller scales \cite{Andrieu:2023buk}. XENONnT and LZ TPCs incorporated the first two designs, respectively, as illustrated in Fig.~\ref{fig:Grids}. The crossed-wire grids in the LZ TPC were produced by a custom-built loom, which provided uniform tension over wires. In the case of XENONnT, individual wires were fixed in parallel to each other into the electrode frame with the use of copper pins. Production of hexagonal etched meshes on scales larger than 1~m is associated with a number of technical challenges. Nonetheless, ongoing R\&D aims to overcome these challenges by welding together smaller mesh sections \cite{Biondi2024}. Materials used for electrode production are selected after screening campaigns and include annealed SS316 (XENONnT) and SS304 (LZ).  Electropolishing of electrode frames and passivation of all electrode components allows to mitigate spurious electron emission from their surfaces~\cite{Tomas:2018pny,Linehan:2021qnb}.

The drift field strength is typically chosen to optimize discrimination between nuclear and electronic recoils while ensuring short drift times to prevent interaction pile-up. 
Studies indicate that values between 240 and \SI{290}{V/cm} provide optimal discrimination between these recoil types~\cite{Akerib:2020lkv}, but effective performance can be achieved for voltages below \SI{100}{V/cm} as shown by LZ and XENONnT~\cite{LZ_2024_DM_results,xenoncollaboration2024fieldcage}.  The gas extraction field is typically held around \si{6-8\,kV/cm} to ensure a large number of extracted electrons and secondary photons for energy and position reconstruction.

\begin{figure*}[h]
    \centering
    \begin{subfigure}[b]{0.4\textwidth}
         \centering
         \includegraphics[width=\textwidth]{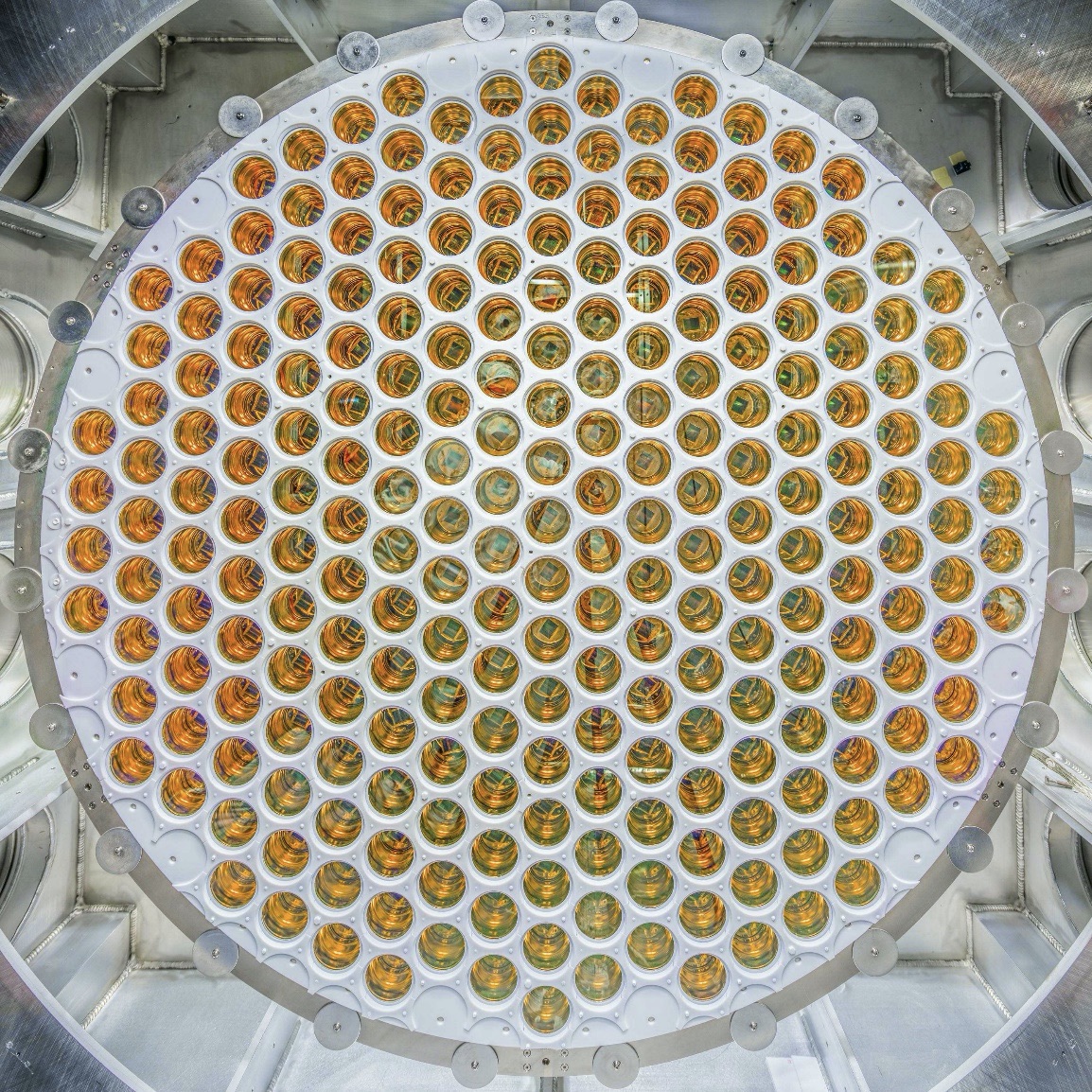}
    \end{subfigure}
    \begin{subfigure}[b]{0.5\textwidth}
         \centering
         \includegraphics[width=\textwidth, trim={150 0 100 0}, clip]{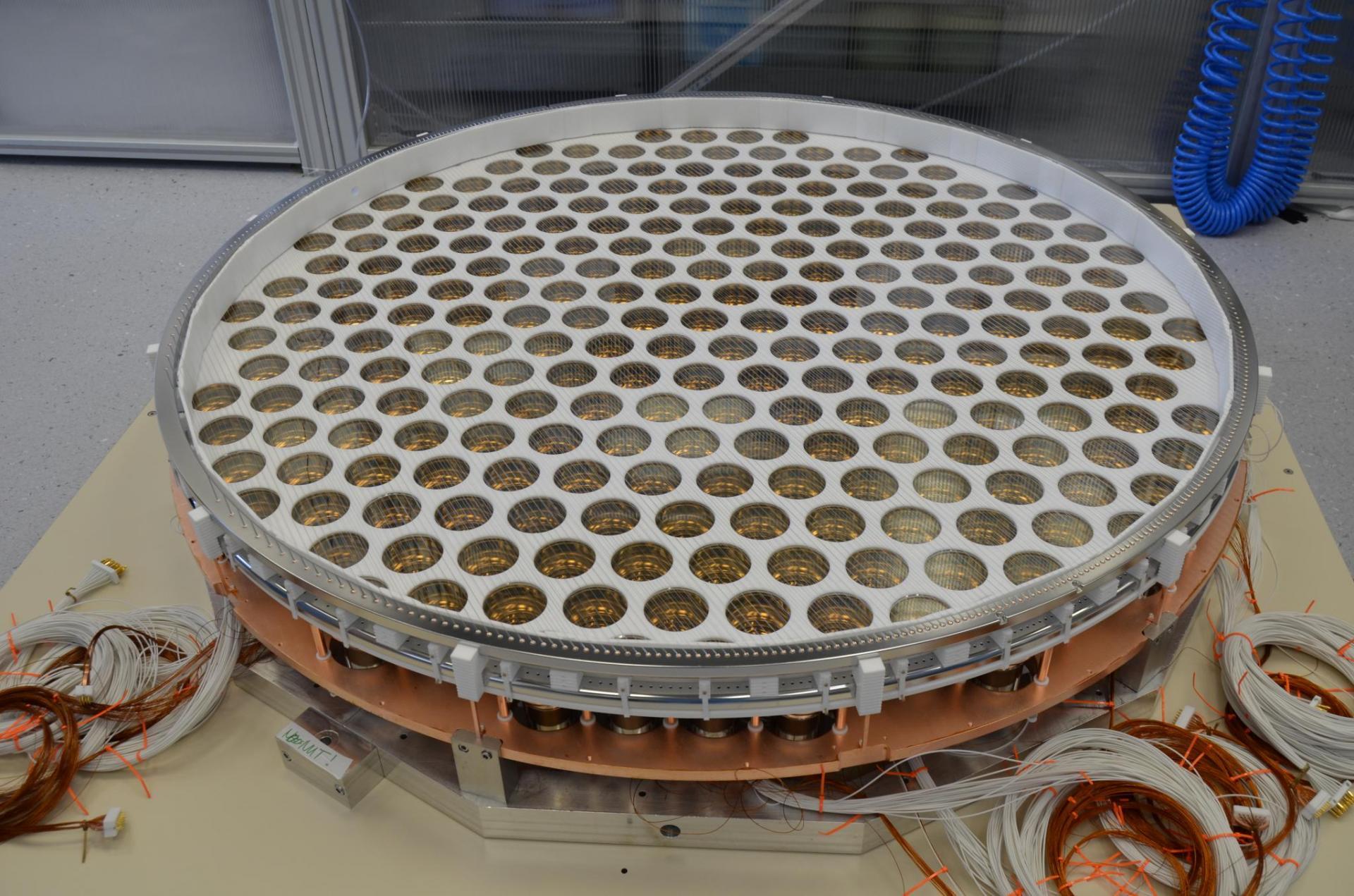}
    \end{subfigure}
    \caption{\small The LZ bottom (left) and XENONnT top (right) PMT arrays. Each experiment has 494 R11410 Hamamatsu PMTs installed. }
    \label{fig:PMTs}
\end{figure*}

The ideal drift field for XLZD requires the delivery of high voltage, of $\mathcal{O}(70)\textrm{kV}$, to the cathode. The high radioactivity in commercially available feedthroughs requires custom-made solutions for LXe TPCs~\cite{Aprile:2011dd,LZ:2019sgr}. The high voltage can be fed through from the top, side, or bottom to the TPC, with different advantages and caveats. The routing of the high voltage conductor from the top, traversing the LXe skin volume, requires the insulation of the conductor with PTFE or Ultra-High Molecular Weight Polyethylene (UHMWP), which is cryofitted in between the conductor and a grounded conductive cylinder, as is the chosen design in the XENONnT TPC~\cite{XENON:2024wpa} - a design that is lightweight and easily connected to the cathode. 
As the HV conductor increases the electric field in its vicinity, clearance with respect to the field cage is required. As LZ uses the space in between the TPC and cryostat as Skin veto, the high voltage is delivered to the cathode from the side of the cryostat. Extensive R\&D was conducted to find a robust mechanical and electrical solution~\cite{LZ:2015kxe}. A high-voltage cable is routed through the vacuum, which allows the use of a commercial cable connected at its end to a xenon-filled umbilical at room temperature outside the water shield. The lower part of the umbilical is filled with LXe. It accommodates a field grading structure that allows for the ground braid of the cable to terminate while the insulation and conductive center of the cable continue, which connects through a compliant spring to the cathode grid ring.

\subsubsection{TPC Photosensor Arrays}\label{sec:photosensors}

The detector baseline design will employ two arrays of low-radioactivity PMTs, similar to the 3-inch diameter PMTs (Hamamatsu R11410-21/22) presently in operation in LZ and XENONnT and shown in Fig.~\ref{fig:PMTs}. A total of 1182 3-inch PMTs would be needed in the top and bottom PMT arrays, respectively, to ensure the required light collection efficiency, which drives the detector's low energy threshold. The top PMT positions will be selected to optimize the position reconstruction using the S2 signal. The currently used PMTs have an average quantum efficiency of 34\% at 20$^{\circ}$C and at a wavelength of 175\,nm, with an average internal photoelectron collection efficiency of 90\%. The quantum efficiency has been measured to show a relative improvement of up to~18\% upon cooling to LXe temperatures~\cite{LopezParedes:2018kzu}.

The VUV-sensitive photomultipliers were jointly developed by Hamamatsu and the two collaborations for operation in liquid xenon. They were optimized for low radioactivity, low spurious light emission, and vacuum tightness in cryogenic conditions. All 988 units currently in operation were characterized at low temperatures~\cite{LopezParedes:2018kzu,Baudis:2013xva,Barrow_2017,Antochi:2021wik} in various test facilities before being installed in the LZ and XENONnT TPCs, respectively. 
The failure rate after LZ and XENONnT commissioning and subsequent three to four years of operation is less than 4\%, with most failures having occurred early in the operational period, which allows for continued good performance of the TPC detector systems. 

The photosensors will be equipped with resistive voltage divider circuits (bases). 
The bases must comply with low radioactivity, high linearity (for the $0\nu\beta\beta$ search) and low power consumption requirements. The expected gains are around $3\times10^6$ at a bias voltage of 1500\,V, with a dark count rate of $\sim$12-24\,Hz/PMT at liquid xenon temperatures \cite{Aprile:2017aty}. Significant PMT channel wiring savings may be possible and are under study. 

The support structures for the two arrays will be made from low-radioactivity copper or titanium, covered by a thin layer of PTFE to maximize VUV reflectance. The design will ensure that the mechanical stress induced by the thermal contraction of the different materials will not affect the PMTs. The high-voltage and signal cables may use low-radioactivity coaxial or Kapton insulated single wire cables as deployed in LZ or XENONnT. The development of other types of transmission lines, such as striplines, may possibly further reduce the Rn emanation from the signal cables. 

The materials used in the PMT variants employed in LZ and XENONnT were selected after extensive radio-assay campaigns. Specific activities for $^{238}$U and $^{232}$Th are $<13.3$\,mBq/PMT and $<0.6$\,mBq/PMT, respectively~\cite{XENON:2015ara,Akerib:2020com}. 
Further optimization of the PMT materials is required to meet the background requirements. Current efforts include screening of different iterations of the stem materials (low-activity glass stems, refined metal stems), with promising results, considering that the ceramic stems in the 21/22 model are dominating their U/Th/K content~\cite{XENON:2015ara}.
Alternative photosensors are being evaluated based on radioactivity, dark counts, and quantum efficiency requirements.

\subsection{Cryostat}\label{sec:cryostat}
\begin{figure*}[t]
    \centering
    \begin{subfigure}[b]{0.5\textwidth}
         \centering
         \includegraphics[width=\textwidth]{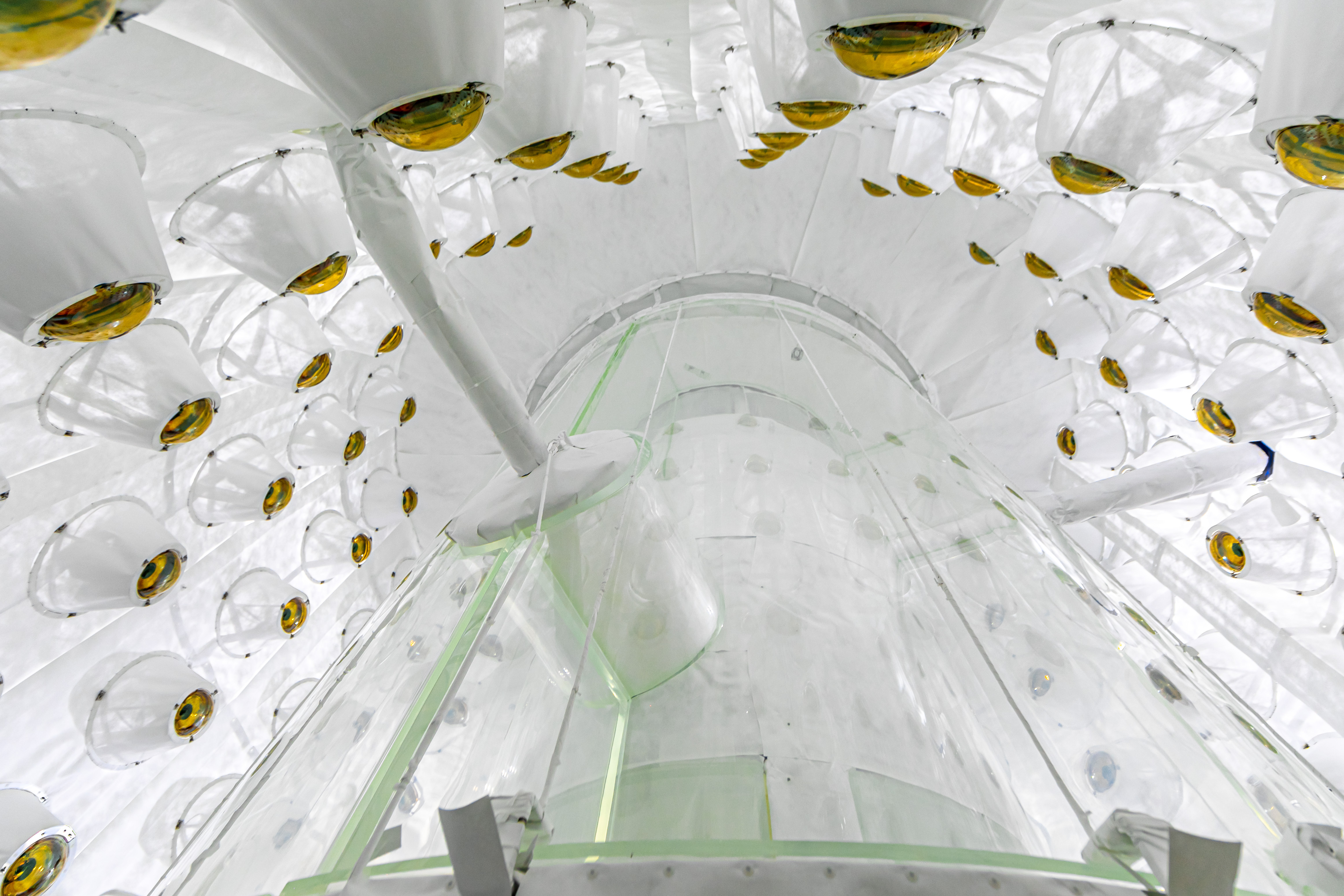}
    \end{subfigure}
    \begin{subfigure}[b]{0.45\textwidth}
         \centering
         \includegraphics[width=\textwidth]{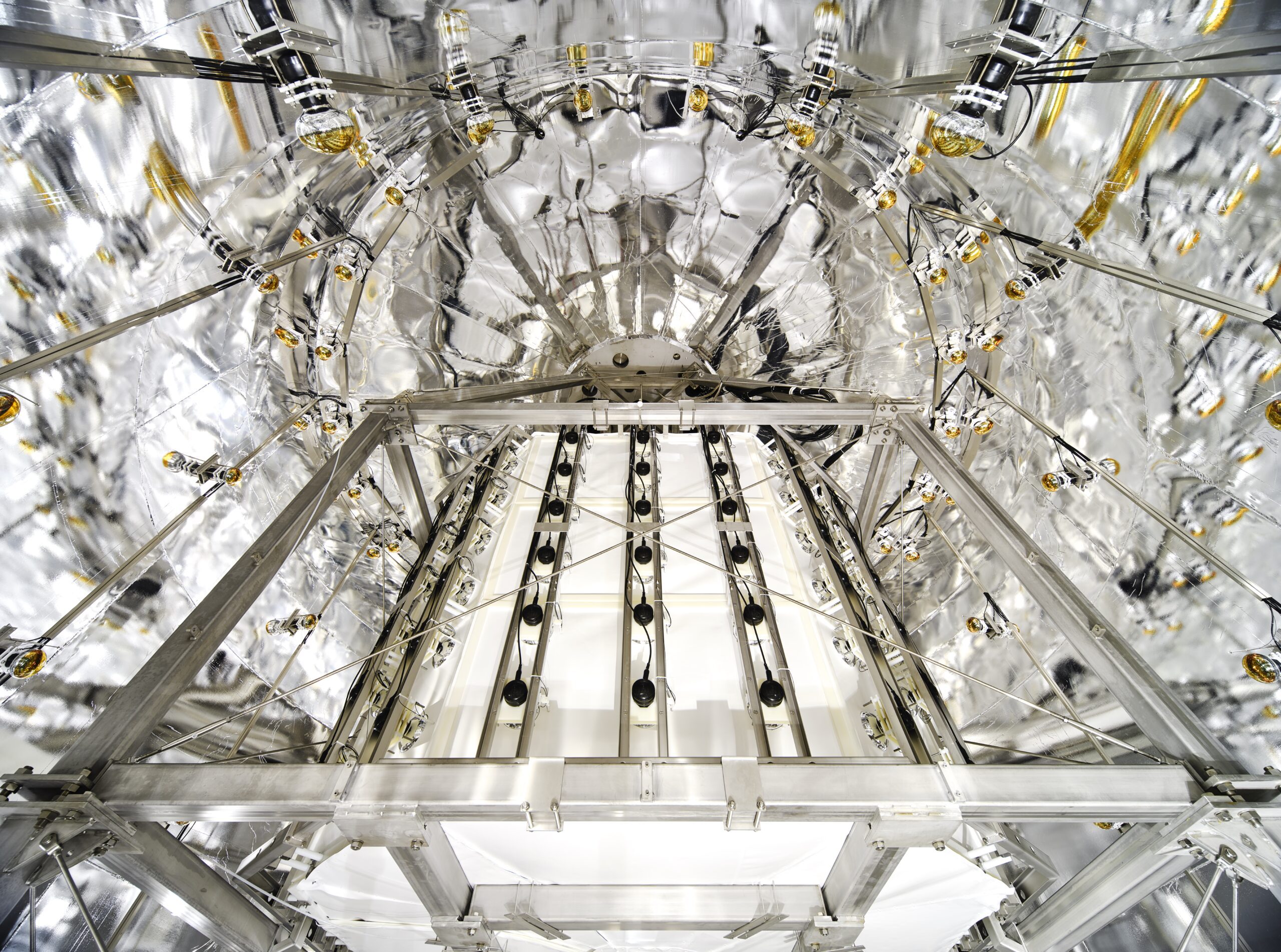}
    \end{subfigure}
    \caption{Left: The LZ outer detector with acrylic tanks monitored by 120 R5912 PMTs in the water tank. Right: The XENONnT water tank is instrumented as muon veto with 84 R5912ASSY PMTs. In the center, the outside of the inner neutron veto is seen, which is monitored by 120 R5920 PMTs. }
    \label{fig:outer_detectors}
\end{figure*}
The cryostat will be a double-walled structure centrally placed in the water tank. The inner vessel holding the TPC and liquid xenon will be nested in the outer vessel, which provides vacuum insulation and is held by the cryostat support. 

As in XENONnT\,\cite{Aprile:2020vtw} and LZ\,\cite{Mount:2017qzi}, both vessels will have conventional cylindrical geometry with shallow torispherical heads. Alternatively, a flat floor cryostat would help to reduce the amount of xenon needed. A first full-scale prototype, the PANCAKE test platform at Freiburg, was successfully commissioned~\cite{Brown_2024}. The shape of the inner vessel will be optimized to minimize passive volumes filled with LXe. The vessels’ design will comply with the pressure code the hosting country requires. 

Due to its mass and direct contact with xenon, the cryostat is one of the main contributors to the experiment’s radioactivity budget.
Therefore, a critical requirement for the cryostat is to limit its contributions from the bulk of the material and its surface to radioactive backgrounds in the sensitive region of the xenon detector. The baseline design for the cryostat assumes the use of commercially pure CP-1-grade titanium obtained from the cold hearth electron beam refining process, which follows successful material search campaigns by the LUX\,\cite{Akerib:2011rr} and LZ\,\cite{LZ:2017iwn} experiments. 

To minimize the impact of radon-emitting surfaces in direct contact with xenon, they will be either electropolished or chemically etched. A novel radon mitigation method showed that a suppression of the $^{222}$Rn emanation rate by up to three orders of magnitude could be achieved using micrometer-thick surface coatings\,\cite{Jorg:2022spz}. Further details about the applicability of this technique are discussed in Section~\ref{sec:radon}.

Alternative welding techniques, such as local vacuum electron beam welding, are being sought to mitigate possible radioactive contamination due to TIG welding with electrodes containing traces of radioactive elements. 

\subsection{Veto Detectors}\label{sec:outerdetectors}
To enhance XLZD's background rejection, the central LXe-TPC will be surrounded by additional veto detectors, such as an instrumented layer of LXe Skin, and dedicated neutron and muon detectors. When run in anti-coincidence with the TPC, veto detectors allow the rejection of neutron and $\gamma$-ray backgrounds, which may mimic a dark matter signal. The additional veto detectors permit the separation of backgrounds based upon timing: signals from $\gamma$-ray background and nuclear recoils from neutrons will occur within tens of nanoseconds of the TPC S1, while signals from neutron captures will be delayed by typically tens of microseconds.

As background events from materials primarily populate the outer regions of the LXe volume, an efficient veto system increases the fraction of LXe which can be fiducialized for the dark matter search. Veto detectors additionally provide a means for \emph{in-situ} characterization of the experiment’s background environment. In particular, it would allow checking whether a putative dark matter signal  could be due to an unexpected flux of background neutrons. The efficient tagging of high- and low-energy $\gamma$'s suppresses backgrounds critical for other new physics searches as well, e.g., $0\nu\beta\beta$ of $^{136}$Xe. 

A LXe Skin veto is achieved by instrumenting the layer of xenon between the TPC field cage and the cryostat with photosensors. PTFE lining of the outside of the field cage and the inside of the cryostat increases the light collection efficiency in this region and makes it an effective gamma veto. Preliminary studies show that a 50~mm thick Skin veto has a 55\% detection efficiency for 2.61~MeV $^{208}Tl$ gammas emitted from the bottom and side walls of the cryostat vessel. For XLZD, the benefits of a potential Skin veto will be evaluated against radon emanation considerations and the design of HV delivery in this region. 

The outermost veto systems for LZ and XENONnT are shown in Fig.~\ref{fig:outer_detectors}. LZ has a passive water shield on the outside, followed by PMTs monitoring both another layer of water and liquid scintillator (LS) tanks for neutron detection.  XENONnT has an instrumented water tank to serve as an active muon veto, while it has an optically insulated inner neutron veto. Both systems veto energetic events, including muons and showers caused by muons in the cavern walls surrounding the detector, with their instrumented water layers.  The optimum configuration for XLZD will be designed by choosing elements of the distinct LZ and XENONnT solutions.

Fast neutrons from detector material impurities are dangerous for nuclear recoil dark matter searches. This background can be significantly suppressed and characterized by surrounding the LXe volume with a neutron detector. Crucial design requirements for a high ($\gtrsim$95\%) neutron tagging efficiency include $\sim4\pi$ detector coverage around the cryostat and minimizing inactive material between the LXe and the neutron-sensitive medium. The rate in the neutron detector must also be kept low to minimize losses in the dark-matter-search livetime, which drives radiopurity requirements for all materials in the outer portions of the experiment. 

Hydrogen-rich liquids such as water or organic liquid scintillators are an elegant solution. They can be made very radiopure and doped with elements with a high neutron capture cross-section, such as boron or gadolinium. In the LZ experiment, gadolinium is dissolved into a linear alkyl benzene (LAB) solvent, which has a high light yield. This allows detection of energy deposited by the 3--5 $\gamma$-rays emitted (on average) after neutron capture on $^{155,157}$Gd or the single 2.2\,MeV $\gamma$ from capture on hydrogen. This permits a threshold as low as 30\,keV for prompt vetos, such as proton recoils and $\gamma$-rays, and a higher threshold adapted for acceptable dead time, which is currently 200\,keV in LZ. 

The XENONnT experiment instead dissolves gadolinium directly into the water of the shield following the technology developed in the EGADS and Super-Kamiokande experiments~\cite{marti2020evaluationgadoliniumsactionwater,Abe_2022}. Cherenkov light is emitted from above-threshold (289\,keV) electrons created by $\gamma$-rays from neutron capture on gadolinium and hydrogen. Although the energy threshold of this configuration is higher, it allows for hermetic coverage around the central cryostat without the need for additional liquid vessels. 

An alternative option that bridges these two media is a water-based liquid scintillator (WbLS), where a scintillator is mixed with water via the addition of a surfactant ``tail" to the liquid scintillator molecule~\cite{Yeh:2011}. A lower energy threshold is possible with WbLS than with water, and the threshold can be adjusted by varying the concentration of the scintillator. Near-future tests at the tens of tonnes scales with a demonstrator at Brookhaven National Lab \cite{Xiang:2024jfp}, the BUTTON experiment at Boulby \cite{button2023} and EOS at LBNL \cite{Anderson:2022lbb} will provide essential information and feedback on the use of this novel technology.

\begin{figure*}[h]
\begin{center}
    \includegraphics[width=0.9\textwidth]{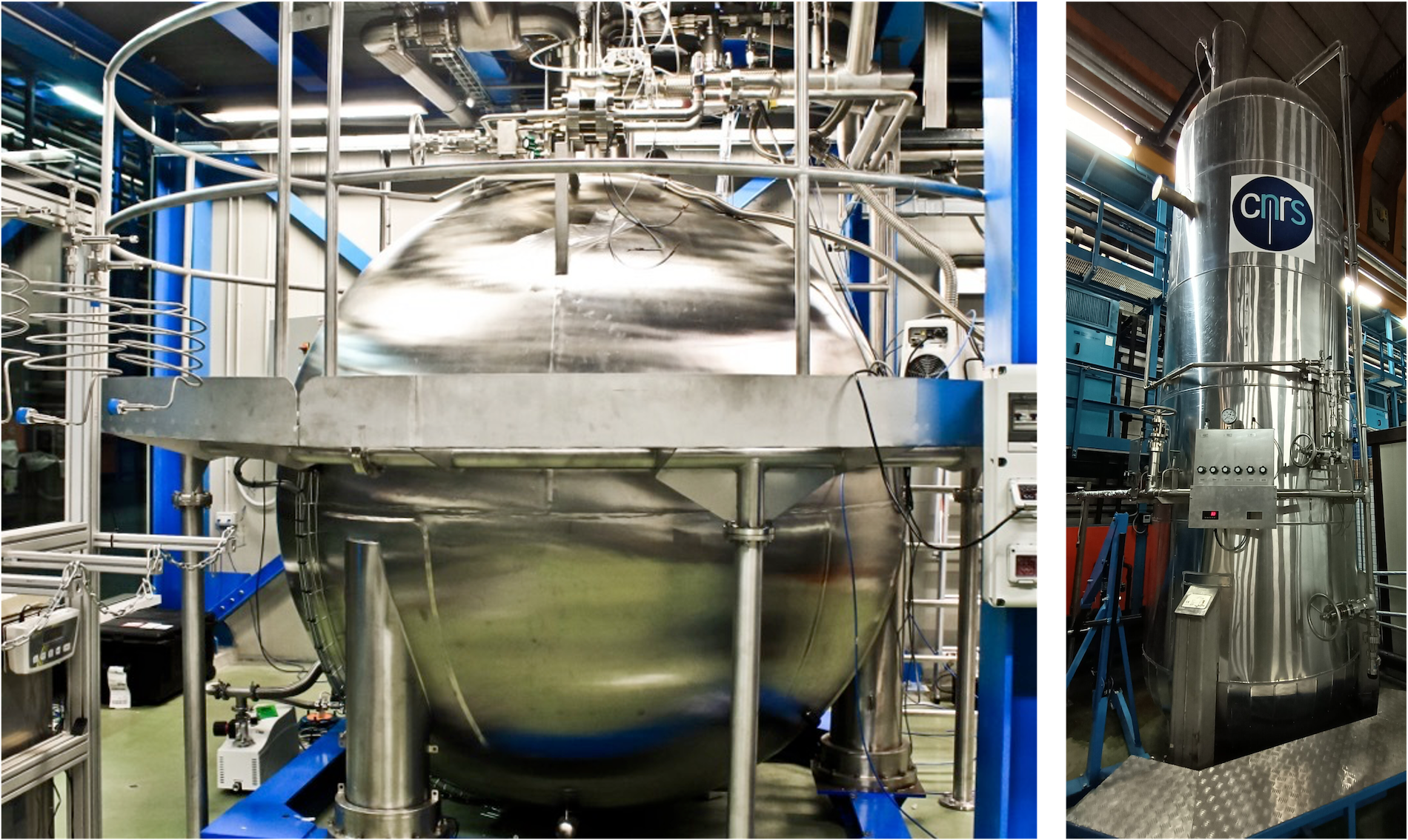}
\caption{\label{fig:restox} The ReStoX I (left, 7.6 tonne capacity) and ReStoX II (right, 10 tonne capacity) on-site cryogenic storage systems for XENONnT have demonstrated high-purity storage of xenon gas and liquid \cite{XENON:2024wpa}.}
\end{center}
\end{figure*}

\subsection{Xenon Handling, Storage and Recuperation}\label{sec:fluid}

The xenon inventory will be delivered to the experimental site in standard high-pressure gas cylinders designed for transport. When the xenon arrives at the experimental site, it will be transferred to a small number of larger monolithic storage vessels; otherwise, an unmanageable number (more than 1000) of gas cylinders will be required to contain the xenon. 

The on-site storage system will be a vital component of the xenon handling system, and a suitable design will optimize various factors, including size, cleanliness, safety, and ease of use. It will consist of several insulated cryogenic storage vessels such that xenon can be loaded via cryo-pumping. The vessels must also be rated for high pressure so that they can return to room temperature, if necessary, while loaded with xenon. The XENONnT experiment has demonstrated safe cryogenic storage for 10 tonnes of xenon with the ReStoX system, shown in Fig.~\ref{fig:restox}~\cite{Aprile:2017aty, XENON:2024wpa}. ReStoX is considered a prototype for the XLZD storage system. The size of the final storage vessels will be impacted by site access (horizontal or vertical), fabrication considerations such as the availability of underground welding and cleaning capabilities, and the dimensions of the underground space.

A simple one-pass removal of electronegative impurities will be done during the initial loading of the Xe into the storage vessels. This pre-purification step, which will be the first in the purification campaign of the experiment, can be done in the gas phase without incurring any phase change energy cost. Conventional and inexpensive purification technologies may be suitable for this treatment, even those that emanate radon, such as Oxysorb \cite{Altenmuller_2021}.

\begin{figure*}[h]
    \centering
    \begin{subfigure}[b]{0.44\textwidth}
         \centering
         \includegraphics[width=\textwidth]{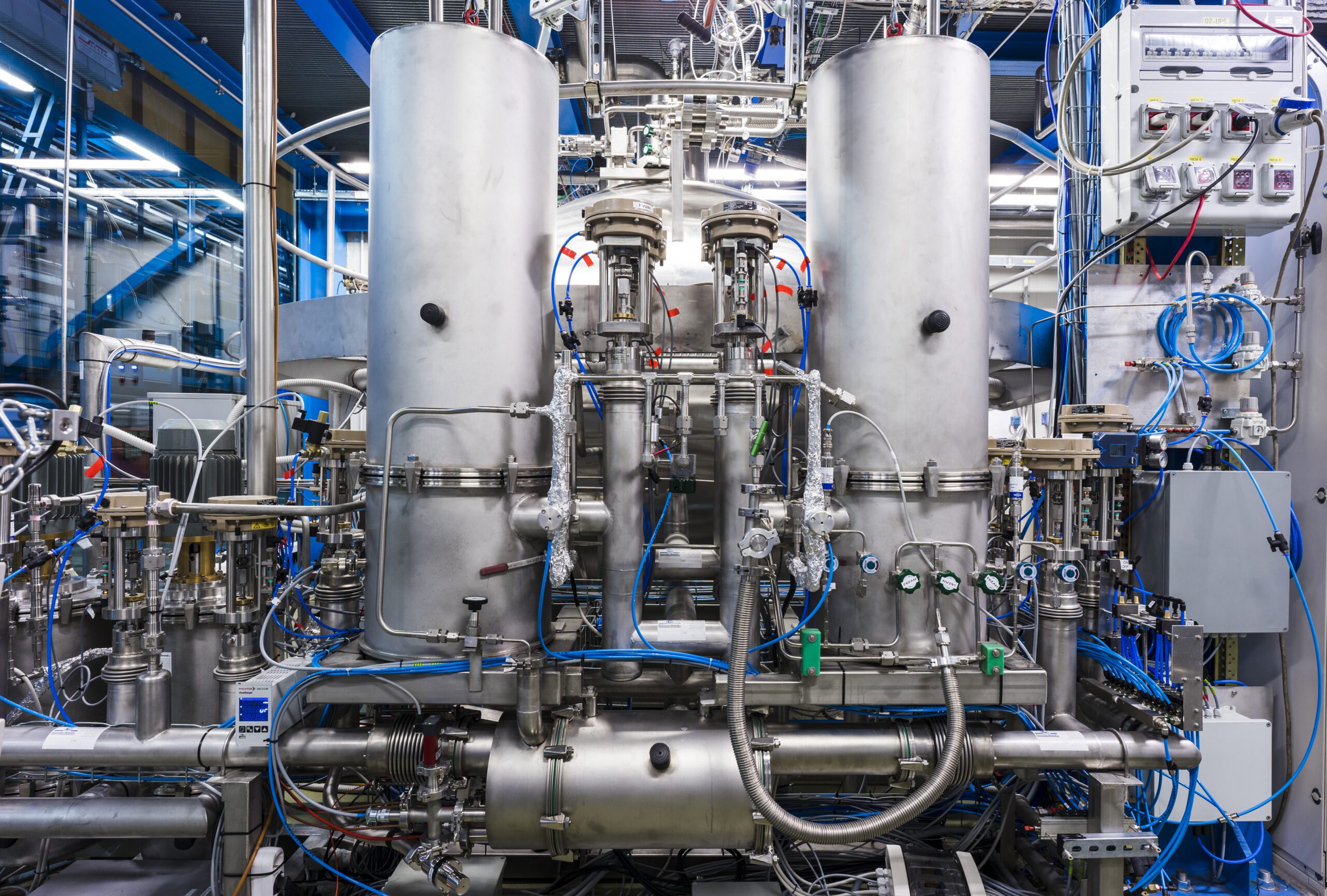}
    \end{subfigure}
    \begin{subfigure}[b]{0.54\textwidth}
         \centering
         \includegraphics[width=\textwidth]{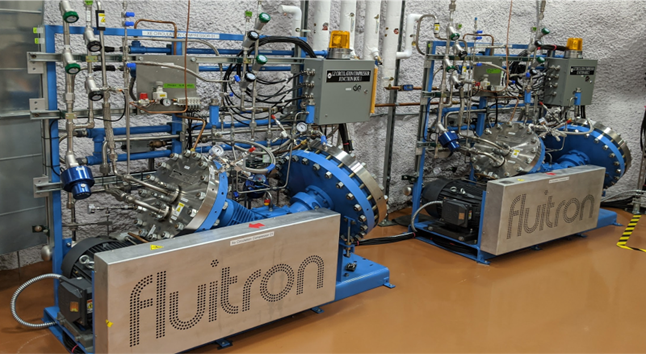}
    \end{subfigure}
    \caption{Left: The XENONnT liquid purification system currently installed at LNGS. Right: The high-purity high-pressure Fluitron gas compressors used for LZ circulation and recovery.}
    \label{fig:liquid-purification}
\end{figure*}

Online purification of electronegative impurities during detector operation will be required to achieve an electron lifetime (an effective quantity directly related to the LXe purity) of better than 10\,ms, largely exceeding the expected electron drift time from the bottom of the detector to the liquid surface. Purifying by a non-evaporate-getter (St707)~\cite{saes}, the LZ experiment has already achieved up to 8\,ms electron lifetime with recirculating gas. XENONnT has operated the liquid recirculation system shown in Fig.~\ref{fig:liquid-purification} and performed purification with 2\,LPM (8.3\,tonnes/day) in the liquid phase, achieving an electron lifetime better than 10\,ms. The system can be scaled up as needed.

The storage system will be connected to the complete purification system, including one or more Kr and Ar removal systems as discussed in Section~\ref{sec:radon}, allowing for the commissioning of the purification systems. The storage system will have the functionality to continuously circulate the xenon inventory in a closed loop for purification purposes as part of the purification campaign. Both the storage and the purification systems will be built and commissioned early so the xenon can be purified during the xenon procurement phase.

The system to recover xenon from the cryostat into the storage vessels must be robust to protect the xenon investment with multiple layers of safeguards. The XENON1T experiment has demonstrated the safe transfer of liquid xenon driven by gravity alone \cite{Aprile:2017aty}. The ReStoX storage vessel is located at a lower elevation than the  XENONnT cryostat to facilitate this recovery mode. Gravity-driven recovery is attractive because it does not require mechanical pumps or energy for the phase transition. The storage system will be designed to accommodate this type of recovery if the underground site architecture allows it. 

Gaseous recovery has been used in many experiments and may also play an important role in the XLZD recovery system. The LZ experiment developed high-purity, high-pressure gas compressors for gaseous circulation and recovery (Fig.~\ref{fig:liquid-purification}). These are required to transfer the xenon inventory back into transport cylinders at the conclusion of the experiment.

\subsection{Cryogenics}\label{sec:cryogenics}
The primary purpose of the cryogenic systems is to provide sufficient cooling power to liquefy gaseous xenon and maintain xenon detector thermodynamics by offsetting the intrinsic heat loads of the system. To date, xenon-based experiments have used bulk liquid nitrogen or electric cryocoolers to provide cooling power and specialized arrangements (e.g. nitrogen-based thermosyphon systems including an evaporator for heat transfer or a cold finger with external heaters for temperature control) to deliver targeted and controlled cooling \cite{LZ:2019sgr,XENON:2024wpa,Aprile:2017aty,LUX:2012kmp}. Cryocoolers (or cryogenerators) are being used by both LZ and XENON to ultimately reject heat from the experiment.

The same systems can be reasonably scaled up for a larger xenon payload. Once other design choices are finalized, a careful engineering analysis will determine the exact specifications for the cryogenic systems. An additional consideration will be the cooling power required to initially condense the xenon payload on a reasonable timescale. For example, condensing 78\,t of xenon, the total required xenon inventory for the 60\,t detector, in 1 month of constant operation, will require $\sim$ 6.1\,kW of concentrated cooling. While cryogenerators can easily provide this power, the delivery mechanisms and cold-heads need to be appropriately sized.  
Details of the xenon circulation and purification systems will have the biggest impact on the requirements for the cryogenic system. The xenon purity goals (see Section~\ref{sec:fluid}) will require faster xenon re-circulation in both the gas and liquid phases and, therefore, greater cooling power to overcome inefficiencies in heat-exchange between xenon gas and liquid and to sub-cool the liquid xenon below saturation temperatures. Other components of the system will also need to be accounted for: radiative heat losses from the xenon vessel, heat from the PMT arrays, heat influx through cables and access ports, increased xenon purge flow through cable conduits, etc. A total operating heat budget of a 100\,t 
xenon system is estimated to be comfortably within the range of commercially available cryogenerator units.   

The cryogenic systems are also responsible for the distribution of liquid nitrogen for operational activities requiring cryogenic temperatures (e.g., cryopumping, xenon gas sampling) or nitrogen gas that can be used for purging detector vessels during commissioning and/or operation. Such distribution systems are in operation for current generation experiments, and can be readily implemented again for the new experiment. The nitrogen gas supply can be sourced by capturing boil-off liquid nitrogen or utilizing systems that generate nitrogen gas from a compressed air supply. 

As with all auxiliary detector systems, redundancy must be embraced to minimize the risk of detector failure or downtime. For the cryogenic systems, this will motivate backup cryocooler units, a liquid nitrogen storage vessel (either integrated with a cryocooler or as a separate backup vessel), and at least two gaseous nitrogen distribution mechanisms.

\subsection{Calibration Systems}\label{sec:calibration}

A thorough calibration of the detector is a crucial step in ensuring accurate interpretation of its data. To this end, the calibration program will focus on extensively calibrating XLZD's core TPC volume and veto detectors. The calibration campaign will build upon the experience gained from successful calibration efforts in previous experiments, including XENON1T, LUX, XENONnT, LZ, and XMASS \cite{Aprile:2019bbb,Akerib:2017vbi,Aalbers_2024_calibration}.

TPC calibrations will serve two main purposes: correcting the position and possibly time-dependent variations for the S1 and S2 signals, such as the light collection, electron lifetime, electron extraction efficiency, and field distortions, and measuring the light and charge yield for specific particle interactions, such as electronic and nuclear recoils and their respective detection efficiencies and energy calibrations. Due to the large size of the detector and the excellent self-shielding capacities of LXe, external sources are not able to calibrate the detector besides the limited volume near the walls. As such, the extensive use of dispersed sources, such as gaseous radioactive sources injected into the calibration system and mixed into the TPC, will be an essential component of the calibration process. Potential calibration sources are ${}^{37}$Ar \cite{Boulton:2017hub}, ${}^{83\mathrm{m}}$Kr \cite{Manalaysay:2009yq}, ${}^{131\mathrm{m}}$Xe, ${}^{220}$Rn \cite{Aprile:2016pmc,Lang:2016zde,J_rg_2023_Rn220}, ${}^{222}$Rn \cite{J_rg_2023_Rn222} and CH$_{4}$-based sources (e.g. $^{14}$C,  $^{3}$H) \cite{Akerib:2015wdi}. These sources will enable corrections to both the S1 and S2 signals with respect to the position of the interaction and the response of the detector to electron recoils induced by $\beta$ and ${\gamma}$ particles. 

Calibration of the nuclear recoil response will still require an external source, as there is currently no efficient and secure method for dispersing them in the liquid. This calibration is typically performed using neutrons from a pulsed D-D generator external to the xenon detector  \cite{Akerib:2016mzi} or sealed radioactive sources exploiting reactions such as $^{241}$Am($\alpha$, n)Be, $^{88}Y$($\gamma$, n)Be~\cite{xenoncollaboration2024lowenergynuclearrecoilcalibration}, or spontaneous fission $^{252}$Cf sources delivered in calibration tubes placed outside the TPC \cite{Collar:2013xva}. While these calibrations are limited to the outer regions of the TPC due to the small interaction length of the neutrons in LXe, they will still be invaluable in verifying our modeling of nuclear recoils in liquid xenon \cite{Szydagis:2022ikv}. In combination with the information on detector response gained from dispersed source electron recoil combinations, the nuclear recoil response in the detector can be precisely predicted.

A D-T neutron generator, emitting neutrons with an energy of 14.1\,MeV, can be used to calibrate the response of nuclear recoils up to 400\,keV, a critical requirement for an Effective Field Theory analysis \cite{LZ:2023lvz}. Additional novel calibration methods are being investigated.

Optical calibration of the light sensors will also be conducted periodically to monitor stability in their gains. Optical calibration options include using LEDs permanently mounted on the top and bottom PMT support arrays or by injecting the LED light into the TPC by means of several optical fibers.

Besides TPC calibrations, the veto detectors (described in Section~\ref{sec:outerdetectors}) will also be thoroughly calibrated, mainly for their external background tagging efficiency for vetoing those backgrounds. Optical calibration of the PMTs in the veto detectors will also be conducted. Currently running detectors have demonstrated the calibration of the neutron detector's tagging efficiency with AmBe or AmLi sources situated outside the cryostat ~\cite{XENON:2023sxq,Aalbers:2022dzr}.

\subsection{Electronics and Data Acquisition}\label{sec:daq}

The individual signals from the light sensors of the TPC and the outer detector need to be digitized, time-stamped, and stored for further analysis. Online reconstruction of peaks (i.e., excursions from the baseline on individual channels) to events induced by particle interactions is required for immediate feedback on the detector performance and for fast data analysis. A single, highly parallelized data acquisition (DAQ) system for all sub-detectors will ensure that the signals are read out in a time-synchronized way, while different digitization speeds might be chosen for the different systems. The baseline digitization frequency for the large TPC is 100\,MHz with a 40\,MHz input bandwidth to prevent aliasing effects. The use of autoencoders on the digitizer's FPGA (Field Programmable Gate Array) to exploit compressive sensing is an interesting option and will be explored. To achieve the lowest possible light (S1) threshold, the digitization threshold for the TPC photosensors needs to be well below a single photoelectron. The DAQ time stamp will be synchronized to a GPS-based time to allow for precise time correlations in case of a supernova event. The DAQ system will also be designed for communication with the Supernova Early Warning System (SNEWS)~\cite{Kharusi:2020ovw}. 

The currently employed DAQ systems~\cite{aalbers2024data,XENON:2019bth,XENON:2022vye} generally fulfill these requirements and can be straightforwardly scaled up for the planned detector. To achieve the lowest possible threshold, the XENON DAQ system~\cite{XENON:2019bth,XENON:2022vye} operates without a global trigger, i.e., every channel is digitized individually if the signal on it exceeds 0.1\,photoelectron. Its currently employed commercial firmware for peak identification on the FPGA~\cite{caen_dpp-daw} could be optimized and better matched to offline hit finding. While such operation is desirable for dark matter searches, the huge amount of data complicates operation during calibration campaigns where large event rates in non-interesting parameter spaces (location, energy) accumulate. Intelligent online veto systems (anti-triggers) that analyze the incoming data stream using machine-learning methods implemented on fast FPGAs and that are adapted to the used digitizers reject events from these parameter spaces and thus ensure that the fraction of relevant signals is highly increased during calibration~\cite{jiang2024ultrafasttransformersfpgas}.

\subsection{Slow Control and Long-Term Stability}\label{sec:slowcontrol}

The slow control (SC) system will provide real-time monitoring and control over all subsystems and allow offline access to the data being monitored. The main goal of the system is to ensure the safety of both the personnel operating the detector and the hardware, including the xenon gas itself, while also maintaining optimal detector performance and enabling quick detection and resolution of any issues that may arise. This will allow the detector to run in a stable mode for extended periods without interruptions. The system will adhere to all relevant laboratory regulations and standards.

Based on experience from current generation detectors~\cite{LZ:2019sgr,XENON:2024wpa}, the SC will be designed as a distributed system of programmable automation controllers (PACs), allowing each subsystem to operate independently while communicating over a dedicated network. To avoid interruptions, all equipment connected to the SC must adhere to high industrial standards.

The central SC interfaces will include a supervisory control and data acquisition (SCADA) system, alert and notification service, as well as an operator log, online monitoring tools, authentication and authorization system,  and SC data archiving. A SC data stream will be integrated with the DAQ system to allow easy access to the data for analysis. Additional design requirements include a high level of network security, expandability, backup network, heartbeat monitoring, and redundancy.

\subsection{Software and Computing Engineering}\label{sec:computing}

Storing, processing, and analyzing data, as well as efficiently simulating interactions in the detector, are essential to ensure that the extensive information collected can be transformed into scientific results. The size and scope of XLZD create strong computing and software design requirements to achieve this goal. These requirements are on scales more similar to other large particle physics experiments than previous dark matter experiments, given the pace of growth in our field. This is explored in detail in the whitepaper of~\cite{Roberts:2022ezy}.

The XLZD software packages benefit from the experience of the current generation detectors. The existing event reconstruction frameworks of LZ and XENON can be effectively scaled to meet our scientific goals. The success of this scaling largely depends on effective integration with the computing infrastructure available. Simulations serve as our blueprint, illuminating the nature of prospective data and the detector's sensitivity, all thanks to the intensive modeling efforts in progress. These simulations, pivotal from the detector's design phase to the conclusive data analysis, rely on tools like Geant4~\cite{GEANT4:2002zbu} and NEST~\cite{Szydagis:2011tk,Szydagis:2013sih}.  However, hurdles remain, especially in simulating intricate detector responses. For instance, with Geant4, it is imperative for the experiment to incorporate and validate updated low-energy models. Geant4-based simulations from earlier projects~\cite{Akerib:2021ap} will be adapted, and co-processors integrated to facilitate optical modeling at this magnitude~\cite{Althueser_2022}.

A primary challenge for XLZD computing infrastructure is the management of the expected data volume with current generation detectors having petabyte-scale datasets. XENONnT and LZ currently employ HEP tools such as Rucio~\cite{Rucio_2019}, DIRAC~\cite{DIRAC_2014}, and Globus~\cite{Globus_2011}, however, tailoring these to the expanded needs of this endeavor necessitates robust community dialogue, crafting best practices, and integrating technologies like containers. This will guarantee efficient data management and reproducibility across various computing systems. 

Therefore, the core foundations of the software and computing project are present and scalable. Still, an R\&D effort is required to determine how to scale these tools to the experiment's requirements using existing computing infrastructure.

\section{Control and Mitigation of Backgrounds}\label{sec:backgrounds}

An ultra-low background environment is essential to the success of XLZD. The background mitigation strategies discussed in this section aim to ensure that backgrounds due to radioactivity fall below the expected level of irreducible neutrino backgrounds in the form of neutrino elastic scattering and neutrino-nucleus scattering. 

Controllable backgrounds in the detector are those that arise from radioactivity present in the detector materials, in particular from the $^{238}$U and the $^{232}$Th primordial decay chains. Given the precedence of radio-purity improvements in the field, accomplished with the aid of state-of-the-art material screening techniques, it is assumed that a reduction by up to a factor of three in material backgrounds compared to LZ/XENONnT can be reached \cite{Aalbers_2023,XENON:2024_backgrounds}. Despite the greater cryostat and photosensor mass, with these radioactivity estimates and considering the self-screening of the monolithic xenon volume, gamma-ray backgrounds will be negligible for a WIMP search with a detector of this target mass. Such a reduction in material backgrounds will also allow us to maximize XLZD's sensitivity to $^{136}$Xe $0\nu\beta\beta$ decays as shown in Section~\ref{sec:doublebeta}.

Radioactive impurities in the xenon target itself, such as $^{39}$Ar, $^{85}$Kr and $^{222}$Rn, can cause beta decays without accompanying gamma ray (often referred to as ``naked"). The resulting electron recoils can leak into the WIMP signal region. Developments in chromatography and distillation techniques to remove trace krypton will allow the experiment to meet a requirement of 0.03 ppt of $^{\text{nat}}$Kr
thus rendering beta backgrounds from $^{85}$Kr insignificant.
The same chromatographic \cite{Ames:2023} and distillation (including online) \cite{XENON:2021fkt} purification techniques simultaneously remove $^{39}$Ar, which even at current purification levels (890~ppt~g/g~$^{\text{nat}}$Ar/Xe achieved in LZ \cite{Aalbers_2023}) represents a subdominant background to $^{85}$Kr due to its smaller abundance in natural argon and its long half-life (269~years). 
$^{214}$Pb decays from the $^{222}$Rn chain are numerically the most significant background and, therefore, a major focus in present-generation experiments. Ongoing and planned R$\&$D as discussed in Section~\ref{sec:radon}, is expected to lead to advancements that can be capitalized on to suppress the $^{222}$Rn activity to a projected 0.1 $\mu$Bq/kg.

Finally, cleanliness protocols will be adopted modeled on the successful measures undertaken during the LZ construction campaign, which were shown to prevent the plate-out of long-lived $^{210}$Pb on potential xenon-wetted surfaces. This, in turn, has lowered the impact of so-called wall backgrounds from misreconstructed $^{206}$Pb recoils on WIMP searches.

\subsection{Low-Background Screening for Detector Materials}

All detector materials contain trace amounts of naturally occurring radioactive isotopes. 
For the detector, the key isotopes to consider are those of uranium ($^{235}$U and $^{238}$U) and thorium ($^{232}$Th) and their subsequent decays within the decay series. Additionally, the isotopes $^{40}$K, $^{60}$Co, and $^{137}$Cs are often present and emit gamma rays which drive the fiducial volume size. The decay chains of uranium and thorium can also produce neutron backgrounds through spontaneous fission and ($\alpha$,n) reactions. Materials are preferentially chosen to limit this residual radioactivity and thus reduce backgrounds within the detector which necessitates a comprehensive screening campaign prior to the detector construction.  

Different technologies are uniquely sensitive to different species, and so a multifaceted approach is necessary. For example, in the case of the uranium and thorium series, various techniques are used. To examine the elemental levels of the parent isotopes, precise inductively-coupled plasma mass spectrometry (ICP-MS) techniques can provide sub-ppt g/g sensitivity\,\cite{LaFerriere:2015}. This destructive technique involves the chemical treatment of samples before injecting them into the machine, which houses a plasma and mass spectrometer. ICP-MS was used heavily in the screening campaigns of LZ and XENON, assaying many hundreds of samples this way. The fast turnaround of samples (within a day) also provided a useful tool for removing more highly contaminated samples before embarking on more time-intensive assays\, \cite{Akerib:2020com}. A Thermo Fisher Element2, Agilent 7900 (as used for LZ) \cite{DOBSON201825}, as well as an Agilent 8900, will be available for a future assay campaign. They offer exceptional throughput and are considered more than sufficient to reach ppt sensitivity. 

Complementary gamma spectroscopy can then provide individual component activities further down the decay chains, where secular equilibrium can be broken due to longer half-lives within the series. Indeed, screening campaigns within LZ\,\cite{Akerib:2020com} and XENONnT\,\cite{XENON:2021mrg} had access to some of the worlds leading gamma spectroscopy devices such as those at BUGS\,\cite{Scovell:2018ap}, BHUC\,\cite{mount:black_2017}, Gator\,\cite{Baudis:2011am,Araujo:2022kip}, GeMSE\,\cite{Garcia:2022jdt}, the GeMPI spectrometers\,\cite{Heusser:2006}, and GIOVE\,\cite{Heusser:2015ifa}. 
During the construction of LZ and XENON1T/nT, hundreds of samples were assayed within gamma counters. Over more than six years, different materials were examined, forming two of the largest and most detailed assay campaigns for a low background experiment ever conducted. The resulting database of materials, including vendor information, is a valuable asset on which future efforts will build. 
In addition to the decays in the radon and thorium chains, these detectors are also sensitive to the other and usually unexpected gamma-emitting isotopes referred to earlier in this section. 

Where there is a greater need for increased sensitivity, such as for components close to the active detector, or where the masses of such materials are very large, it may be worth surveying additional analytical techniques that are not heavily used in the rare event field. Neutron activation analysis (NAA), which several groups in XLZD have access to, can be used in combination with gamma screening to achieve ultra-low background results. Routine NAA sensitivity for K, Th, and U is possible at 1\,ppb, 1\,ppt, and 1\,ppt, respectively\cite{Akerib:2020com}. However, for critical samples, fine-tuned analyses and sample preparations have resulted in even greater sensitivity levels, such as Th and U reaching 0.02\,ppt (EXO cryogen)\cite{LEONARD2017169,LEONARD2008490}. In addition, activation measurements have achieved the sensitivity required for LZ, detecting $^{232}$Th at concentrations ranging from 10 to 20\,ppt, and placing limits on $^{238}$U concentrations ranging from 0.4 to 4.6\,ppt, respectively \cite{Akerib:2020com}. 
Another interesting technique is the Accelerator Mass Spectroscopy (AMS) which is orders of magnitude more sensitive than ICP-MS for long-lived radionuclides (e.g. $^{210}$Pb)\,\cite{stenstrom2020identifying}.

\subsection{Control of Krypton and Radon}
\label{sec:radon}
\begin{figure*}[h]
    \centering
    \begin{subfigure}[b]{0.2\textwidth}
         \centering
         \includegraphics[height=8.25cm]{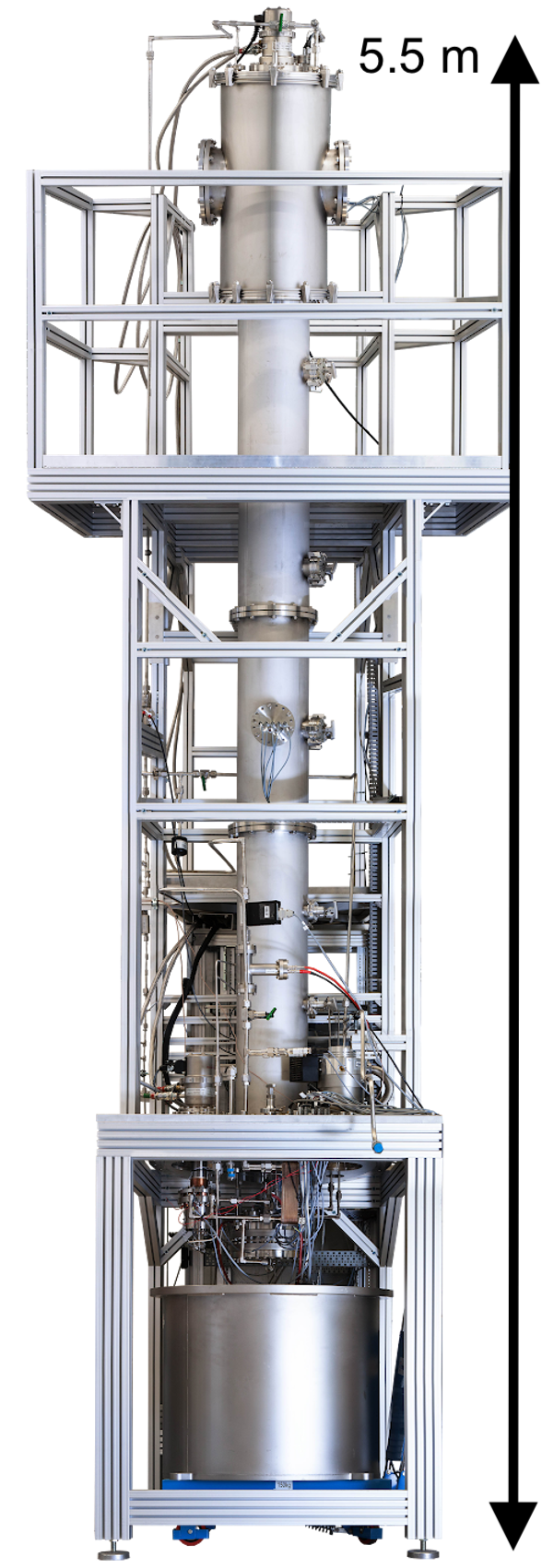}
    \end{subfigure}
    \begin{subfigure}[b]{0.3\textwidth}
         \centering
         \includegraphics[height=5.7cm]{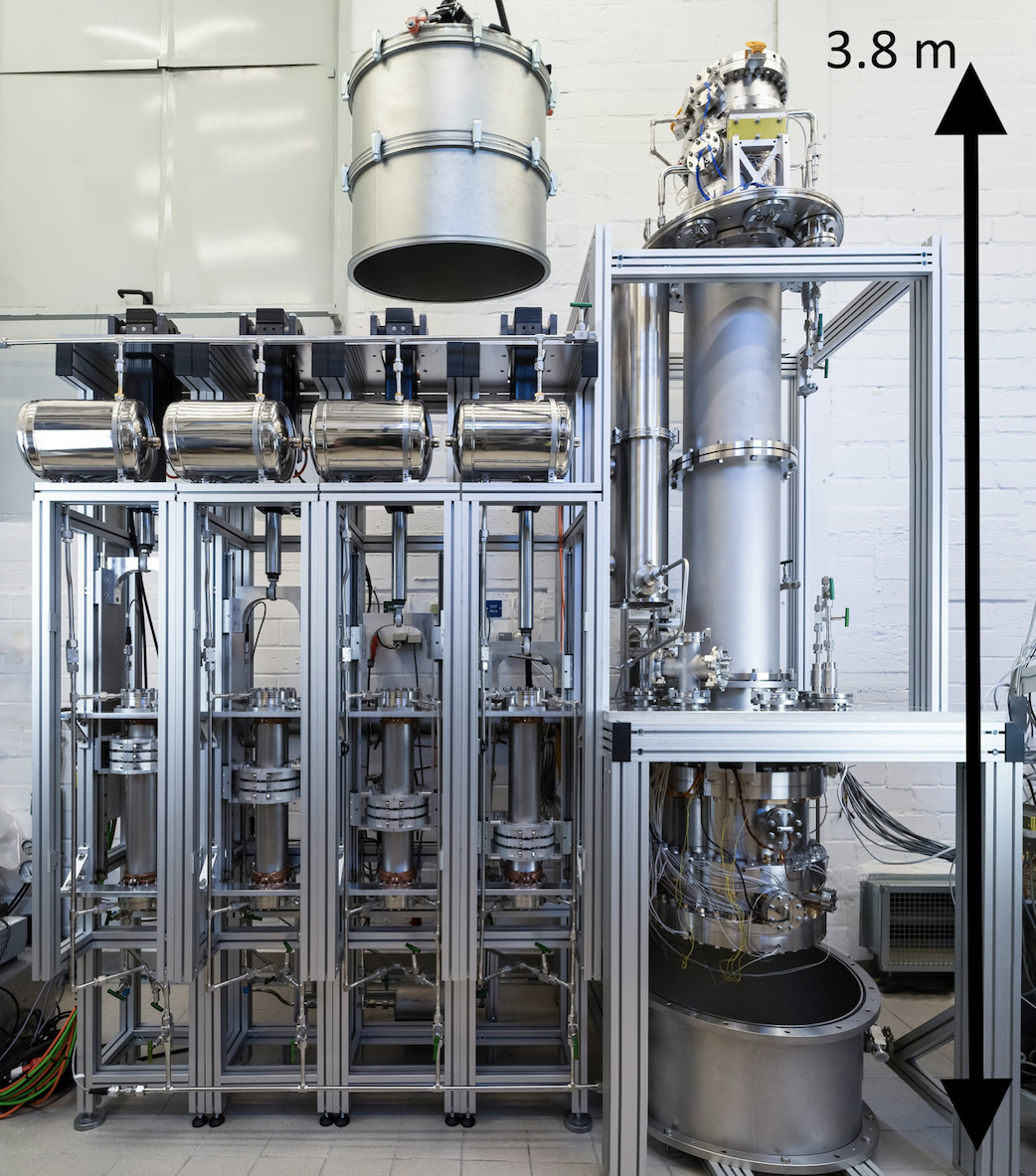}
    \end{subfigure}
    \begin{subfigure}[b]{0.45\textwidth}
        \centering
        \includegraphics[height=5.7cm]{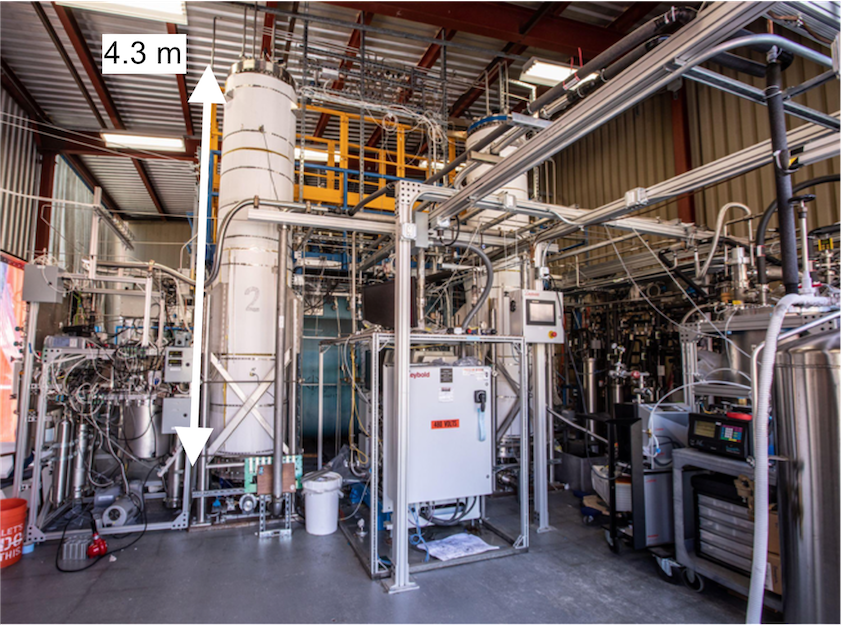}
    \end{subfigure}
    \caption{\small The online krypton~\cite{XENON:2016bmq} (left) and radon removal systems~\cite{Murra:2022mlr} (middle) developed for the XENON1T/nT experiments and the LZ offline krypton removal system located at SLAC (right).}
    \label{fig:kryptonremoval}
\end{figure*}
Small amounts of radioactive noble gases are mixed in the liquid xenon target and can neither be removed by getters or LXe purification, nor by fiducialization. Krypton contains the anthropogenic beta emitter $^{85}$Kr and is typically present at the few ppb levels in commercial xenon. Therefore, the xenon has to be purified either by gas chromatography \cite{Ames:2023} or by cryogenic distillation \cite{XENON:2016bmq} to the required purity level of less than 0.03\,ppt. Purification systems of current generation detectors are shown in Fig.~\ref{fig:kryptonremoval}. There are no sources of $^{85}$Kr in the detector, so in principle only one purification cycle is needed. However, tiny air leaks or outgassing from plastic components (mostly from PTFE) in the detector or an operation failure may increase the Kr concentration. A variation in the krypton level can only be discovered by regular monitoring with ppq sensitivity. This can be realized by a commercial RGA analyzer with enhanced sensitivity for krypton \cite{Dobi:2011vc} or by a dedicated rare gas mass spectrometer \cite{Lindemann:2013kna}, which is currently undergoing a major upgrade to allow fully automated continuous monitoring with improved sensitivity and robustness~\cite{guida2025}. Both techniques require an efficient removal of the bulk xenon prior to the measurements, which may be realized with a custom-made gas chromatography system, by crogenic distillation, or by cryo-trapping of xenon at a well-defined temperature. The two former technologies are already available within XLZD, and the latter is being explored. Should the krypton level rise above a certain threshold, online xenon purification for krypton removal as demonstrated in XENON1T \cite{XENON:2021fkt} will be applied. During this online cryogenic distillation, krypton is concentrated in the offgas, which is taken out of the system. To keep the corresponding xenon losses for such a large detector minimal a dual stage distillation system with very little offgas losses (e.g., $10^{-5}$) needs to be constructed.
Therefore, krypton removal phases will not disturb the continuous data taking.

Radon, in particular $^{222}$Rn, is considered to be the more challenging dispersed contaminant because $^{222}$Rn sources will unavoidably exist in the detector materials due to traces of $^{226}$Ra from the natural uranium decay chain. Therefore, $^{222}$Rn will continuously emanate from detector materials into the xenon target. Avoiding $^{222}$Rn sources is the best strategy to mitigate radon backgrounds. This will be addressed by a thorough material screening program similar to the one applied for bulk trace radioactivity. Electrostatic radon monitors, scintillation counting, and ultralow background proportional counters \cite{Akerib:2020com,XENON:2020fbs} are available with sensitivities down to 20\,$\mu$Bq. While most emanation measurements are done at room temperature, a setup to study the radon emanation at cryogenic temperature is also being developed. The Cold Radon Emanation Facility (CREF), based at the Rutherford Appleton Laboratory in the UK, aims to measure the emanation rates for large samples at cryogenic temperatures to better than 0.1\,mBq sensitivity, furthering the knowledge of these processes.  However, given the ubiquitous nature of $^{226}$Ra, not all $^{222}$Rn sources can be removed by screening. Further mitigation strategies include online radon removal, special surface treatments, radon tagging during data analysis, and TPC design.

Since there are radon sources inside the detector, any radon removal strategy requires continuous operation during the lifetime of the experiment. The radon removal from xenon is done either via gas chromatography, making use of the different diffusion times of radon and xenon in cold charcoal filters \cite{Arthurs_2021}, or using a cryogenic distillation column \cite{Murra:2022mlr,Aprile:2017kop} profiting from the factor 10 lower vapor pressure of radon with respect to xenon. 
In contrast to cryogenic online distillation for krypton removal, for radon removal, there is no offgas loss because the radon is kept in the system until disintegration \cite{Murra:2022mlr}. The experience from previous detectors showed that, typically, the majority of radon sources are located in the gaseous xenon part of the experiment. By a smart design of the xenon flow paths, it will be possible to selectively purify those parts that contain only a small fraction of the entire xenon mass. 
 
At XENONnT the cryogenic online distillation system has demonstrated a radon reduction factor of two when extracting xenon gas from the most relevant points.
In addition, radon already inside the active LXe volume can be removed by online purification with LXe extraction if the purification time for the detector's entire xenon mass is of the same order of magnitude as the $^{222}$Rn half-life \cite{Murra:2022mlr}. This can be met by cryogenic distillation as demonstrated by XENONnT, which operates such an online radon removal system running at a typical purification speed of nearly 2\,tonnes per day. This reduces the Rn concentration by another factor of two and yields a Rn concentration below 1\,µBq/kg \cite{XENON:2024wpa,aprile2025radonremovalxenonntsolar}. The larger XLZD detector requires a challenging xenon purification speed of $\mathcal{O}$(10 tonnes per day).  R\&D for constructing a larger system with efficient use of heat pumps to obtain the enormous cooling power is underway.  

For the cryostat containing the liquid xenon itself, novel surface treatment techniques are under development. A thin, clean, and tight coating may block recoil- and diffusion-driven $^{222}$Rn emanation. It was recently demonstrated that electro-deposited copper coating on a small stainless steel sample can reduce the radon emanation rate by more than a factor of 1500 \cite{Jorg:2022spz}. To become applicable in XLZD, up-scaling is necessary, as well as strict purity control, as tiny traces of $^{226}$Ra in the coating would spoil the achieved reduction. Another promising method to reduce the radon level is an almost gas-tight TPC design, which separates the active target from the more Rn-contaminated outer volume \cite{Dierle_2023,Sato2020development}. Finally, the remaining radon-induced background may be reduced by smart data analysis techniques \cite{XENON:2024qvh, LZ_2024_DM_results}. The approach makes use of characteristic coincidences in the radon decay chains, which may be used to tag the event if the convective tracks of radon daughters in the detector are sufficiently well understood.
LZ has demonstrated that by adjusting the temperature and relative flow of LXe into the cryostat, distinct mixing states can be established, including states where radon-induced backgrounds can be reliably flow-tagged, reducing their impact on final physics analyses. Additionally, even in the absence of successful flow tagging, LZ's low-mixing states have shown lower overall radon activity in the central fiducial volume, offering an additional radon-background reduction~\cite{LZ_2024_DM_results}.

\subsection{Surface Contaminants}\label{sec:surface}

Despite thorough material screening and selection efforts, surfaces may be contaminated during parts machining and the subsequent detector construction. This is predominantly from two sources: firstly, the plate-out of long-lived radon daughters such as $^{210}$Pb\,($\tau_{1/2}=22.3\, \textrm{yrs}$) \cite{jacobi_activity_1972} and from the deposits of dust and debris onto the detector parts. In the latter case, the dust can then become mobile within the LXe volume, which can cause instabilities in the HV should this dust later attach to electrodes or feed-throughs. 
In addition, this dust also emanates radon, thus adding to the total detector background.

In the case of radon daughters that have been plated out upon the surfaces of detector materials during detector construction, there are multiple contributions to detector backgrounds. When radon daughters decay on the TPC wall, charge loss near the wall can lead to reduced S2 signals, creating a background that may leak into the WIMP signal region.  This is especially problematic in the case where daughters of radon become mobile, and hence, decays occur within the fiducial volume. Additionally, the alpha decays of $^{210}$Po within the $^{222}$Rn chain can undergo ($\alpha,n$) processes within the detector walls, which, depending on the interaction topology of the neutron, can be indistinguishable from WIMPs. To mitigate this as much as possible, the experiment can utilize and further develop techniques used in previous experiments. For example, LZ used a reduced radon cleanroom in which air is passed through a cold carbon filter to reduce the quantity of radon in the air by several orders of magnitude~\cite{Akerib:2020com}. This will be particularly important within an underground setting where radon levels can be higher than on the surface. Air-ionizer units were additionally utilized to prevent charge accumulation on PTFE, limiting the plate-out of charged daughters onto the detector materials~\cite{Akerib:2020com}. Chemical surface treatment of individual detector parts like etching, pickling, leaching, and passivation could also be used to remove radon daughters that have plated out before the components reach the cleanroom. This process includes placing components into acid to remove a thin layer of material (with any contamination) on the surface. This was utilized in both XENONnT~\cite{XENON:2021mrg} and LZ~\cite{LZ:2019sgr}. It has also been shown that the bagging of detector parts within nylon or storing in nitrogen-flushed or evacuated containers for extended periods of time can further reduce any plate-out \cite{XENON:2021mrg,meng_new_2020}. While these mitigations limit the plate-out, any exposure should still be tracked during construction in order to help form an estimated background rate. For example, in LZ, the average plate-out of $^{210}$Pb was calculated to be ($158\pm13$)\,$\mu$Bq/m$^2$, less than a third of the requirement \cite{Akerib:2020com}. Measurement of the plate-out radon progeny on material surfaces such as that of PTFE and electrodes is possible, typically by examining the surface alpha activity from the $^{210}$Po isotope measured via XIA UltraLo-1800 or Si-PIN detectors~\cite{XENON:2021mrg}. It is also possible to measure the plate out in detectors such as BetaCage, which is a neon drift chamber that is extremely sensitive to low energy electrons and alpha particles~\cite{schnee_screening_2007}. The $^{210}$Pb plate-out rate can be inferred by repeated measurements over an extended time period while parts are exposed to the detector construction conditions.

\begin{figure*}[h]
    \centering
    \includegraphics[width=0.49\textwidth]{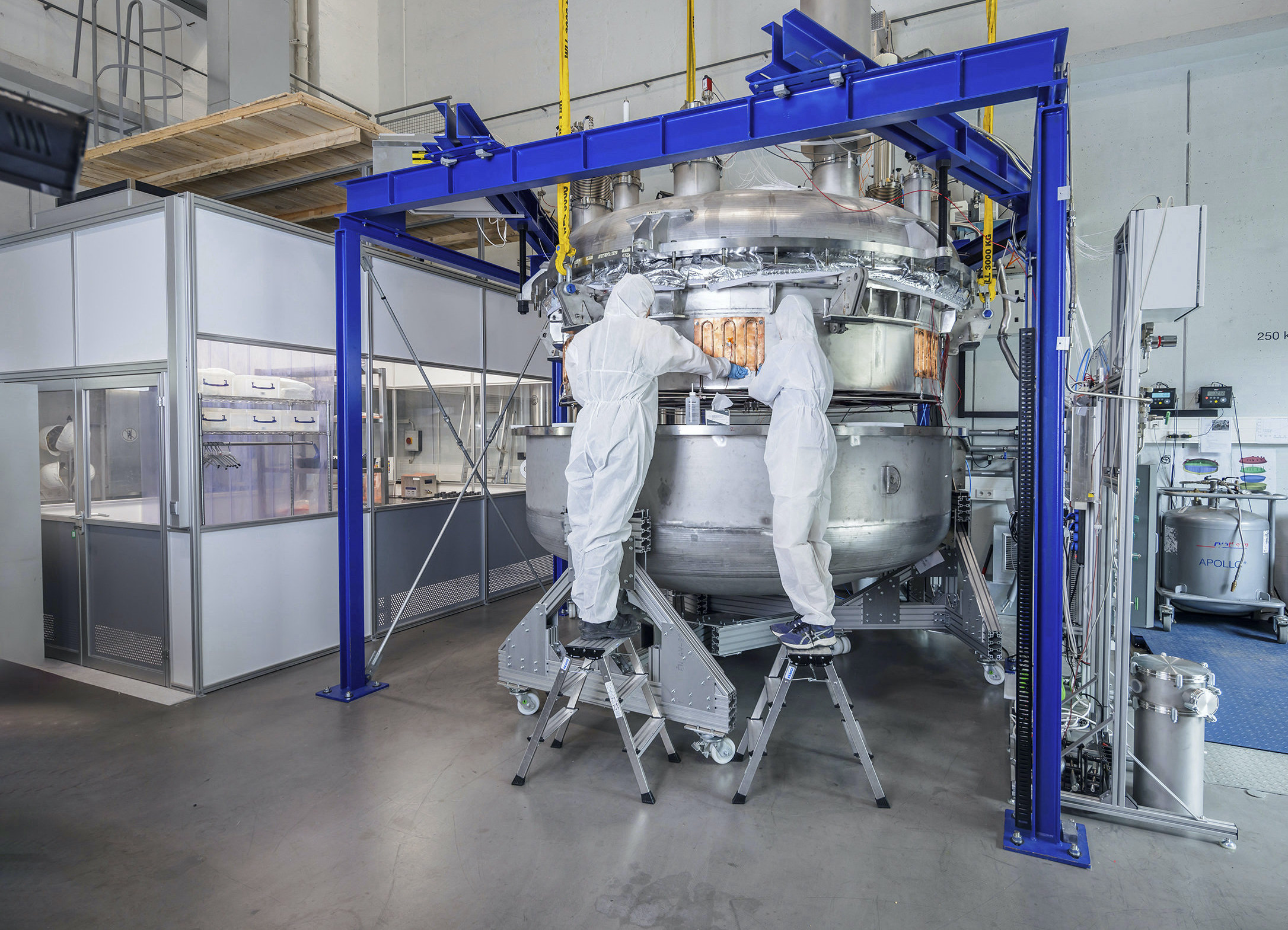}
    \includegraphics[width=0.49\textwidth]{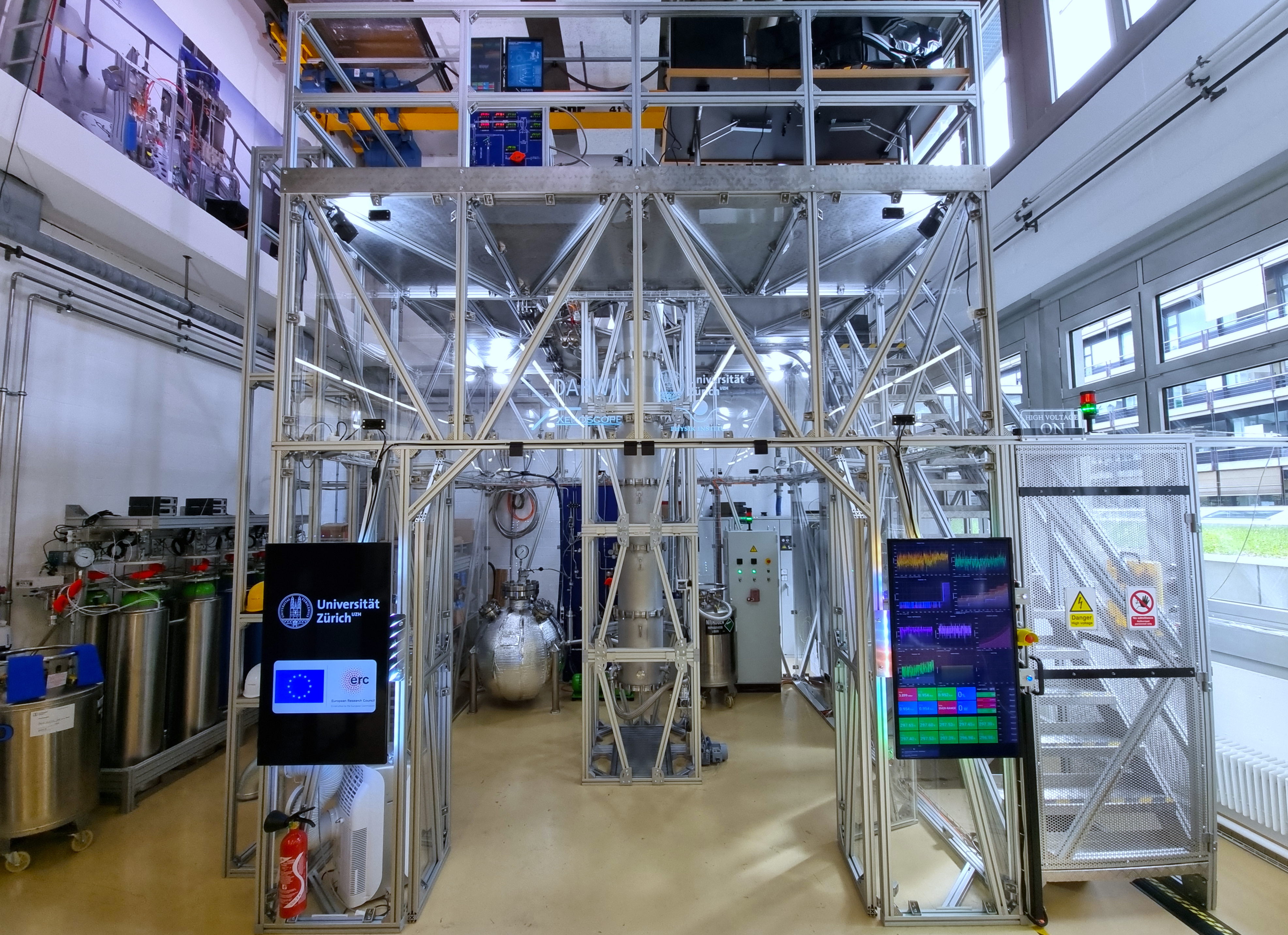}
    \includegraphics[width=0.49\textwidth]{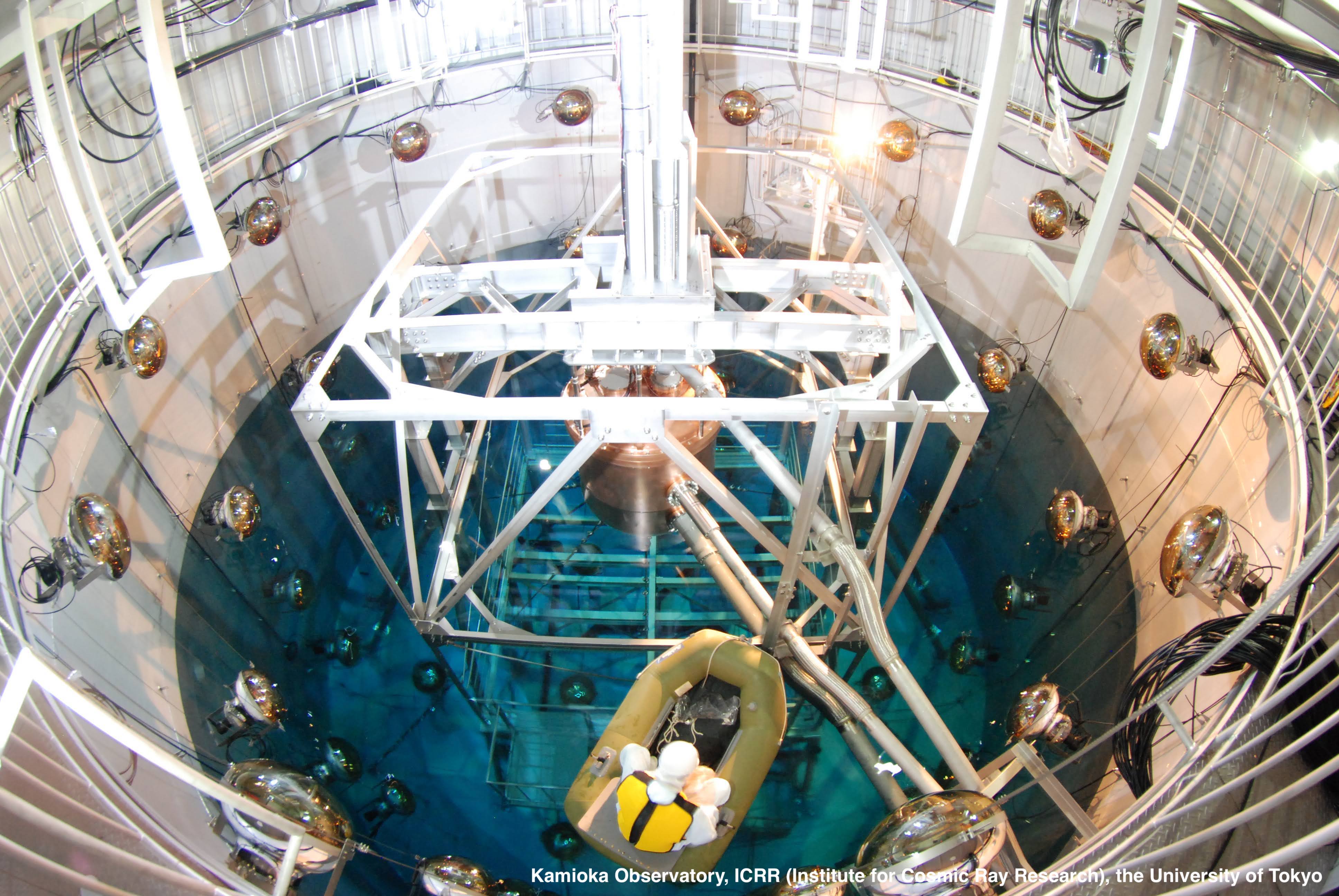}
    \includegraphics[width=0.49\textwidth, trim = {1.8cm, 0, 0.8cm ,0},clip]{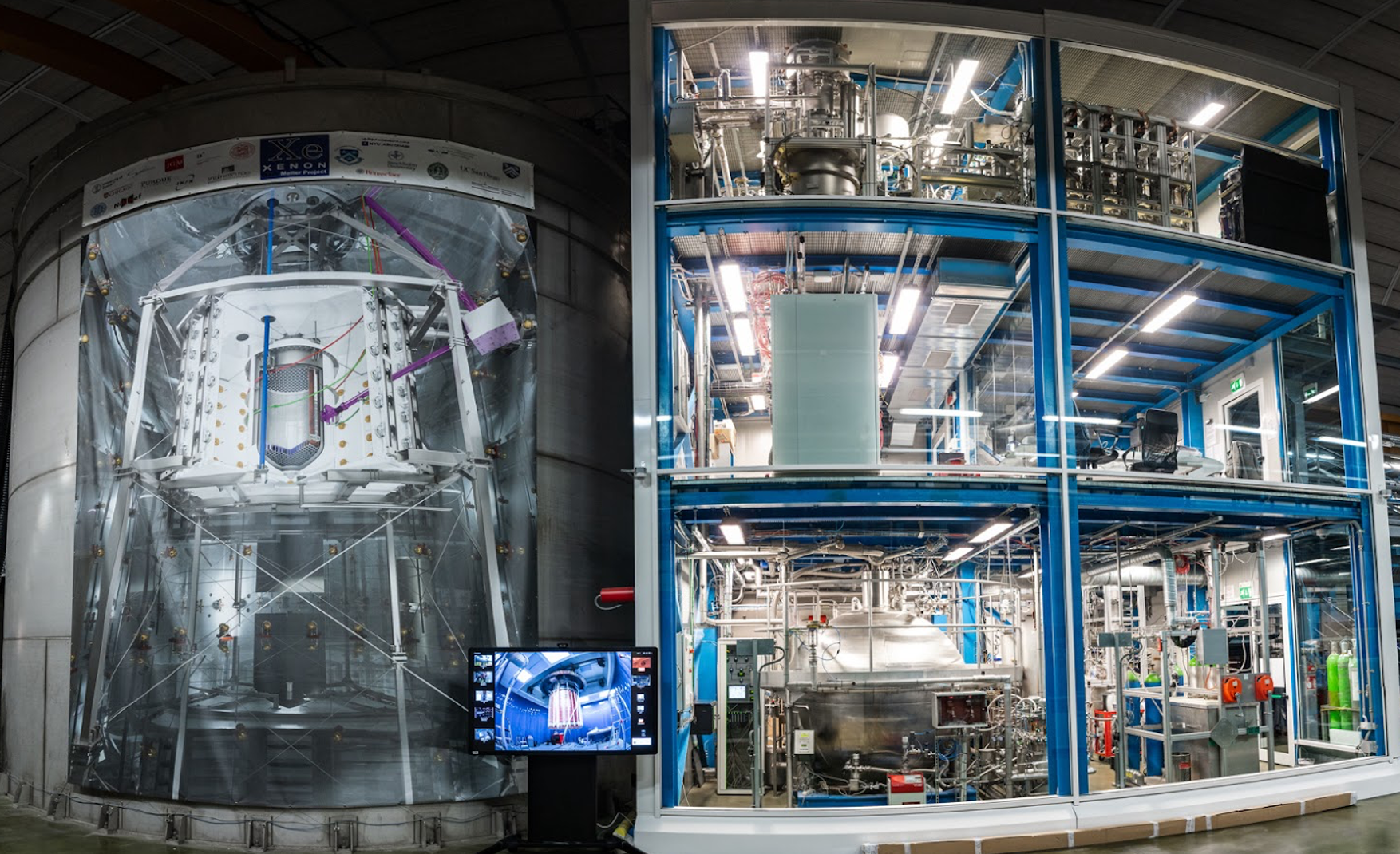}
    \caption{\small Existing large test facilities at surface labs (top left: Freiburg \cite{Brown_2024}, top right: Zurich \cite{Baudis:2021ipf}), and underground facility (bottom left: XMASS at Kamioka \cite{Abe_2013}) within the XLZD collaboration. The LZ and XENONnT (bottom right) experimental infrastructures at SURF and LNGS offer additional testing capabilities after completing their science programs.}
    \label{fig:testfacilities}
\end{figure*}

To limit dust deposition, the detector must be constructed within a cleanroom environment. Both XENONnT and LZ used class 1000 (ISO-6) clean rooms~\cite{Akerib:2020com,XENON:2021mrg}. During operations, the levels of dust need to be monitored while working within the space. This can be performed using sensitive dust counters and witness coupons, which can estimate the deposition and hence provide an indication of the contamination. It was shown with previous experiments that the dominant contribution was from personnel working within the clean spaces by bringing in dust on clean room suits or equipment and by disturbing the laminar air flow~\cite{Akerib:2020com}.  Therefore, methods to reduce quantities of dust and debris brought into the clean spaces will be examined more closely. The use of the air-ionizer fans also ensured that the dust was not attracted to charged surfaces, further limiting this deposition. Deposition levels of (214$\pm$22)\,ng/cm$^2$ were estimated from measurements on LZ, less than half of the target contamination~\cite{Akerib:2020com}.

\subsection{Accidental Coincidence Backgrounds}\label{sec:acbkg}
Although not a radioactive background, accidental coincidence events from lone-S1 and S2 signals of low intensity are observed in the XENONnT, LZ, and other similar LXe-TPC detectors. `Lone' refers to the observation of the S1 (S2) part of an event without the accompanying S2 (S1) signal. Typically, the lone-S1 pulses originate from the charge insensitive volume in or below the target and dark counts from PMTs. Lone-S2 pulses come from multiple sources such as low energy events with S1s too low to be detected, low energy $\beta$ decays from the cathode~\cite{Aprile:2019xxb} and delayed electrons~\cite{XENON:2021myl} attached on impurities or trapped below the liquid surface. Uncorrelated lone-S1 and S2 signals can combine to form accidental coincidence events, forming a background located largely in the low-energy region relevant to the study of low-mass WIMPs, CE$\nu$NS, and axions.  
The current generation detectors XENONnT and LZ, with their low backgrounds, are among the best resources available to characterize these backgrounds in detail. With a larger and taller target volume in XLZD, critical R\&D will be carried out to reduce the lone-S1 and lone-S2 pulses to control and reduce the accidental coincidence background to improve the sensitivity of the detector at the lowest energy threshold possible.

\section{Critical R\&D, Risk Mitigation, and Opportunities}
\label{sec:rnd}

The next-generation liquid xenon observatory XLZD will adopt most of the mature technologies that have already been demonstrated and rigorously tested in the LZ and XENON experiments. Further R\&D efforts will provide solutions to lessons learned from the current generation experiments. New and redundant technology choices provide risk mitigation and alternative solutions during the design stage.  A complete program of quality assurance and control for the production hardware will be required to meet the needs of scaling up and the longevity of the experiment. R\&D infrastructure will be used to test operational parameters and background mitigation strategies 
to employ during detector running. In addition, opportunities for detector upgrades and improved performance to increase physics sensitivity shall be explored. 

The areas of critical R\&D study link back to the fundamental operations of the liquid xenon TPC, as well as the veto systems: reducing external and internal backgrounds to unprecedented levels, enhancing light and charge collection to lower energy thresholds, and aiding recoil discrimination. These primary physics goals lead to studies of low background materials, radon reduction and removal, photosensors and surface reflectivity, high voltage and electrodes, xenon purification, reduction of spurious charge and light noises, Cherenkov and scintillator veto. Although in situ calibrations can never be replaced, continued studies of xenon microphysics, the charge and light yield, and their physical dependencies will also aid the larger project.

Many of these R\&D studies will be based on experience and mature technologies from the current generation of running experiments, which can be further improved at individual institutions, harnessing the established expertise of the collaboration. A key development for this scaled-up experiment will be larger test facilities both at the surface and underground. Large and multiple testing facilities allow for testing the integration of subcomponents, as well as accessing the unknowns that arrive in scaling up to mitigate the risks. Maintaining a suitable system to test issues that may arise in the main detector operations will also provide key information for the operational team. Large testing facilities within the collaboration (Fig.~\ref{fig:testfacilities}) already exist at surface and underground laboratories. Surface testing facilities will allow performance verification of critical R\&D components. Dedicated underground facilities will be used to investigate and certify the scaled-up components in a low background environment before installing them in the final detector.  

Finally, the XLZD program will allow opportunities to improve the physics sensitivity in the search for light and low-mass dark matter by doping light elements in the bulk liquid and/or in a single-phase operation, where the proportional S2 signal is produced in the liquid xenon phase \cite{Kuger:2021sxn}. Photosensors with higher photon detection efficiency and lower intrinsic radioactivity would further enhance the detector's capability with lower threshold, higher position and energy resolutions, and enhanced background discrimination to continue improving the sensitivities to dark matter. Filling the detector with a xenon target enriched in $^{136}$Xe would improve the sensitivity to neutrino-less double beta decays.

\section{Summary}

In this report, we present the experimental strategy and technology choices of the next-generation liquid xenon observatory XLZD for dark matter and neutrino physics with a nominal 60-80\,tonnes active liquid xenon target mass. Thanks to the development and extended operations of the current generation experiments XENONnT and LZ, we now have mature, and in various areas redundant, technology choices to embark on such a massive next-generation experiment. Further improvement and R\&D will continue to push the technological frontier, such as reducing the trace radioactivity background below the irreducible neutrino backgrounds to maximize the sensitivity for physics. Modest risks will be mitigated through extensive tests at existing large-scale testing facilities at surface and underground labs.

\section*{Acknowledgements}
This work was supported by the UKRI's Science \& Technology Facilities Council under awards ST/W000636/1, ST/Y003594/1, ST/Z000866/1, ST/W000547/1, ST/Y003586/1, ST/Z000807/1, ST/Y003527/1, ST/Z00084X/1, ST/W00058X/1, and ST/Z000831/1; the U.S. Department of Energy under contract numbers DE-SC0014223, DE-SC0012447, DE-AC02-76SF00515, and DE-SC0012704; the U.S. National Science Foundation under award number 2112802; the German Federal Ministry of Education and Research (BMBF) under the ErUM-Pro framework grant no. 05A23UM1, 05A23VF1 (CRESST-XENON-DARWIN), 05A23PM1, and 05A23VK3; the German Research Foundation (DFG) through the Research Training Group 2149 and the PRISMA\textsuperscript{+} Excellence Cluster project no. 390831469; the program ``Matter and the Universe" of the Helmholtz Association, through the Helmholtz Initiative and Networking Fund (grant no. W2/W3-118); the European Research Council (ERC) by the ERC Advanced Grants Xenoscope (no. 742789) and LowRad (no. 101055063), as well as by the European Union’s Horizon 2020 research and innovation programme under grant agreements 724320 (ULTIMATE), 101020842 (EUSTRONG), and Marie Skłodowska-Curie grant agreement No 860881-HIDDeN; the MCIN/AEI/10.13039/501100011033 from the following grants: PID2020-118758GB-I00, CNS2022-135716 funded by the ``European Union NextGenerationEU/PRTR" and CEX2019-000918-M to the ``Unit of Excellence Mar\'ia de Maeztu 2020-2023" award to the Institute of Cosmos Sciences; the Swiss National Science Foundation under grant numbers PCEFP2\_181117, 20FL20-216573, 200020-219290; the Next Generation EU, Italian National Recovery and Resilience Plan, Investment PE1–``Future Artificial Intelligence Research''; the Fondazione ICSC, Spoke 3 ``Astrophysics and Cosmos Observations'', the Italian National Recovery and Resilience Plan (Project ID CN00000013); the Italian Research Center on High-Performance Computing, Big Data and Quantum Computing funded by MUR Missione 4 Componente 2 Investimento 1.4.; the Istituto Nazionale di Fisica Nucleare (Italy); the Portuguese Foundation for Science and Technology (FCT) under award number PTDC/FIS-PAR/2831/2020; the JSPS Kakenhi under contract numbers 24H02240, 24H02236, 24K00659, and 24H00223 and by JST FOREST Program under contract number JPMJFR212Q; the Dutch Research Council (NWO) and the Dutch national e-infrastructure with the support of SURF Cooperative; the Israeli Science Foundation 1859/22 and the Weizmann Institute; the Australian Government through the Australian Research Council Centre of Excellence for Dark Matter Particle Physics (CDM, CE200100008); the Region des Pays de la Loire and the Australia-France Network of Doctoral Excellence (AUFRANDE) program. 
For the purpose of open access, the authors have applied a Creative Commons Attribution (CC BY) licence to any Author Accepted Manuscript version arising from this submission.

\bibliographystyle{JHEP}
\bibliography{xlzd_design_book}

\providecommand{\href}[2]{#2}\begingroup\raggedright\begin{thebibliography}{100}

\bibitem{Alner:2007ja}
{\scshape ZEPLIN} collaboration, G.~J. Alner et~al., \emph{{First limits on
  WIMP nuclear recoil signals in ZEPLIN-II: A two phase xenon detector for dark
  matter detection}},
  \href{http://dx.doi.org/10.1016/j.astropartphys.2007.06.002}{\emph{Astropart.
  Phys.} {\bfseries 28} (2007) 287--302},
  [\href{https://arxiv.org/abs/astro-ph/0701858}{{\ttfamily
  astro-ph/0701858}}].

\bibitem{Lebedenko:2008gb}
{\scshape ZEPLIN} collaboration, V.~N. Lebedenko et~al., \emph{{Result from the
  First Science Run of the ZEPLIN-III Dark Matter Search Experiment}},
  \href{http://dx.doi.org/10.1103/PhysRevD.80.052010}{\emph{Phys. Rev. D}
  {\bfseries 80} (2009) 052010},
  [\href{https://arxiv.org/abs/0812.1150}{{\ttfamily 0812.1150}}].

\bibitem{Angle:2007uj}
{\scshape XENON} collaboration, J.~Angle et~al., \emph{{First Results from the
  XENON10 Dark Matter Experiment at the Gran Sasso National Laboratory}},
  \href{http://dx.doi.org/10.1103/PhysRevLett.100.021303}{\emph{Phys. Rev.
  Lett.} {\bfseries 100} (2008) 021303},
  [\href{https://arxiv.org/abs/0706.0039}{{\ttfamily 0706.0039}}].

\bibitem{Aprile:2016swn}
{\scshape XENON} collaboration, E.~Aprile et~al., \emph{{XENON100 Dark Matter
  Results from a Combination of 477 Live Days}},
  \href{http://dx.doi.org/10.1103/PhysRevD.94.122001}{\emph{Phys. Rev.}
  {\bfseries D94} (2016) 122001},
  [\href{https://arxiv.org/abs/1609.06154}{{\ttfamily 1609.06154}}].

\bibitem{Aprile:2017iyp}
{\scshape XENON} collaboration, E.~Aprile et~al., \emph{{First Dark Matter
  Search Results from the XENON1T Experiment}},
  \href{http://dx.doi.org/10.1103/PhysRevLett.119.181301}{\emph{Phys. Rev.
  Lett.} {\bfseries 119} (2017) 181301},
  [\href{https://arxiv.org/abs/1705.06655}{{\ttfamily 1705.06655}}].

\bibitem{Akerib:2016vxi}
{\scshape LUX} collaboration, D.~S. Akerib et~al., \emph{{Results from a search
  for dark matter in the complete LUX exposure}},
  \href{http://dx.doi.org/10.1103/PhysRevLett.118.021303}{\emph{Phys. Rev.
  Lett.} {\bfseries 118} (2017) 021303},
  [\href{https://arxiv.org/abs/1608.07648}{{\ttfamily 1608.07648}}].

\bibitem{Xiao:2014xyn}
{\scshape PandaX} collaboration, M.~Xiao et~al., \emph{{First dark matter
  search results from the PandaX-I experiment}},
  \href{http://dx.doi.org/10.1007/s11433-014-5598-7}{\emph{Sci. China Phys.
  Mech. Astron.} {\bfseries 57} (2014) 2024--2030},
  [\href{https://arxiv.org/abs/1408.5114}{{\ttfamily 1408.5114}}].

\bibitem{Wang:2020coa}
{\scshape PandaX} collaboration, Q.~Wang et~al., \emph{{Results of dark matter
  search using the full PandaX-II exposure}},
  \href{http://dx.doi.org/10.1088/1674-1137/abb658}{\emph{Chin. Phys. C}
  {\bfseries 44} (2020) 125001},
  [\href{https://arxiv.org/abs/2007.15469}{{\ttfamily 2007.15469}}].

\bibitem{PandaX-4T:2021bab}
{\scshape PandaX} collaboration, Y.~Meng et~al., \emph{{Dark Matter Search
  Results from the PandaX-4T Commissioning Run}},
  \href{http://dx.doi.org/10.1103/PhysRevLett.127.261802}{\emph{Phys. Rev.
  Lett.} {\bfseries 127} (2021) 261802},
  [\href{https://arxiv.org/abs/2107.13438}{{\ttfamily 2107.13438}}].

\bibitem{LZ:2022ufs}
{\scshape LZ} collaboration, J.~Aalbers et~al., \emph{{First Dark Matter Search
  Results from the LUX-ZEPLIN (LZ) Experiment}},
  \href{http://dx.doi.org/10.1103/PhysRevLett.131.041002}{\emph{Phys. Rev.
  Lett.} {\bfseries 131} (2023) 041002},
  [\href{https://arxiv.org/abs/2207.03764}{{\ttfamily 2207.03764}}].

\bibitem{LZ_2024_DM_results}
{\scshape LZ} collaboration, J.~Aalbers et~al., \emph{{Dark Matter Search
  Results from $4.2\text{ }\text{
  }\text{Tonne}\text{\ensuremath{-}}\text{Years}$ of Exposure of the LUX-ZEPLIN
  (LZ) Experiment}}, \href{http://dx.doi.org/10.1103/4dyc-z8zf}{\emph{Phys.
  Rev. Lett.} {\bfseries 135} (2025) 011802},
  [\href{https://arxiv.org/abs/2410.17036}{{\ttfamily 2410.17036}}].

\bibitem{XENON:2023sxq}
{\scshape XENON} collaboration, E.~Aprile et~al., \emph{{First Dark Matter
  Search with Nuclear Recoils from the XENONnT Experiment}},
  \href{http://dx.doi.org/10.1103/PhysRevLett.131.041003}{\emph{Phys. Rev.
  Lett.} {\bfseries 131} (2023) 041003},
  [\href{https://arxiv.org/abs/2303.14729}{{\ttfamily 2303.14729}}].

\bibitem{aprile2025wimpdarkmattersearch}
{\scshape XENON Collaboration} collaboration, E.~Aprile et~al., \emph{{WIMP
  Dark Matter Search using a 3.1 tonne $\times$ year Exposure of the XENONnT
  Experiment}}, {\emph{arXiv} (2025) },
  [\href{https://arxiv.org/abs/2502.18005}{{\ttfamily 2502.18005}}].

\bibitem{PandaX-4T:2024dm}
{\scshape {PandaX}} collaboration, Z.~Bo et~al., \emph{{Dark Matter Search
  Results from 1.54 Tonne$\cdot$Year Exposure of PandaX-4T}},
  \href{http://dx.doi.org/10.1103/physrevlett.134.011805}{\emph{Physical Review
  Letters} {\bfseries 134} (2025) },
  [\href{https://arxiv.org/abs/2408.00664}{{\ttfamily 2408.00664}}].

\bibitem{Billard:2013qya}
J.~Billard, L.~Strigari and E.~Figueroa-Feliciano, \emph{{Implication of
  neutrino backgrounds on the reach of next generation dark matter direct
  detection experiments}},
  \href{http://dx.doi.org/10.1103/PhysRevD.89.023524}{\emph{Phys. Rev. D}
  {\bfseries 89} (2014) 023524},
  [\href{https://arxiv.org/abs/1307.5458}{{\ttfamily 1307.5458}}].

\bibitem{O'Hare:2015mda}
C.~A.~J. O'Hare, A.~M. Green, J.~Billard, E.~Figueroa-Feliciano and L.~E.
  Strigari, \emph{{Readout strategies for directional dark matter detection
  beyond the neutrino background}},
  \href{http://dx.doi.org/10.1103/PhysRevD.92.063518}{\emph{Phys. Rev. D}
  {\bfseries 92} (2015) 063518},
  [\href{https://arxiv.org/abs/1505.08061}{{\ttfamily 1505.08061}}].

\bibitem{OHare:2021utq}
C.~A.~J. O'Hare, \emph{{Fog on the horizon: a new definition of the neutrino
  floor for direct dark matter searches}},
  \href{http://dx.doi.org/10.1103/physrevlett.127.251802}{\emph{Physical Review
  Letters} {\bfseries 127} (9, 2021) },
  [\href{https://arxiv.org/abs/2109.03116}{{\ttfamily 2109.03116}}].

\bibitem{Aalbers:2022dzr}
J.~Aalbers et~al., \emph{{A next-generation liquid xenon observatory for dark
  matter and neutrino physics}},
  \href{http://dx.doi.org/10.1088/1361-6471/ac841a}{\emph{J. Phys. G}
  {\bfseries 50} (2023) 013001},
  [\href{https://arxiv.org/abs/2203.02309}{{\ttfamily 2203.02309}}].

\bibitem{Kopec:2023uii}
A.~Kopec, \emph{{Design Challenges for a Future Liquid Xenon Observatory}},
  \href{http://dx.doi.org/10.31526/jais.2024.480}{\emph{JAIS} {\bfseries 2024}
  (2024) 480}, [\href{https://arxiv.org/abs/2310.00722}{{\ttfamily
  2310.00722}}].

\bibitem{Clowe:2006eq}
D.~Clowe et~al., \emph{{A direct empirical proof of the existence of dark
  matter}}, \href{http://dx.doi.org/10.1086/508162}{\emph{Astrophys. J. Lett.}
  {\bfseries 648} (2006) L109--L113},
  [\href{https://arxiv.org/abs/astro-ph/0608407}{{\ttfamily
  astro-ph/0608407}}].

\bibitem{aghanim2020planck}
N.~Aghanim et~al., \emph{{Planck2018 results: VI. Cosmological parameters}},
  \href{http://dx.doi.org/10.1051/0004-6361/201833910}{\emph{Astronomy \&
  Astrophysics} {\bfseries 641} (2020) A6},
  [\href{https://arxiv.org/abs/1807.06209}{{\ttfamily 1807.06209}}].

\bibitem{gelmini2010dm}
G.~Gelmini and P.~Gondolo, \emph{{DM Production Mechanisms}},  in
  \emph{Particle Dark Matter: Observations, Models and Searches} ({G. Bertone},
  ed.), pp.~121--141.
\newblock Cambridge U. Press, 2010.
\newblock \href{https://arxiv.org/abs/1009.3690}{{\ttfamily 1009.3690}}.

\bibitem{Jungman:1995df}
G.~Jungman, M.~Kamionkowski and K.~Griest, \emph{{Supersymmetric dark matter}},
  \href{http://dx.doi.org/10.1016/0370-1573(95)00058-5}{\emph{Phys. Rept.}
  {\bfseries 267} (1996) 195--373},
  [\href{https://arxiv.org/abs/9506380}{{\ttfamily 9506380}}].

\bibitem{Bertone:2010zza}
G.~Bertone, ed., \emph{{Particle Dark Matter: Observations, Models and
  Searches}}.
\newblock Cambridge Univ. Press, Cambridge, 2010,
  \href{http://dx.doi.org/10.1017/CBO9780511770739}{10.1017/CBO9780511770739}.

\bibitem{AKIMOV201214}
D.~Akimov et~al., \emph{{WIMP-nucleon cross-section results from the second
  science run of ZEPLIN-III}},
  \href{http://dx.doi.org/https://doi.org/10.1016/j.physletb.2012.01.064}{\emph{Physics
  Letters B} {\bfseries 709} (2012) 14--20},
  [\href{https://arxiv.org/abs/1110.4769}{{\ttfamily 1110.4769}}].

\bibitem{Aprile:2018dbl}
{\scshape XENON} collaboration, E.~Aprile et~al., \emph{{Dark Matter Search
  Results from a One Ton-Year Exposure of XENON1T}},
  \href{http://dx.doi.org/10.1103/PhysRevLett.121.111302}{\emph{Phys. Rev.
  Lett.} {\bfseries 121} (2018) 111302},
  [\href{https://arxiv.org/abs/1805.12562}{{\ttfamily 1805.12562}}].

\bibitem{Bloch:2024}
I.~M. Bloch, S.~Bottaro, D.~Redigolo and L.~Vittorio, \emph{{Looking for WIMPs
  through the neutrino fogs}}, {\emph{arXiv} (2024) },
  [\href{https://arxiv.org/abs/2410.02723}{{\ttfamily 2410.02723}}].

\bibitem{Blanco:2019hah}
C.~Blanco, M.~Escudero, D.~Hooper and S.~J. Witte, \emph{{$Z'$ mediated WIMPs:
  dead, dying, or soon to be detected?}},
  \href{http://dx.doi.org/10.1088/1475-7516/2019/11/024}{\emph{JCAP} {\bfseries
  11} (2019) 024}, [\href{https://arxiv.org/abs/1907.05893}{{\ttfamily
  1907.05893}}].

\bibitem{ATLAS2024}
{\scshape ATLAS} collaboration, G.~Aad et~al., \emph{{ATLAS Run 2 searches for
  electroweak production of supersymmetric particles interpreted within the
  pMSSM}}, \href{http://dx.doi.org/10.1007/jhep05(2024)106}{\emph{Journal of
  High Energy Physics} {\bfseries 2024} (2024) }.

\bibitem{Ellis_2023}
J.~Ellis, K.~A. Olive, V.~C. Spanos and I.~D. Stamou, \emph{{The CMSSM survives
  Planck, the LHC, LUX-ZEPLIN, Fermi-LAT, H.E.S.S. and IceCube}},
  \href{http://dx.doi.org/10.1140/epjc/s10052-023-11405-1}{\emph{The European
  Physical Journal C} {\bfseries 83} (2023) },
  [\href{https://arxiv.org/abs/2210.16337}{{\ttfamily 2210.16337}}].

\bibitem{Fieguth:2018vob}
A.~Fieguth, M.~Hoferichter, P.~Klos, J.~Men{\'e}ndez, A.~Schwenk and
  C.~Weinheimer, \emph{{Discriminating WIMP-nucleus response functions in
  present and future XENON-like direct detection experiments}},
  \href{http://dx.doi.org/10.1103/PhysRevD.97.103532}{\emph{Phys. Rev. D}
  {\bfseries 97} (2018) 103532},
  [\href{https://arxiv.org/abs/1802.04294}{{\ttfamily 1802.04294}}].

\bibitem{Aprile:2018cxk}
{\scshape XENON} collaboration, E.~Aprile et~al., \emph{{First results on the
  scalar WIMP-pion coupling, using the XENON1T experiment}},
  \href{http://dx.doi.org/10.1103/PhysRevLett.122.071301}{\emph{Phys. Rev.
  Lett.} {\bfseries 122} (2019) 071301},
  [\href{https://arxiv.org/abs/1811.12482}{{\ttfamily 1811.12482}}].

\bibitem{Baudis:2013bba}
L.~Baudis et~al., \emph{{Signatures of Dark Matter Scattering Inelastically Off
  Nuclei}}, \href{http://dx.doi.org/10.1103/PhysRevD.88.115014}{\emph{Phys.
  Rev. D} {\bfseries 88} (2013) 115014},
  [\href{https://arxiv.org/abs/1309.0825}{{\ttfamily 1309.0825}}].

\bibitem{LZ:2023lvz}
{\scshape LZ} collaboration, J.~Aalbers et~al., \emph{{First constraints on
  WIMP-nucleon effective field theory couplings in an extended energy region
  from LUX-ZEPLIN}},
  \href{http://dx.doi.org/10.1103/PhysRevD.109.092003}{\emph{Phys. Rev. D}
  {\bfseries 109} (2024) 092003},
  [\href{https://arxiv.org/abs/2312.02030}{{\ttfamily 2312.02030}}].

\bibitem{Hoferichter:2016nvd}
M.~Hoferichter, P.~Klos, J.~Men{\'e}ndez and A.~Schwenk, \emph{{Analysis
  strategies for general spin-independent WIMP-nucleus scattering}},
  \href{http://dx.doi.org/10.1103/PhysRevD.94.063505}{\emph{Phys. Rev. D}
  {\bfseries 94} (2016) 063505},
  [\href{https://arxiv.org/abs/1605.08043}{{\ttfamily 1605.08043}}].

\bibitem{Hoferichter:2018acd}
M.~Hoferichter, P.~Klos, J.~Men\'endez and A.~Schwenk, \emph{{Nuclear structure
  factors for general spin-independent WIMP-nucleus scattering}},
  \href{http://dx.doi.org/10.1103/PhysRevD.99.055031}{\emph{Phys. Rev. D}
  {\bfseries 99} (2019) 055031},
  [\href{https://arxiv.org/abs/1812.05617}{{\ttfamily 1812.05617}}].

\bibitem{Essig:2012yx}
R.~Essig, A.~Manalaysay, J.~Mardon, P.~Sorensen and T.~Volansky, \emph{{First
  Direct Detection Limits on sub-GeV Dark Matter from XENON10}},
  \href{http://dx.doi.org/10.1103/PhysRevLett.109.021301}{\emph{Phys. Rev.
  Lett.} {\bfseries 109} (2012) 021301},
  [\href{https://arxiv.org/abs/1206.2644}{{\ttfamily 1206.2644}}].

\bibitem{Pospelov:2008jk}
M.~Pospelov, A.~Ritz and M.~B. Voloshin, \emph{{Bosonic super-WIMPs as
  keV-scale dark matter}},
  \href{http://dx.doi.org/10.1103/PhysRevD.78.115012}{\emph{Phys. Rev. D}
  {\bfseries 78} (2008) 115012},
  [\href{https://arxiv.org/abs/0807.3279}{{\ttfamily 0807.3279}}].

\bibitem{Essig:2013lka}
R.~Essig et~al., \emph{{Working Group Report: New Light Weakly Coupled
  Particles}},  in \emph{{Community Summer Study 2013: Snowmass on the
  Mississippi (CSS2013) Minneapolis, MN, USA, July 29-August 6, 2013}}, 2013.
\newblock \href{https://arxiv.org/abs/1311.0029}{{\ttfamily 1311.0029}}.

\bibitem{Abbott:1982af}
L.~Abbott and P.~Sikivie, \emph{{A Cosmological Bound on the Invisible Axion}},
  \href{http://dx.doi.org/10.1016/0370-2693(83)90638-X}{\emph{Phys. Lett. B}
  {\bfseries 120} (1983) 133--136}.

\bibitem{raffelt2002axions}
G.~Raffelt, \emph{Axions},
  \href{http://dx.doi.org/10.1023/A:1015822212542}{\emph{Space science reviews}
  {\bfseries 100} (2002) 153--158}.

\bibitem{Preskill:1982cy}
J.~Preskill, M.~B. Wise and F.~Wilczek, \emph{{Cosmology of the Invisible
  Axion}}, \href{http://dx.doi.org/10.1016/0370-2693(83)90637-8}{\emph{Phys.
  Lett. B} {\bfseries 120} (1983) 127--132}.

\bibitem{duffy2009axions}
L.~D. Duffy and K.~Van~Bibber, \emph{Axions as dark matter particles},
  {\emph{New Journal of Physics} {\bfseries 11} (2009) 105008},
  [\href{https://arxiv.org/abs/0904.3346}{{\ttfamily 0904.3346}}].

\bibitem{XENON:2023iku}
{\scshape XENON} collaboration, E.~Aprile et~al., \emph{{Searching for Heavy
  Dark Matter near the Planck Mass with XENON1T}},
  \href{http://dx.doi.org/10.1103/PhysRevLett.130.261002}{\emph{Phys. Rev.
  Lett.} {\bfseries 130} (2023) 261002},
  [\href{https://arxiv.org/abs/2304.10931}{{\ttfamily 2304.10931}}].

\bibitem{Akerib:2015rjg}
{\scshape LUX} collaboration, D.~Akerib et~al., \emph{{Improved Limits on
  Scattering of Weakly Interacting Massive Particles from Reanalysis of 2013
  LUX Data}},
  \href{http://dx.doi.org/10.1103/PhysRevLett.116.161301}{\emph{Phys. Rev.
  Lett.} {\bfseries 116} (2016) 161301},
  [\href{https://arxiv.org/abs/1512.03506}{{\ttfamily 1512.03506}}].

\bibitem{Schumann:2015cpa}
M.~Schumann, L.~Baudis, L.~B\"utikofer, A.~Kish and M.~Selvi, \emph{{Dark
  matter sensitivity of multi-ton liquid xenon detectors}},
  \href{http://dx.doi.org/10.1088/1475-7516/2015/10/016}{\emph{JCAP} {\bfseries
  10} (2015) 016}, [\href{https://arxiv.org/abs/1506.08309}{{\ttfamily
  1506.08309}}].

\bibitem{Agostini:2023}
M.~Agostini, G.~Benato, J.~A. Detwiler, J.~Men\'endez and F.~Vissani,
  \emph{{Toward the discovery of matter creation with neutrinoless
  $\ensuremath{\beta}\ensuremath{\beta}$ decay}},
  \href{http://dx.doi.org/10.1103/RevModPhys.95.025002}{\emph{Rev. Mod. Phys.}
  {\bfseries 95} (May, 2023) 025002},
  [\href{https://arxiv.org/abs/2202.01787}{{\ttfamily 2202.01787}}].

\bibitem{nEXO:2021ujk}
{\scshape nEXO} collaboration, G.~Adhikari et~al., \emph{{nEXO: neutrinoless
  double beta decay search beyond 10$^{28}$ year half-life sensitivity}},
  \href{http://dx.doi.org/10.1088/1361-6471/ac3631}{\emph{J. Phys. G}
  {\bfseries 49} (2022) 015104},
  [\href{https://arxiv.org/abs/2106.16243}{{\ttfamily 2106.16243}}].

\bibitem{NEXT:2020amj}
{\scshape NEXT} collaboration, C.~Adams et~al., \emph{{Sensitivity of a
  tonne-scale NEXT detector for neutrinoless double beta decay searches}},
  \href{http://dx.doi.org/10.1007/JHEP08(2021)164}{\emph{JHEP} {\bfseries 2021}
  (2021) 164}, [\href{https://arxiv.org/abs/2005.06467}{{\ttfamily
  2005.06467}}].

\bibitem{Agostini:2017fom}
M.~Agostini, G.~Benato and J.~A. Detwiler, \emph{Discovery probability of
  next-generation neutrinoless double-$\ensuremath{\beta}$ decay experiments},
  \href{http://dx.doi.org/10.1103/PhysRevD.96.053001}{\emph{Phys. Rev. D}
  {\bfseries 96} (Sep, 2017) 053001}.

\bibitem{xlzd:2024}
{\scshape {XLZD}} collaboration, J.~Aalbers et~al., \emph{{Neutrinoless Double
  Beta Decay Sensitivity of the XLZD Rare Event Observatory}},
  \href{http://dx.doi.org/10.1088/1361-6471/adb900}{\emph{Journal of Physics G:
  Nuclear and Particle Physics} {\bfseries 52} (2025) 045102},
  [\href{https://arxiv.org/abs/2410.19016}{{\ttfamily 2410.19016}}].

\bibitem{Pereira_2023}
{\scshape LZ} collaboration, G.~Pereira, C.~Silva and V.~N. Solovov,
  \emph{{Energy resolution of the LZ detector for high-energy electronic
  recoils}},
  \href{http://dx.doi.org/10.1088/1748-0221/18/04/C04007}{\emph{JINST}
  {\bfseries 18} (2023) C04007}.

\bibitem{Aprile:2020yad}
{\scshape XENON} collaboration, E.~Aprile et~al., \emph{{Energy resolution and
  linearity of XENON1T in the MeV energy range}},
  \href{http://dx.doi.org/10.1140/epjc/s10052-020-8284-0}{\emph{Eur. Phys. J.
  C} {\bfseries 80} (2020) 785},
  [\href{https://arxiv.org/abs/2003.03825}{{\ttfamily 2003.03825}}].

\bibitem{DARWIN:2023uje}
{\scshape DARWIN} collaboration, M.~Adrover et~al., \emph{{Cosmogenic
  background simulations for the DARWIN observatory at different underground
  locations}},
  \href{http://dx.doi.org/10.1140/epjc/s10052-023-12298-w}{\emph{Eur. Phys. J.
  C} {\bfseries 84} (2024) 88},
  [\href{https://arxiv.org/abs/2306.16340}{{\ttfamily 2306.16340}}].

\bibitem{Baudis:2013qla}
L.~Baudis, A.~Ferella, A.~Kish, A.~Manalaysay, T.~Marrodan~Undagoitia and
  M.~Schumann, \emph{{Neutrino physics with multi-ton scale liquid xenon
  detectors}},
  \href{http://dx.doi.org/10.1088/1475-7516/2014/01/044}{\emph{JCAP} {\bfseries
  1401} (2014) 044}, [\href{https://arxiv.org/abs/1309.7024}{{\ttfamily
  1309.7024}}].

\bibitem{Agostini:2018uly}
{\scshape BOREXINO} collaboration, M.~Agostini et~al., \emph{{Comprehensive
  measurement of $pp$-chain solar neutrinos}},
  \href{http://dx.doi.org/10.1038/s41586-018-0624-y}{\emph{Nature} {\bfseries
  562} (2018) 505--510}.

\bibitem{Agostini_2020}
M.~Agostini et~al., \emph{{Improved measurement of $^{8}$B solar neutrinos with
  1.5 kt y of Borexino exposure}},
  \href{http://dx.doi.org/10.1103/physrevd.101.062001}{\emph{Physical Review D}
  {\bfseries 101} (2020) }, [\href{https://arxiv.org/abs/1709.00756}{{\ttfamily
  1709.00756}}].

\bibitem{KamLAND:2011fld}
{\scshape KamLAND} collaboration, S.~Abe et~al., \emph{{Measurement of the
  $^{8}B$ Solar Neutrino Flux with the KamLAND Liquid Scintillator Detector}},
  \href{http://dx.doi.org/10.1103/PhysRevC.84.035804}{\emph{Phys. Rev. C}
  {\bfseries 84} (2011) 035804},
  [\href{https://arxiv.org/abs/1106.0861}{{\ttfamily 1106.0861}}].

\bibitem{Aharmim:2011vm}
{\scshape SNO} collaboration, B.~Aharmim et~al., \emph{{Combined Analysis of
  all Three Phases of Solar Neutrino Data from the Sudbury Neutrino
  Observatory}},
  \href{http://dx.doi.org/10.1103/PhysRevC.88.025501}{\emph{Phys. Rev. C}
  {\bfseries 88} (2013) 025501},
  [\href{https://arxiv.org/abs/1109.0763}{{\ttfamily 1109.0763}}].

\bibitem{Capozzi:2017}
F.~Capozzi, E.~Di~Valentino, E.~Lisi, A.~Marrone, A.~Melchiorri and A.~Palazzo,
  \emph{{Global constraints on absolute neutrino masses and their ordering}},
  \href{http://dx.doi.org/10.1103/PhysRevD.95.096014}{\emph{Phys. Rev. D}
  {\bfseries 95} (May, 2017) 096014},
  [\href{https://arxiv.org/abs/1703.04471}{{\ttfamily 1703.04471}}].

\bibitem{Freedman:1973yd}
D.~Z. Freedman, \emph{{Coherent Neutrino Nucleus Scattering as a Probe of the
  Weak Neutral Current}},
  \href{http://dx.doi.org/10.1103/PhysRevD.9.1389}{\emph{Phys. Rev. D}
  {\bfseries 9} (1974) 1389--1392}.

\bibitem{aprile2024measurementsolar8bneutrinos}
{\scshape {XENON}} collaboration, E.~Aprile et~al., \emph{{First Indication of
  Solar $^8$B Neutrinos via Coherent Elastic Neutrino-Nucleus Scattering with
  XENONnT}},
  \href{http://dx.doi.org/10.1103/physrevlett.133.191002}{\emph{Physical Review
  Letters} {\bfseries 133} (2024) },
  [\href{https://arxiv.org/abs/2408.02877}{{\ttfamily 2408.02877}}].

\bibitem{pandaxcollaboration2024indicationsolar8bneutrino}
{\scshape {PandaX}} collaboration, Z.~Bo et~al., \emph{{First Indication of
  Solar $^8$B Neutrino Flux through Coherent Elastic Neutrino-Nucleus
  Scattering in PandaX-4T}},
  \href{http://dx.doi.org/10.1103/PhysRevLett.133.191001}{\emph{Phys. Rev.
  Lett.} {\bfseries 133} (2024) 191001},
  [\href{https://arxiv.org/abs/2407.10892}{{\ttfamily 2407.10892}}].

\bibitem{Aalbers:2016jon}
{\scshape DARWIN} collaboration, J.~Aalbers et~al., \emph{{DARWIN: towards the
  ultimate dark matter detector}},
  \href{http://dx.doi.org/10.1088/1475-7516/2016/11/017}{\emph{JCAP} {\bfseries
  11} (2016) 017}, [\href{https://arxiv.org/abs/1606.07001}{{\ttfamily
  1606.07001}}].

\bibitem{Hoferichter:2020osn}
M.~Hoferichter, J.~Men\'endez and A.~Schwenk, \emph{{Coherent elastic
  neutrino-nucleus scattering: EFT analysis and nuclear responses}},
  \href{http://dx.doi.org/10.1103/PhysRevD.102.074018}{\emph{Phys. Rev. D}
  {\bfseries 102} (2020) 074018},
  [\href{https://arxiv.org/abs/2007.08529}{{\ttfamily 2007.08529}}].

\bibitem{Aalbers:2020gsn}
{\scshape DARWIN} collaboration, J.~Aalbers et~al., \emph{{Solar neutrino
  detection sensitivity in DARWIN via electron scattering}},
  \href{http://dx.doi.org/10.1140/epjc/s10052-020-08602-7}{\emph{Eur. Phys. J.
  C} {\bfseries 80} (2020) 1133},
  [\href{https://arxiv.org/abs/2006.03114}{{\ttfamily 2006.03114}}].

\bibitem{PhysRevD.106.096017}
A.~de~Gouv\^ea, E.~McGinness, I.~Martinez-Soler and Y.~F. Perez-Gonzalez,
  \emph{{$pp$ solar neutrinos at DARWIN}},
  \href{http://dx.doi.org/10.1103/PhysRevD.106.096017}{\emph{Phys. Rev. D}
  {\bfseries 106} (Nov, 2022) 096017},
  [\href{https://arxiv.org/abs/2111.02421}{{\ttfamily 2111.02421}}].

\bibitem{Lang:2016zhv}
R.~F. Lang, C.~McCabe, S.~Reichard, M.~Selvi and I.~Tamborra, \emph{{Supernova
  neutrino physics with xenon dark matter detectors: A timely perspective}},
  \href{http://dx.doi.org/10.1103/PhysRevD.94.103009}{\emph{Phys. Rev. D}
  {\bfseries 94} (2016) 103009},
  [\href{https://arxiv.org/abs/1606.09243}{{\ttfamily 1606.09243}}].

\bibitem{Antonioli:2004zb}
{\scshape SNEWS} collaboration, P.~Antonioli et~al., \emph{{SNEWS: The
  Supernova Early Warning System}},
  \href{http://dx.doi.org/10.1088/1367-2630/6/1/114}{\emph{New J. Phys.}
  {\bfseries 6} (2004) 114},
  [\href{https://arxiv.org/abs/astro-ph/0406214}{{\ttfamily
  astro-ph/0406214}}].

\bibitem{Kharusi:2020ovw}
{\scshape SNEWS} collaboration, S.~Al~Kharusi et~al., \emph{{SNEWS 2.0: a
  next-generation supernova early warning system for multi-messenger
  astronomy}}, \href{http://dx.doi.org/10.1088/1367-2630/abde33}{\emph{New J.
  Phys.} {\bfseries 23} (2021) 031201},
  [\href{https://arxiv.org/abs/2011.00035}{{\ttfamily 2011.00035}}].

\bibitem{Ong:2017ihp}
{\scshape CTA} collaboration, R.~A. Ong, \emph{{Cherenkov Telescope Array: The
  Next Generation Gamma-ray Observatory}},
  \href{http://dx.doi.org/10.22323/1.301.1071}{\emph{PoS} {\bfseries ICRC2017}
  (2018) 1071}, [\href{https://arxiv.org/abs/1709.05434}{{\ttfamily
  1709.05434}}].

\bibitem{FermiLAT_2015}
M.~Ackermann et~al., \emph{{Searching for Dark Matter Annihilation from Milky
  Way Dwarf Spheroidal Galaxies with Six Years of Fermi Large Area Telescope
  Data}},
  \href{http://dx.doi.org/10.1103/physrevlett.115.231301}{\emph{Physical Review
  Letters} {\bfseries 115} (2015) },
  [\href{https://arxiv.org/abs/1503.02641}{{\ttfamily 1503.02641}}].

\bibitem{abreu2019_swgo}
P.~Abreu et~al., \emph{{The Southern Wide-Field Gamma-Ray Observatory (SWGO): A
  Next-Generation Ground-Based Survey Instrument for VHE Gamma-Ray Astronomy}},
  {\emph{arXiv} (2019) }, [\href{https://arxiv.org/abs/1907.07737}{{\ttfamily
  1907.07737}}].

\bibitem{Arduini:2016xsb}
G.~Arduini et~al., \emph{{High Luminosity LHC: challenges and plans}},
  \href{http://dx.doi.org/10.1088/1748-0221/11/12/C12081}{\emph{JINST}
  {\bfseries 11} (2016) C12081}.

\bibitem{PandaX:2024oxq}
{\scshape PandaX} collaboration, A.~Abdukerim et~al., \emph{{PandaX-xT: A deep
  underground multi-ten-tonne liquid xenon observatory}},
  \href{http://dx.doi.org/10.1007/s11433-024-2539-y}{\emph{Sci. China Phys.
  Mech. Astron.} {\bfseries 68} (2025) 221011},
  [\href{https://arxiv.org/abs/2402.03596}{{\ttfamily 2402.03596}}].

\bibitem{Agnes_2021}
P.~Agnes et~al., \emph{{Sensitivity of future liquid argon dark matter search
  experiments to core-collapse supernova neutrinos}},
  \href{http://dx.doi.org/10.1088/1475-7516/2021/03/043}{\emph{Journal of
  Cosmology and Astroparticle Physics} {\bfseries 2021} (2021) 043},
  [\href{https://arxiv.org/abs/2011.07819}{{\ttfamily 2011.07819}}].

\bibitem{Pato:2010zk}
M.~Pato, L.~Baudis, G.~Bertone, R.~Ruiz~de Austri, L.~E. Strigari and
  R.~Trotta, \emph{{Complementarity of Dark Matter Direct Detection Targets}},
  \href{http://dx.doi.org/10.1103/PhysRevD.83.083505}{\emph{Phys. Rev. D}
  {\bfseries 83} (2011) 083505},
  [\href{https://arxiv.org/abs/1012.3458}{{\ttfamily 1012.3458}}].

\bibitem{Bozorgnia_2018}
N.~Bozorgnia, D.~G. Cerdeño, A.~Cheek and B.~Penning, \emph{{Opening the
  energy window on direct dark matter detection}},
  \href{http://dx.doi.org/10.1088/1475-7516/2018/12/013}{\emph{JCAP} {\bfseries
  1812} (2018) 013}, [\href{https://arxiv.org/abs/1810.05576}{{\ttfamily
  1810.05576}}].

\bibitem{Chepel:2012sj}
V.~Chepel and H.~Araujo, \emph{{Liquid noble gas detectors for low energy
  particle physics}},
  \href{http://dx.doi.org/10.1088/1748-0221/8/04/R04001}{\emph{JINST}
  {\bfseries 8} (2013) R04001},
  [\href{https://arxiv.org/abs/1207.2292}{{\ttfamily 1207.2292}}].

\bibitem{Nakamura:2020szx}
R.~Nakamura, H.~Sambonsugi, K.~Shiraishi and Y.~Wada, \emph{{Research and
  development toward KamLAND2-Zen}},
  \href{http://dx.doi.org/10.1088/1742-6596/1468/1/012256}{\emph{J. Phys. Conf.
  Ser.} {\bfseries 1468} (2020) 012256}.

\bibitem{Albanese_2021}
V.~Albanese et~al., \emph{{The SNO+ experiment}},
  \href{http://dx.doi.org/10.1088/1748-0221/16/08/p08059}{\emph{Journal of
  Instrumentation} {\bfseries 16} (2021) P08059},
  [\href{https://arxiv.org/abs/2104.11687}{{\ttfamily 2104.11687}}].

\bibitem{legendcollaboration2021legend1000}
{\scshape LEGEND} collaboration, N.~Abgrall et~al., \emph{Legend-1000
  preconceptual design report}, {\emph{arXiv} (2021) },
  [\href{https://arxiv.org/abs/2107.11462}{{\ttfamily 2107.11462}}].

\bibitem{CUPIDInterestGroup:2019inu}
{\scshape CUPID} collaboration, W.~Armstrong et~al., \emph{{CUPID pre-CDR}},
  {\emph{arXiv} (2019) }, [\href{https://arxiv.org/abs/1907.09376}{{\ttfamily
  1907.09376}}].

\bibitem{amore2021}
Y.~Oh and the AMoRE~Collaboration, \emph{{Status of AMoRE}},
  \href{http://dx.doi.org/10.1088/1742-6596/2156/1/012146}{\emph{J. Phys.:
  Conf. Ser.} (2021) }.

\bibitem{Arnold_2010}
R.~Arnold et~al., \emph{{Probing new physics models of neutrinoless double beta
  decay with SuperNEMO}},
  \href{http://dx.doi.org/10.1140/epjc/s10052-010-1481-5}{\emph{The European
  Physical Journal C} {\bfseries 70} (2010) 927–943},
  [\href{https://arxiv.org/abs/1005.1241}{{\ttfamily 1005.1241}}].

\bibitem{Akimov:2017ade}
{\scshape COHERENT} collaboration, D.~Akimov et~al., \emph{{Observation of
  Coherent Elastic Neutrino-Nucleus Scattering}},
  \href{http://dx.doi.org/10.1126/science.aao0990}{\emph{Science} {\bfseries
  357} (2017) 1123--1126}, [\href{https://arxiv.org/abs/1708.01294}{{\ttfamily
  1708.01294}}].

\bibitem{Akimov:2020pdx}
{\scshape COHERENT} collaboration, D.~Akimov et~al., \emph{{First Measurement
  of Coherent Elastic Neutrino-Nucleus Scattering on Argon}},
  \href{http://dx.doi.org/10.1103/PhysRevLett.126.012002}{\emph{Phys. Rev.
  Lett.} {\bfseries 126} (2021) 012002},
  [\href{https://arxiv.org/abs/2003.10630}{{\ttfamily 2003.10630}}].

\bibitem{Angloher:2019}
{\scshape NUCLEUS} collaboration, G.~Angloher et~al., \emph{{Exploring $\hbox
  {CE}\nu \hbox {NS}$ with NUCLEUS at the Chooz nuclear power plant}},
  \href{http://dx.doi.org/10.1140/epjc/s10052-019-7454-4}{\emph{Eur. Phys. J.
  C} {\bfseries 79} (2019) 1018},
  [\href{https://arxiv.org/abs/1905.10258}{{\ttfamily 1905.10258}}].

\bibitem{ackermann2025observationreactorantineutrinoscoherent}
{N. Ackermann and others}, \emph{{First observation of reactor antineutrinos by
  coherent scattering}}, {\emph{arXiv} (2025) },
  [\href{https://arxiv.org/abs/2501.05206}{{\ttfamily 2501.05206}}].

\bibitem{Abdullah:2022zue}
M.~Abdullah et~al., \emph{{Coherent elastic neutrino-nucleus scattering:
  Terrestrial and astrophysical applications}}, {\emph{arXiv} (2022) },
  [\href{https://arxiv.org/abs/2203.07361}{{\ttfamily 2203.07361}}].

\bibitem{Strigari:2023}
N.~Mishra and L.~E. Strigari, \emph{Solar neutrinos with
  $\mathrm{CE}\ensuremath{\nu}\mathrm{NS}$ and flavor-dependent radiative
  corrections},
  \href{http://dx.doi.org/10.1103/PhysRevD.108.063023}{\emph{Phys. Rev. D}
  {\bfseries 108} (2023) 063023},
  [\href{https://arxiv.org/abs/2305.17827}{{\ttfamily 2305.17827}}].

\bibitem{Smirnov:2015lxy}
{\scshape Borexino} collaboration, O.~Smirnov et~al., \emph{{Measurement of
  neutrino flux from the primary proton--proton fusion process in the Sun with
  Borexino detector}},
  \href{http://dx.doi.org/10.1134/S106377961606023X}{\emph{Phys. Part. Nucl.}
  {\bfseries 47} (2016) 995--1002},
  [\href{https://arxiv.org/abs/1507.02432}{{\ttfamily 1507.02432}}].

\bibitem{Liao:2017awz}
J.~Liao, D.~Marfatia and K.~Whisnant, \emph{{Nonstandard interactions in solar
  neutrino oscillations with Hyper-Kamiokande and JUNO}},
  \href{http://dx.doi.org/10.1016/j.physletb.2017.05.054}{\emph{Phys. Lett. B}
  {\bfseries 771} (2017) 247--253},
  [\href{https://arxiv.org/abs/1704.04711}{{\ttfamily 1704.04711}}].

\bibitem{Aprile:2008bga}
E.~Aprile, A.~E. Bolotnikov, A.~L. Bolozdynya and T.~Doke, \emph{{Noble Gas
  Detectors}}.
\newblock Wiley, 2008,
  \href{http://dx.doi.org/10.1002/9783527610020}{10.1002/9783527610020}.

\bibitem{Aprile:2006kx}
E.~Aprile et~al., \emph{{Simultaneous measurement of ionization and
  scintillation from nuclear recoils in liquid xenon as target for a dark
  matter experiment}},
  \href{http://dx.doi.org/10.1103/PhysRevLett.97.081302}{\emph{Phys. Rev.
  Lett.} {\bfseries 97} (2006) 081302},
  [\href{https://arxiv.org/abs/astro-ph/0601552}{{\ttfamily
  astro-ph/0601552}}].

\bibitem{Doke:2002oab}
T.~Doke, A.~Hitachi, J.~Kikuchi, K.~Masuda, H.~Okada and E.~Shibamura,
  \emph{{Absolute Scintillation Yields in Liquid Argon and Xenon for Various
  Particles}}, \href{http://dx.doi.org/10.1143/JJAP.41.1538}{\emph{Jap. J.
  Appl. Phys.} {\bfseries 41} (2002) 1538--1545}.

\bibitem{Akerib:2016qlr}
{\scshape LUX} collaboration, D.~Akerib et~al., \emph{{Signal yields, energy
  resolution, and recombination fluctuations in liquid xenon}},
  \href{http://dx.doi.org/10.1103/PhysRevD.95.012008}{\emph{Phys. Rev. D}
  {\bfseries 95} (2017) 012008},
  [\href{https://arxiv.org/abs/1610.02076}{{\ttfamily 1610.02076}}].

\bibitem{Aprile:2017xxh}
{\scshape XENON} collaboration, E.~Aprile et~al., \emph{{Signal Yields of keV
  Electronic Recoils and Their Discrimination from Nuclear Recoils in Liquid
  Xenon}}, \href{http://dx.doi.org/10.1103/PhysRevD.97.092007}{\emph{Phys. Rev.
  D} {\bfseries 97} (2018) 092007},
  [\href{https://arxiv.org/abs/1709.10149}{{\ttfamily 1709.10149}}].

\bibitem{Akerib:2020lkv}
{\scshape LUX} collaboration, D.~S. Akerib et~al., \emph{{Discrimination of
  electronic recoils from nuclear recoils in two-phase xenon time projection
  chambers}}, \href{http://dx.doi.org/10.1103/PhysRevD.102.112002}{\emph{Phys.
  Rev. D} {\bfseries 102} (2020) 112002},
  [\href{https://arxiv.org/abs/2004.06304}{{\ttfamily 2004.06304}}].

\bibitem{Szydagis:2022ikv}
M.~Szydagis et~al., \emph{{A review of NEST models for liquid xenon and an
  exhaustive comparison with other approaches}},
  \href{http://dx.doi.org/10.3389/fdest.2024.1480975}{\emph{Frontiers in
  Detector Science and Technology} {\bfseries 2} (2025) },
  [\href{https://arxiv.org/abs/2211.10726}{{\ttfamily 2211.10726}}].

\bibitem{Araujo:2020rwg}
H.~Ara\'ujo, \emph{{Revised performance parameters of the ZEPLIN-III dark
  matter experiment}},  \href{https://arxiv.org/abs/2007.01683}{{\ttfamily
  2007.01683}}.

\bibitem{Akerib_2018}
{\scshape LUX} collaboration, D.~Akerib et~al., \emph{{Liquid xenon
  scintillation measurements and pulse shape discrimination in the LUX dark
  matter detector}},
  \href{http://dx.doi.org/10.1103/physrevd.97.112002}{\emph{Physical Review D}
  {\bfseries 97} (2018) }, [\href{https://arxiv.org/abs/1802.06162}{{\ttfamily
  1802.06162}}].

\bibitem{LZ:2019sgr}
{\scshape LZ} collaboration, D.~S. Akerib et~al., \emph{{The LUX-ZEPLIN (LZ)
  Experiment}}, \href{http://dx.doi.org/10.1016/j.nima.2019.163047}{\emph{Nucl.
  Instrum. Meth. A} {\bfseries 953} (2020) 163047},
  [\href{https://arxiv.org/abs/1910.09124}{{\ttfamily 1910.09124}}].

\bibitem{XENON:2024wpa}
{\scshape XENON} collaboration, E.~Aprile et~al., \emph{{The XENONnT Dark
  Matter Experiment}},
  \href{http://dx.doi.org/10.1140/epjc/s10052-024-12982-5}{\emph{Eur. Phys. J.
  C} {\bfseries 84} (2024) 784},
  [\href{https://arxiv.org/abs/2402.10446}{{\ttfamily 2402.10446}}].

\bibitem{Aprile:2017aty}
{\scshape XENON} collaboration, E.~Aprile et~al., \emph{{The XENON1T Dark
  Matter Experiment}},
  \href{http://dx.doi.org/10.1140/epjc/s10052-017-5326-3}{\emph{Eur. Phys. J.
  C} {\bfseries 77} (2017) 881},
  [\href{https://arxiv.org/abs/1708.07051}{{\ttfamily 1708.07051}}].

\bibitem{Aprile:2011dd}
{\scshape XENON} collaboration, E.~Aprile et~al., \emph{{The XENON100 Dark
  Matter Experiment}},
  \href{http://dx.doi.org/10.1016/j.astropartphys.2012.01.003}{\emph{Astropart.
  Phys.} {\bfseries 35} (2012) 573--590},
  [\href{https://arxiv.org/abs/1107.2155}{{\ttfamily 1107.2155}}].

\bibitem{xenoncollaboration2024neutronvetoxenonntexperiment}
{\scshape {XENON Collaboration}} collaboration, E.~Aprile et~al., \emph{{The
  neutron veto of the XENONnT experiment: Results with demineralized water}},
  {\emph{arXiv} (2024) }, [\href{https://arxiv.org/abs/2412.05264}{{\ttfamily
  2412.05264}}].

\bibitem{Plante:2022}
G.~Plante, E.~Aprile, J.~Howlett and et~al., \emph{Liquid-phase purification
  for multi-tonne xenon detectors},
  \href{http://dx.doi.org/10.1140/epjc/s10052-022-10832-w}{\emph{Eur. Phys. J.
  C} {\bfseries 82} (2022) 860},
  [\href{https://arxiv.org/abs/2205.07336}{{\ttfamily 2205.07336}}].

\bibitem{XENON:2021fkt}
{\scshape XENON} collaboration, E.~Aprile et~al., \emph{{Application and
  modeling of an online distillation method to reduce krypton and argon in
  XENON1T}}, \href{http://dx.doi.org/10.1093/ptep/ptac074}{\emph{Progress of
  Theoretical and Experimental Physics} (2021) },
  [\href{https://arxiv.org/abs/2112.12231}{{\ttfamily 2112.12231}}].

\bibitem{Murra:2022mlr}
M.~Murra, D.~Schulte, C.~Huhmann and C.~Weinheimer, \emph{{Design, construction
  and commissioning of a high-flow radon removal system for XENONnT}},
  \href{http://dx.doi.org/10.1140/epjc/s10052-022-11001-9}{\emph{Eur. Phys. J.
  C} {\bfseries 82} (2022) 1104},
  [\href{https://arxiv.org/abs/2205.11492}{{\ttfamily 2205.11492}}].

\bibitem{XENON:2022ltv}
{\scshape XENON} collaboration, E.~Aprile et~al., \emph{{Search for New Physics
  in Electronic Recoil Data from XENONnT}},
  \href{http://dx.doi.org/10.1103/PhysRevLett.129.161805}{\emph{Phys. Rev.
  Lett.} {\bfseries 129} (2022) 161805},
  [\href{https://arxiv.org/abs/2207.11330}{{\ttfamily 2207.11330}}].

\bibitem{Andrieu:2023buk}
B.~Andrieu et~al., \emph{{XeLab: a test platform for xenon TPC
  instrumentation}}, \href{http://dx.doi.org/10.22323/1.444.1420}{\emph{PoS}
  {\bfseries ICRC2023} (2023) 1420}.

\bibitem{Biondi2024}
Y.~Biondi et~al., \emph{High voltage in large structures: challenges and
  progress},  in \emph{Nagoya Workshop on Technology and Instrumentation in
  Future Liquid Noble Gas Detectors}, 2024.

\bibitem{Tomas:2018pny}
A.~Tom\'as et~al., \emph{{Study and mitigation of spurious electron emission
  from cathodic wires in noble liquid time projection chambers}},
  \href{http://dx.doi.org/10.1016/j.astropartphys.2018.07.001}{\emph{Astropart.
  Phys.} {\bfseries 103} (2018) 49--61},
  [\href{https://arxiv.org/abs/1801.07231}{{\ttfamily 1801.07231}}].

\bibitem{Linehan:2021qnb}
R.~Linehan et~al., \emph{{Design and production of the high voltage electrode
  grids and electron extraction region for the LZ dual-phase xenon time
  projection chamber}},
  \href{http://dx.doi.org/10.1016/j.nima.2021.165955}{\emph{Nucl. Instrum.
  Meth. A} {\bfseries 1031} (2022) 165955},
  [\href{https://arxiv.org/abs/2106.06622}{{\ttfamily 2106.06622}}].

\bibitem{xenoncollaboration2024fieldcage}
{\scshape XENON Collaboration} collaboration, E.~Aprile et~al., \emph{{Design
  and performance of the field cage for the XENONnT experiment}},
  \href{http://dx.doi.org/10.1140/epjc/s10052-023-12296-y}{\emph{The European
  Physical Journal C} {\bfseries 84} (2024) },
  [\href{https://arxiv.org/abs/2309.11996}{{\ttfamily 2309.11996}}].

\bibitem{LZ:2015kxe}
{\scshape LZ} collaboration, D.~S. Akerib et~al., \emph{{LUX-ZEPLIN (LZ)
  Conceptual Design Report}}, {\emph{arXiv} (2015) },
  [\href{https://arxiv.org/abs/1509.02910}{{\ttfamily 1509.02910}}].

\bibitem{LopezParedes:2018kzu}
B.~L\'opez~Paredes et~al., \emph{{Response of photomultiplier tubes to xenon
  scintillation light}},
  \href{http://dx.doi.org/10.1016/j.astropartphys.2018.04.006}{\emph{Astropart.
  Phys.} {\bfseries 102} (2018) 56--66},
  [\href{https://arxiv.org/abs/1801.01597}{{\ttfamily 1801.01597}}].

\bibitem{Baudis:2013xva}
L.~Baudis et~al., \emph{{Performance of the Hamamatsu R11410 Photomultiplier
  Tube in cryogenic Xenon Environments}},
  \href{http://dx.doi.org/10.1088/1748-0221/8/04/P04026}{\emph{JINST}
  {\bfseries 8} (2013) P04026},
  [\href{https://arxiv.org/abs/1303.0226}{{\ttfamily 1303.0226}}].

\bibitem{Barrow_2017}
P.~Barrow et~al., \emph{{Qualification tests of the R11410-21 photomultiplier
  tubes for the {XENON}1T detector}},
  \href{http://dx.doi.org/10.1088/1748-0221/12/01/p01024}{\emph{Journal of
  Instrumentation} {\bfseries 12} (jan, 2017) P01024--P01024}.

\bibitem{Antochi:2021wik}
V.~C. Antochi et~al., \emph{{Improved quality tests of R11410-21
  photomultiplier tubes for the XENONnT experiment}},
  \href{http://dx.doi.org/10.1088/1748-0221/16/08/P08033}{\emph{JINST}
  {\bfseries 16} (2021) P08033},
  [\href{https://arxiv.org/abs/2104.15051}{{\ttfamily 2104.15051}}].

\bibitem{XENON:2015ara}
{\scshape XENON} collaboration, E.~Aprile et~al., \emph{{Lowering the
  radioactivity of the photomultiplier tubes for the XENON1T dark matter
  experiment}},
  \href{http://dx.doi.org/10.1140/epjc/s10052-015-3657-5}{\emph{Eur. Phys. J.
  C} {\bfseries 75} (2015) 546},
  [\href{https://arxiv.org/abs/1503.07698}{{\ttfamily 1503.07698}}].

\bibitem{Akerib:2020com}
{\scshape LZ} collaboration, D.~S. Akerib et~al., \emph{{The LUX-ZEPLIN (LZ)
  radioactivity and cleanliness control programs}},
  \href{http://dx.doi.org/10.1140/epjc/s10052-020-8420-x}{\emph{Eur. Phys. J.
  C} {\bfseries 80} (2020) 1044},
  [\href{https://arxiv.org/abs/2006.02506}{{\ttfamily 2006.02506}}].

\bibitem{Aprile:2020vtw}
{\scshape XENON} collaboration, E.~Aprile et~al., \emph{{Projected WIMP
  sensitivity of the XENONnT dark matter experiment}},
  \href{http://dx.doi.org/10.1088/1475-7516/2020/11/031}{\emph{JCAP} {\bfseries
  11} (2020) 031}, [\href{https://arxiv.org/abs/2007.08796}{{\ttfamily
  2007.08796}}].

\bibitem{Mount:2017qzi}
B.~J. Mount et~al., \emph{{LUX-ZEPLIN (LZ) Technical Design Report}},
  {\emph{arXiv} (2017) }, [\href{https://arxiv.org/abs/1703.09144}{{\ttfamily
  1703.09144}}].

\bibitem{Brown_2024}
A.~Brown et~al., \emph{{PANCAKE: a large-diameter cryogenic test platform with
  a flat floor for next generation multi-tonne liquid xenon detectors}},
  \href{http://dx.doi.org/10.1088/1748-0221/19/05/p05018}{\emph{JINST}
  {\bfseries 19} (2024) P05018},
  [\href{https://arxiv.org/abs/2312.14785}{{\ttfamily 2312.14785}}].

\bibitem{Akerib:2011rr}
{\scshape LUX} collaboration, D.~S. Akerib et~al., \emph{{Radio-assay of
  Titanium samples for the LUX Experiment}}, {\emph{arXiv} (2011) },
  [\href{https://arxiv.org/abs/1112.1376}{{\ttfamily 1112.1376}}].

\bibitem{LZ:2017iwn}
{\scshape LZ} collaboration, D.~S. Akerib et~al., \emph{{Identification of
  Radiopure Titanium for the LZ Dark Matter Experiment and Future Rare Event
  Searches}},
  \href{http://dx.doi.org/10.1016/j.astropartphys.2017.09.002}{\emph{Astropart.
  Phys.} {\bfseries 96} (2017) 1--10},
  [\href{https://arxiv.org/abs/1702.02646}{{\ttfamily 1702.02646}}].

\bibitem{Jorg:2022spz}
F.~J\"org, \emph{{From $^{222}$Rn measurements in XENONnT and HeXe to radon
  mitigation in future liquid xenon experiments}}.
\newblock PhD thesis, Heidelberg U., 2022.
\newblock 10.11588/heidok.00031915.

\bibitem{marti2020evaluationgadoliniumsactionwater}
L.~Marti et~al., \emph{{Evaluation of Gadolinium's Action on Water Cherenkov
  Detector Systems with EGADS}}, {\emph{arXiv} (2020) },
  [\href{https://arxiv.org/abs/1908.11532}{{\ttfamily 1908.11532}}].

\bibitem{Abe_2022}
K.~Abe et~al., \emph{{First gadolinium loading to Super-Kamiokande}},
  \href{http://dx.doi.org/10.1016/j.nima.2021.166248}{\emph{Nuclear Instruments
  and Methods in Physics Research Section A: Accelerators, Spectrometers,
  Detectors and Associated Equipment} {\bfseries 1027} (2022) 166248},
  [\href{https://arxiv.org/abs/2109.00360}{{\ttfamily 2109.00360}}].

\bibitem{Yeh:2011}
M.~Yeh, S.~Hans, R.~Beriguete, R.~Rosero and L.~Hu, \emph{{A new water-based
  liquid scintillator and potential applications}},
  \href{http://dx.doi.org/10.1016/j.nima.2011.08.040}{\emph{Nucl. Instrum.
  Meth. A} {\bfseries 660} (2011) 51--56}.

\bibitem{Xiang:2024jfp}
X.~Xiang et~al., \emph{{Design, construction, and operation of a 1-ton
  Water-based Liquid scintillator detector at Brookhaven National Laboratory}},
  \href{http://dx.doi.org/10.1088/1748-0221/19/06/P06033}{\emph{JINST}
  {\bfseries 19} (2024) P06033},
  [\href{https://arxiv.org/abs/2403.13231}{{\ttfamily 2403.13231}}].

\bibitem{button2023}
L.~Kneale, \emph{Button: A technology testbed for future (anti)neutrino
  detection},  in \emph{IoP HEPP \& APP Conference}, April, 2023.

\bibitem{Anderson:2022lbb}
T.~Anderson et~al., \emph{{EOS: conceptual design for a demonstrator of hybrid
  optical detector technology}},
  \href{http://dx.doi.org/10.1088/1748-0221/18/02/P02009}{\emph{JINST}
  {\bfseries 18} (2023) P02009},
  [\href{https://arxiv.org/abs/2211.11969}{{\ttfamily 2211.11969}}].

\bibitem{Altenmuller_2021}
K.~Altenmuller et~al., \emph{{Purification Efficiency and Radon Emanation of
  Gas Purifiers used with Pure and Binary Gas Mixtures for Gaseous Dark Matter
  Detectors}},
  \href{http://dx.doi.org/10.1109/nss/mic44867.2021.9875870}{\emph{2021 IEEE
  Nuclear Science Symposium and Medical Imaging Conference (NSS/MIC)} (2021) },
  [\href{https://arxiv.org/abs/2211.10148}{{\ttfamily 2211.10148}}].

\bibitem{saes}
{SAES Pure Gas}. http://www.saespuregas. com/Library/specifications-brochures/.

\bibitem{LUX:2012kmp}
{\scshape LUX} collaboration, D.~S. Akerib et~al., \emph{{The Large Underground
  Xenon (LUX) Experiment}},
  \href{http://dx.doi.org/10.1016/j.nima.2012.11.135}{\emph{Nucl. Instrum.
  Meth. A} {\bfseries 704} (2013) 111--126},
  [\href{https://arxiv.org/abs/1211.3788}{{\ttfamily 1211.3788}}].

\bibitem{Aprile:2019bbb}
{\scshape XENON} collaboration, E.~Aprile et~al., \emph{{XENON1T Dark Matter
  Data Analysis: Signal Reconstruction, Calibration and Event Selection}},
  \href{http://dx.doi.org/10.1103/PhysRevD.100.052014}{\emph{Phys. Rev. D}
  {\bfseries 100} (2019) 052014},
  [\href{https://arxiv.org/abs/1906.04717}{{\ttfamily 1906.04717}}].

\bibitem{Akerib:2017vbi}
{\scshape LUX} collaboration, D.~S. Akerib et~al., \emph{{Calibration, event
  reconstruction, data analysis, and limit calculation for the LUX dark matter
  experiment}}, \href{http://dx.doi.org/10.1103/PhysRevD.97.102008}{\emph{Phys.
  Rev. D} {\bfseries 97} (2018) 102008},
  [\href{https://arxiv.org/abs/1712.05696}{{\ttfamily 1712.05696}}].

\bibitem{Aalbers_2024_calibration}
{\scshape LZ} collaboration, J.~Aalbers et~al., \emph{{The design,
  implementation, and performance of the LZ calibration systems}},
  \href{http://dx.doi.org/10.1088/1748-0221/19/08/p08027}{\emph{Journal of
  Instrumentation} {\bfseries 19} (2024) P08027},
  [\href{https://arxiv.org/abs/2406.12874}{{\ttfamily 2406.12874}}].

\bibitem{Boulton:2017hub}
E.~Boulton et~al., \emph{{Calibration of a two-phase xenon time projection
  chamber with a $^{37}$Ar source}},
  \href{http://dx.doi.org/10.1088/1748-0221/12/08/P08004}{\emph{JINST}
  {\bfseries 12} (2017) P08004},
  [\href{https://arxiv.org/abs/1705.08958}{{\ttfamily 1705.08958}}].

\bibitem{Manalaysay:2009yq}
A.~Manalaysay et~al., \emph{{Spatially uniform calibration of a liquid xenon
  detector at low energies using 83m-Kr}},
  \href{http://dx.doi.org/10.1063/1.3436636}{\emph{Rev. Sci. Instrum.}
  {\bfseries 81} (2010) 073303},
  [\href{https://arxiv.org/abs/0908.0616}{{\ttfamily 0908.0616}}].

\bibitem{Aprile:2016pmc}
{\scshape XENON} collaboration, E.~Aprile et~al., \emph{{Results from a
  Calibration of XENON100 Using a Source of Dissolved Radon-220}},
  \href{http://dx.doi.org/10.1103/PhysRevD.95.072008}{\emph{Phys. Rev. D}
  {\bfseries 95} (2017) 072008},
  [\href{https://arxiv.org/abs/1611.03585}{{\ttfamily 1611.03585}}].

\bibitem{Lang:2016zde}
R.~F. Lang et~al., \emph{{A $^{220}$Rn source for the calibration of
  low-background experiments}},
  \href{http://dx.doi.org/10.1088/1748-0221/11/04/P04004}{\emph{JINST}
  {\bfseries 11} (2016) P04004},
  [\href{https://arxiv.org/abs/1602.01138}{{\ttfamily 1602.01138}}].

\bibitem{J_rg_2023_Rn220}
F.~Jörg, S.~Li, J.~Schreiner, H.~Simgen and R.~F. Lang,
  \emph{{Characterization of a 220Rn source for low-energy electronic recoil
  calibration of the XENONnT detector}},
  \href{http://dx.doi.org/10.1088/1748-0221/18/11/p11009}{\emph{Journal of
  Instrumentation} {\bfseries 18} (Nov., 2023) P11009},
  [\href{https://arxiv.org/abs/2306.05673}{{\ttfamily 2306.05673}}].

\bibitem{J_rg_2023_Rn222}
F.~Jörg, G.~Eurin and H.~Simgen, \emph{{Production and characterization of a
  222Rn-emanating stainless steel source}},
  \href{http://dx.doi.org/10.1016/j.apradiso.2023.110666}{\emph{Applied
  Radiation and Isotopes} {\bfseries 194} (Apr., 2023) 110666},
  [\href{https://arxiv.org/abs/2205.15926}{{\ttfamily 2205.15926}}].

\bibitem{Akerib:2015wdi}
{\scshape LUX} collaboration, D.~Akerib et~al., \emph{{Tritium calibration of
  the LUX dark matter experiment}},
  \href{http://dx.doi.org/10.1103/PhysRevD.93.072009}{\emph{Phys. Rev. D}
  {\bfseries 93} (2016) 072009},
  [\href{https://arxiv.org/abs/1512.03133}{{\ttfamily 1512.03133}}].

\bibitem{Akerib:2016mzi}
{\scshape LUX} collaboration, D.~Akerib et~al., \emph{{Low-energy (0.7-74 keV)
  nuclear recoil calibration of the LUX dark matter experiment using D-D
  neutron scattering kinematics}}, {\emph{arXiv} (2016) },
  [\href{https://arxiv.org/abs/1608.05381}{{\ttfamily 1608.05381}}].

\bibitem{xenoncollaboration2024lowenergynuclearrecoilcalibration}
{\scshape XENON Collaboration} collaboration, E.~Aprile et~al.,
  \emph{{Low-Energy Nuclear Recoil Calibration of XENONnT with a $^{88}$YBe
  Photoneutron Source}},  \href{https://arxiv.org/abs/2412.10451}{{\ttfamily
  2412.10451}}.

\bibitem{Collar:2013xva}
J.~I. Collar, \emph{{Applications of an $^{88}Y/Be$ photo-neutron calibration
  source to Dark Matter and Neutrino Experiments}},
  \href{http://dx.doi.org/10.1103/PhysRevLett.110.211101}{\emph{Phys. Rev.
  Lett.} {\bfseries 110} (2013) 211101},
  [\href{https://arxiv.org/abs/1303.2686}{{\ttfamily 1303.2686}}].

\bibitem{aalbers2024data}
{\scshape LZ} collaboration, J.~Aalbers et~al., \emph{{The Data Acquisition
  System of the LZ Dark Matter Detector: FADR}},
  \href{http://dx.doi.org/https://doi.org/10.1016/j.nima.2024.169712}{\emph{Nuclear
  Instruments and Methods in Physics Research Section A: Accelerators,
  Spectrometers, Detectors and Associated Equipment} {\bfseries 1068} (2024)
  169712}, [\href{https://arxiv.org/abs/2405.14732}{{\ttfamily 2405.14732}}].

\bibitem{XENON:2019bth}
{\scshape XENON} collaboration, E.~Aprile et~al., \emph{{The XENON1T Data
  Acquisition System}},
  \href{http://dx.doi.org/10.1088/1748-0221/14/07/P07016}{\emph{JINST}
  {\bfseries 14} (2019) P07016},
  [\href{https://arxiv.org/abs/1906.00819}{{\ttfamily 1906.00819}}].

\bibitem{XENON:2022vye}
{\scshape XENON} collaboration, E.~Aprile et~al., \emph{{The Triggerless Data
  Acquisition System of the XENONnT Experiment}},
  \href{http://dx.doi.org/10.1088/1748-0221/18/07/p07054}{\emph{Journal of
  Instrumentation} {\bfseries 18} (12, 2022) },
  [\href{https://arxiv.org/abs/2212.11032}{{\ttfamily 2212.11032}}].

\bibitem{caen_dpp-daw}
{CAEN}, ``{DPP-DAW - Digital Pulse Processing with Dynamic Acquisition
  Window}.'' https://www.caen.it/products/dpp-daw/.

\bibitem{jiang2024ultrafasttransformersfpgas}
Z.~Jiang et~al., \emph{{Ultra Fast Transformers on FPGAs for Particle Physics
  Experiments}}, {\emph{Machine Learning and the Physical Sciences Workshop}
  (2024) }, [\href{https://arxiv.org/abs/2402.01047}{{\ttfamily 2402.01047}}].

\bibitem{Roberts:2022ezy}
A.~Roberts et~al., \emph{{Dark-matter And Neutrino Computation Explored (DANCE)
  Community Input to Snowmass}},  in \emph{{Snowmass 2021}}, 2022.
\newblock \href{https://arxiv.org/abs/2203.08338}{{\ttfamily 2203.08338}}.

\bibitem{GEANT4:2002zbu}
{\scshape GEANT4} collaboration, S.~Agostinelli et~al., \emph{{GEANT4--a
  simulation toolkit}},
  \href{http://dx.doi.org/10.1016/S0168-9002(03)01368-8}{\emph{Nucl. Instrum.
  Meth. A} {\bfseries 506} (2003) 250--303}.

\bibitem{Szydagis:2011tk}
M.~Szydagis et~al., \emph{{NEST: A Comprehensive Model for Scintillation Yield
  in Liquid Xenon}},
  \href{http://dx.doi.org/10.1088/1748-0221/6/10/P10002}{\emph{JINST}
  {\bfseries 6} (2011) P10002},
  [\href{https://arxiv.org/abs/1106.1613}{{\ttfamily 1106.1613}}].

\bibitem{Szydagis:2013sih}
M.~Szydagis, A.~Fyhrie, D.~Thorngren and M.~Tripathi, \emph{{Enhancement of
  NEST Capabilities for Simulating Low-Energy Recoils in Liquid Xenon}},
  \href{http://dx.doi.org/10.1088/1748-0221/8/10/C10003}{\emph{JINST}
  {\bfseries 8} (2013) C10003},
  [\href{https://arxiv.org/abs/1307.6601}{{\ttfamily 1307.6601}}].

\bibitem{Akerib:2021ap}
{\scshape LUX-ZEPLIN} collaboration, D.~S. Akerib et~al., \emph{{Simulations of
  Events for the LUX-ZEPLIN (LZ) Dark Matter Experiment}},
  \href{http://dx.doi.org/10.1016/j.astropartphys.2020.102480}{\emph{Astropart.
  Phys.} {\bfseries 125} (2021) 102480},
  [\href{https://arxiv.org/abs/2001.09363}{{\ttfamily 2001.09363}}].

\bibitem{Althueser_2022}
L.~Althueser et~al., \emph{{GPU-based optical simulation of the DARWIN
  detector}},
  \href{http://dx.doi.org/10.1088/1748-0221/17/07/p07018}{\emph{Journal of
  Instrumentation} {\bfseries 17} (July, 2022) P07018},
  [\href{https://arxiv.org/abs/2203.14354}{{\ttfamily 2203.14354}}].

\bibitem{Rucio_2019}
M.~Barisits et~al., \emph{{Rucio: Scientific Data Management}},
  \href{http://dx.doi.org/10.1007/s41781-019-0026-3}{\emph{Computing and
  Software for Big Science} {\bfseries 3} (2019) },
  [\href{https://arxiv.org/abs/1902.09857}{{\ttfamily 1902.09857}}].

\bibitem{DIRAC_2014}
A.~Tsaregorodtsev, \emph{{DIRAC distributed computing services}},
  \href{http://dx.doi.org/10.1088/1742-6596/513/3/032096}{\emph{Journal of
  Physics: Conference Series} {\bfseries 513} (06, 2014) }.

\bibitem{Globus_2011}
I.~Foster, \emph{{Globus Online: Accelerating and Democratizing Science through
  Cloud-Based Services}},
  \href{http://dx.doi.org/10.1109/MIC.2011.64}{\emph{IEEE Internet Computing}
  {\bfseries 15} (2011) 70--73}.

\bibitem{Aalbers_2023}
{\scshape LZ} collaboration, J.~Aalbers et~al., \emph{{Background determination
  for the LUX-ZEPLIN dark matter experiment}},
  \href{http://dx.doi.org/10.1103/physrevd.108.012010}{\emph{Physical Review D}
  {\bfseries 108} (2023) }, [\href{https://arxiv.org/abs/2211.17120}{{\ttfamily
  2211.17120}}].

\bibitem{XENON:2024_backgrounds}
{\scshape {XENON}} collaboration, E.~Aprile et~al., \emph{{XENONnT WIMP Search:
  Signal \& Background Modeling and Statistical Inference}}, {\emph{arXiv}
  (2024) }, [\href{https://arxiv.org/abs/2406.13638}{{\ttfamily 2406.13638}}].

\bibitem{Ames:2023}
A.~Ames, \emph{{Krypton removal via gas chromatography for the LZ experiment}},
  {\emph{AIP Conf. Proc.} {\bfseries 2908} (2023) 070001}.

\bibitem{LaFerriere:2015}
B.~LaFerriere, T.~Maiti, I.~Arnquist and E.~Hoppe, \emph{{A novel assay method
  for the trace determination of Th and U in copper and lead using inductively
  coupled plasma mass spectrometry}},
  \href{http://dx.doi.org/https://doi.org/10.1016/j.nima.2014.11.052}{\emph{NIM-A}
  {\bfseries 775} (2015) 93--98}.

\bibitem{DOBSON201825}
J.~Dobson, C.~Ghag and L.~Manenti, \emph{{Ultra-low background mass
  spectrometry for rare-event searches}},
  \href{http://dx.doi.org/https://doi.org/10.1016/j.nima.2017.10.014}{\emph{NIM-A}
  {\bfseries 879} (2018) 25--30},
  [\href{https://arxiv.org/abs/1708.08860}{{\ttfamily 1708.08860}}].

\bibitem{XENON:2021mrg}
{\scshape XENON} collaboration, E.~Aprile et~al., \emph{{Material radiopurity
  control in the XENONnT experiment}},
  \href{http://dx.doi.org/10.1140/epjc/s10052-022-10345-6}{\emph{Eur. Phys. J.
  C} {\bfseries 82} (2022) 599},
  [\href{https://arxiv.org/abs/2112.05629}{{\ttfamily 2112.05629}}].

\bibitem{Scovell:2018ap}
P.~Scovell et~al., \emph{{Low-background gamma spectroscopy at the Boulby
  Underground Laboratory}},
  \href{http://dx.doi.org/https://doi.org/10.1016/j.astropartphys.2017.11.006}{\emph{Astroparticle
  Physics} {\bfseries 97} (2018) 160 -- 173},
  [\href{https://arxiv.org/abs/1708.06086}{{\ttfamily 1708.06086}}].

\bibitem{mount:black_2017}
B.~J. Mount et~al., \emph{Black {Hills} {State} {University} {Underground}
  {Campus}},
  \href{http://dx.doi.org/10.1016/j.apradiso.2017.02.025}{\emph{Applied
  Radiation and Isotopes} {\bfseries 126} (Aug., 2017) 130--133}.

\bibitem{Baudis:2011am}
L.~Baudis et~al., \emph{{Gator: a low-background counting facility at the Gran
  Sasso Underground Laboratory}},
  \href{http://dx.doi.org/10.1088/1748-0221/6/08/P08010}{\emph{JINST}
  {\bfseries 6} (2011) P08010},
  [\href{https://arxiv.org/abs/1103.2125}{{\ttfamily 1103.2125}}].

\bibitem{Araujo:2022kip}
G.~R. Araujo, L.~Baudis, Y.~Biondi, A.~Bismark and M.~Galloway, \emph{{The
  upgraded low-background germanium counting facility Gator for
  high-sensitivity \ensuremath{\gamma}-ray spectrometry}},
  \href{http://dx.doi.org/10.1088/1748-0221/17/08/P08010}{\emph{JINST}
  {\bfseries 17} (2022) P08010},
  [\href{https://arxiv.org/abs/2204.12478}{{\ttfamily 2204.12478}}].

\bibitem{Garcia:2022jdt}
D.~R. Garc\'\i{}a et~al., \emph{{GeMSE: a low-background facility for
  gamma-spectrometry at moderate rock overburden}},
  \href{http://dx.doi.org/10.1088/1748-0221/17/04/P04005}{\emph{JINST}
  {\bfseries 17} (2022) P04005},
  [\href{https://arxiv.org/abs/2202.06540}{{\ttfamily 2202.06540}}].

\bibitem{Heusser:2006}
G.~Heusser, M.~Laubenstein and H.~Neder, \emph{{Low-level germanium gamma-ray
  spectrometry at the $\mu$Bq/kg level and future developments towards higher
  sensitivity}},
  \href{http://dx.doi.org/https://doi.org/10.1016/S1569-4860(05)08039-3}{\emph{Radioactivity
  in the Environment} {\bfseries 8} (2006) 495--510}.

\bibitem{Heusser:2015ifa}
G.~Heusser et~al., \emph{{GIOVE - A new detector setup for high sensitivity
  germanium spectroscopy at shallow depth}},
  \href{http://dx.doi.org/10.1140/epjc/s10052-015-3704-2}{\emph{Eur. Phys. J.
  C} {\bfseries 75} (2015) 531},
  [\href{https://arxiv.org/abs/1507.03319}{{\ttfamily 1507.03319}}].

\bibitem{LEONARD2017169}
D.~Leonard et~al., \emph{{Trace radioactive impurities in final construction
  materials for EXO-200}},
  \href{http://dx.doi.org/https://doi.org/10.1016/j.nima.2017.04.049}{\emph{NIM-A}
  {\bfseries 871} (2017) 169--179},
  [\href{https://arxiv.org/abs/1703.10799}{{\ttfamily 1703.10799}}].

\bibitem{LEONARD2008490}
D.~Leonard et~al., \emph{{Systematic study of trace radioactive impurities in
  candidate construction materials for EXO-200}},
  \href{http://dx.doi.org/https://doi.org/10.1016/j.nima.2008.03.001}{\emph{NIM-A}
  {\bfseries 591} (2008) 490--509},
  [\href{https://arxiv.org/abs/0709.4524}{{\ttfamily 0709.4524}}].

\bibitem{stenstrom2020identifying}
K.~E. Stenstr{\"o}m et~al., \emph{{Identifying radiologically important
  ESS-specific radionuclides and relevant detection methods}}, {\emph{Technical
  Report SSM 2020: 08} (2020) }.

\bibitem{XENON:2016bmq}
{\scshape XENON} collaboration, E.~Aprile et~al., \emph{{Removing krypton from
  xenon by cryogenic distillation to the ppq level}},
  \href{http://dx.doi.org/10.1140/epjc/s10052-017-4757-1}{\emph{Eur. Phys. J.
  C} {\bfseries 77} (2017) 275},
  [\href{https://arxiv.org/abs/1612.04284}{{\ttfamily 1612.04284}}].

\bibitem{Dobi:2011vc}
A.~Dobi, C.~G. Davis, C.~Hall, T.~Langford, S.~Slutsky and Y.-R. Yen,
  \emph{{Detection of krypton in xenon for dark matter applications}},
  \href{http://dx.doi.org/10.1016/j.nima.2011.11.043}{\emph{Nucl. Instrum.
  Meth. A} {\bfseries 665} (2011) 1--6},
  [\href{https://arxiv.org/abs/1103.2714}{{\ttfamily 1103.2714}}].

\bibitem{Lindemann:2013kna}
S.~Lindemann and H.~Simgen, \emph{{Krypton assay in xenon at the ppq level
  using a gas chromatographic system and mass spectrometer}},
  \href{http://dx.doi.org/10.1140/epjc/s10052-014-2746-1}{\emph{Eur. Phys. J.
  C} {\bfseries 74} (2014) 2746},
  [\href{https://arxiv.org/abs/1308.4806}{{\ttfamily 1308.4806}}].

\bibitem{guida2025}
M.~Guida, Y.-T. Lin and H.~Simgen, \emph{{Improved and automated krypton assay
  for low-background xenon detectors with Auto-RGMS}},
  \href{http://dx.doi.org/10.1140/epjc/s10052-025-14262-2}{\emph{Eur. Phys. J.
  C} {\bfseries 85} (2025) 576},
  [\href{https://arxiv.org/abs/2501.10993}{{\ttfamily 2501.10993}}].

\bibitem{XENON:2020fbs}
{\scshape XENON} collaboration, E.~Aprile et~al., \emph{{$^{222}$Rn emanation
  measurements for the XENON1T experiment}},
  \href{http://dx.doi.org/10.1140/epjc/s10052-020-08777-z}{\emph{Eur. Phys. J.
  C} {\bfseries 81} (2021) 337},
  [\href{https://arxiv.org/abs/2009.13981}{{\ttfamily 2009.13981}}].

\bibitem{Arthurs_2021}
M.~Arthurs, D.~Huang, C.~Amarasinghe, E.~Miller and W.~Lorenzon,
  \emph{{Performance study of charcoal-based radon reduction systems for
  ultraclean rare event detectors}},
  \href{http://dx.doi.org/10.1088/1748-0221/16/07/P07047}{\emph{Journal of
  Instrumentation} {\bfseries 16} (2021) P07047},
  [\href{https://arxiv.org/abs/2009.06069}{{\ttfamily 2009.06069}}].

\bibitem{Aprile:2017kop}
{\scshape XENON} collaboration, E.~Aprile et~al., \emph{{Online $^{222}$Rn
  removal by cryogenic distillation in the XENON100 experiment}},
  \href{http://dx.doi.org/10.1140/epjc/s10052-017-4902-x}{\emph{Eur. Phys. J.
  C} {\bfseries 77} (2017) 358},
  [\href{https://arxiv.org/abs/1702.06942}{{\ttfamily 1702.06942}}].

\bibitem{aprile2025radonremovalxenonntsolar}
{\scshape XENON} collaboration, E.~Aprile et~al., \emph{{Radon Removal in
  XENONnT down to the Solar Neutrino Level}}, {\emph{arXiv} (2025) },
  [\href{https://arxiv.org/abs/2502.04209}{{\ttfamily 2502.04209}}].

\bibitem{Dierle_2023}
J.~Dierle et~al., \emph{{Reduction of $^{222}Rn$-induced backgrounds in a
  hermetic dual-phase xenon time projection chamber}},
  \href{http://dx.doi.org/10.1140/epjc/s10052-022-11151-w}{\emph{The European
  Physical Journal C} {\bfseries 83} (2023) }.

\bibitem{Sato2020development}
K.~Sato et~al., \emph{{Development of Dual-phase Xenon TPC with a Quartz
  Chamber for Direct Dark Matter Search}}, {\emph{arXiv} (2020) },
  [\href{https://arxiv.org/abs/1910.13831}{{\ttfamily 1910.13831}}].

\bibitem{XENON:2024qvh}
{\scshape XENON} collaboration, E.~Aprile et~al., \emph{{Offline tagging of
  radon-induced backgrounds in XENON1T and applicability to other liquid xenon
  detectors}}, \href{http://dx.doi.org/10.1103/PhysRevD.110.012011}{\emph{Phys.
  Rev. D} {\bfseries 110} (2024) 012011},
  [\href{https://arxiv.org/abs/2403.14878}{{\ttfamily 2403.14878}}].

\bibitem{jacobi_activity_1972}
W.~Jacobi, \emph{{Activity and potential alpha-energy of 222 radon and 220
  radon-daughters in different air atmospheres}},
  \href{http://dx.doi.org/10.1097/00004032-197205000-00002}{\emph{Health
  Physics} {\bfseries 22} (1972) 441--450}.

\bibitem{meng_new_2020}
Y.~Meng, J.~Busenitz and A.~Piepke, \emph{A new method for evaluating the
  effectiveness of plastic packaging against radon penetration},
  \href{http://dx.doi.org/10.1016/j.apradiso.2019.108963}{\emph{Applied
  Radiation and Isotopes} {\bfseries 156} (2020) 108963},
  [\href{https://arxiv.org/abs/1903.02643}{{\ttfamily 1903.02643}}].

\bibitem{schnee_screening_2007}
R.~W. Schnee, Z.~Ahmed, S.~R. Golwala, D.~R. Grant and K.~Poinar,
  \emph{Screening {Surface} {Contamination} with {BetaCage}},
  \href{http://dx.doi.org/10.1063/1.2722063}{\emph{AIP Conference Proceedings}
  {\bfseries 897} (2007) 20--25}.

\bibitem{Baudis:2021ipf}
L.~Baudis et~al., \emph{{Design and construction of Xenoscope \textemdash{} a
  full-scale vertical demonstrator for the DARWIN observatory}},
  \href{http://dx.doi.org/10.1088/1748-0221/16/08/P08052}{\emph{JINST}
  {\bfseries 16} (2021) P08052},
  [\href{https://arxiv.org/abs/2105.13829}{{\ttfamily 2105.13829}}].

\bibitem{Abe_2013}
K.~Abe et~al., \emph{{XMASS detector}},
  \href{http://dx.doi.org/10.1016/j.nima.2013.03.059}{\emph{Nuclear Instruments
  and Methods in Physics Research Section A: Accelerators, Spectrometers,
  Detectors and Associated Equipment} {\bfseries 716} (2013) 78–85},
  [\href{https://arxiv.org/abs/1301.2815}{{\ttfamily 1301.2815}}].

\bibitem{Aprile:2019xxb}
{\scshape XENON} collaboration, E.~Aprile et~al., \emph{{Light Dark Matter
  Search with Ionization Signals in XENON1T}},
  \href{http://dx.doi.org/10.1103/PhysRevLett.123.251801}{\emph{Phys. Rev.
  Lett.} {\bfseries 123} (2019) 251801},
  [\href{https://arxiv.org/abs/1907.11485}{{\ttfamily 1907.11485}}].

\bibitem{XENON:2021myl}
{\scshape XENON} collaboration, E.~Aprile et~al., \emph{{Emission of Single and
  Few Electrons in XENON1T and Limits on Light Dark Matter}},
  \href{http://dx.doi.org/10.1103/physrevd.106.022001}{\emph{Physical Review D}
  {\bfseries 106} (12, 2021) },
  [\href{https://arxiv.org/abs/2112.12116}{{\ttfamily 2112.12116}}].

\bibitem{Kuger:2021sxn}
F.~Kuger, J.~Dierle, H.~Fischer, M.~Schumann and F.~Toschi, \emph{{Prospects of
  charge signal analyses in liquid xenon TPCs with proportional scintillation
  in the liquid phase}},
  \href{http://dx.doi.org/10.1088/1748-0221/17/03/P03027}{\emph{JINST}
  {\bfseries 17} (2022) P03027},
  [\href{https://arxiv.org/abs/2112.11844}{{\ttfamily 2112.11844}}].

\end{thebibliography}\endgroup

\end{document}